\documentclass[journal]{IEEEtran}
\usepackage[utf8]{inputenc}
\usepackage[T1]{fontenc}
\usepackage{stfloats}
\usepackage{url}
\usepackage{bibentry}  
\usepackage{cite}
\usepackage{verbatim}
\usepackage{grffile}
\usepackage{algorithm}
\usepackage{algpseudocode}
\usepackage{tabularx}
\usepackage{xcolor,soul,framed} 
\colorlet{shadecolor}{yellow}
\usepackage{amsmath,amssymb,amsfonts}
\usepackage{graphicx}
\usepackage{bm}
\usepackage{textcomp}
\usepackage{mathtools}

\usepackage{mathrsfs}
\usepackage{enumitem}
\usepackage{epstopdf}
\usepackage{subcaption}
\usepackage{float}
\usepackage{multirow}
\usepackage{multicol}
\usepackage{array}
\usepackage{pgfplots}
\usepackage{hyperref}
\usepackage{cancel}
\usepackage{titlesec}
\usepackage{bbm}
\usepackage{tikz}
\usetikzlibrary{shapes,calc}
\usepackage{anyfontsize}
\usepackage{booktabs}
\usepackage{siunitx}

\newtheorem{remark}{\textbf{Remark}}

\begin{document}

\title{Fronthaul Network Planning for Hierarchical and Radio-Stripes-Enabled CF-mMIMO in O-RAN}

\author{Anas~S.~Mohammed, Krishnendu~S.~Tharakan, \IEEEmembership{Member, IEEE}, Hussein~A.~Ammar, \IEEEmembership{Member, IEEE}, Hesham~ElSawy, \IEEEmembership{Senior Member, IEEE}, and Hossam~S.~Hassanein, \IEEEmembership{Fellow, IEEE}\thanks{A. S. Mohammed is with the Department of Electrical and Computer Engineering, Queen’s University, Kingston, Canada. E-mail: anas.m@queensu.ca.} \thanks{K. S. Tharakan is with the School of Electrical Engineering and Computer Science, KTH Royal Institute of Technology, Stockholm, Sweden. E-mail: tharakan@kth.se.}
\thanks{H. A. Ammar is with the Department of Electrical and Computer Engineering, Royal Military College of Canada, Kingston, Canada. E-mail: hussein.ammar@rmc.ca.}
\thanks{H. ElSawy and H. S. Hassanein are with the School of Computing, Queen’s University, Kingston, Canada. E-mail: hesham.elsawy@queensu.ca, hossam.hassanein@queensu.ca.}}
\markboth{%Journal of \LaTeX\ Class Files,~Vol.~14, No.~8, August~2015}%
}
{Shell \MakeLowercase{\textit{et al.}}: Bare Demo of IEEEtran.cls for IEEE Journals}

\maketitle

\maketitle
\begin{abstract}
The deployment of ultra-dense networks (UDNs), particularly cell-free massive MIMO (CF-mMIMO), is mainly hindered by costly and capacity-limited fronthaul links. This work proposes a two-tiered optimization framework for cost-effective hybrid fronthaul planning, comprising a Near-Optimal Fronthaul Association and Configuration (NOFAC) algorithm in the first tier and an Integer Linear Program (ILP) in the second, integrating fiber optics, millimeter-wave (mmWave), and free-space optics (FSO) technologies. The proposed framework accommodates various functional split (FS) options (7.2x and 8), decentralized processing levels, and network configurations. We introduce the hierarchical scheme (HS) as a resilient, cost-effective fronthaul solution for CF-mMIMO and compare its performance with radio-stripes (RS)-enabled CF-mMIMO, validating both across diverse dense topologies within the open radio access network (O-RAN) architecture. Results show that the proposed framework achieves better cost-efficiency and higher capacity compared to traditional benchmark schemes such as all-fiber fronthaul network. Our key findings reveal fiber dominance in highly decentralized deployments, mmWave suitability in moderately centralized scenarios, and FSO complements both by bridging deployment gaps. Additionally, FS7.2x consistently outperforms FS8, offering greater capacity at lower cost, affirming its role as the preferred O-RAN functional split. Most importantly, our study underscores the importance of hybrid fronthaul effective planning for UDNs in minimizing infrastructural redundancy, and ensuring scalability to meet current and future traffic demands.
\end{abstract}
\begin{IEEEkeywords}
 Fronthaul, Planning, Optimization, Ultra-Dense Network (UDN), Cell-Free massive MIMO (CF-mMIMO), Radio-Stripes, mmWave, Fiber Optics, Free Space Optics (FSO).
\end{IEEEkeywords}

\section{Introduction}\label{sec:introduction}

\IEEEPARstart{A}{s} mobile networks evolve toward beyond-5G (B5G) and 6G, demand for connectivity, high throughput, and ubiquitous coverage continues to rise \cite{4P2H}. This trend has driven the adoption of ultra-dense networks (UDNs), such as cell-free massive MIMO (CF-mMIMO) and small cells, which rely on dense deployment of Access Points (APs) and Central Processing Units (CPUs) to extend coverage and improve capacity \cite{UDN,10P}. Unlike small cells, CF-mMIMO employs many simple cooperative APs that simultaneously serve users, removing cell boundaries and ensuring seamless service \cite{14P}. However, the \textit{fronthaul links} connecting APs to CPUs remain a major deployment bottleneck, making the design of cost-effective and scalable fronthaul a critical priority for 6G networks \cite{6GTakesShape}.

The transition from Distributed RAN (D-RAN) to Cloud/Open RAN (C-/O-RAN) highlights virtualization and functional split options (e.g., FS7.2x), which aim to balance fronthaul capacity and latency requirements \cite{C-RAN, O-RAN}. Still, UDNs, particularly CF-mMIMO, face challenges in scalable and cost-efficient fronthaul design \cite{10P,UDN}. Conventional wired solutions, such as fiber, provide reliability and capacity but entail high costs and poor scalability \cite{RadioStripes3}. Alternatives like Ericsson’s \textit{radio stripes (RS)} offer cost-effective wired serial connections, while wireless options such as mmWave and Free-Space Optics (FSO) promise flexibility and fast deployment but remain limited by environmental factors and capacity constraints \cite{patentericsson,mmwave_source,FSOandFiberCFMIMO}.

In this work, we propose the hierarchical scheme (HS) as an alternative fronthaul solution for CF-mMIMO, leveraging a hierarchical topology. Similar to the RS scheme, HS significantly reduces the number of required fronthaul links. However, unlike RS, HS enhances network resilience by eliminating single points of failure inherent in the RS serial connections, while effectively addressing architectural limitations in UDNs.

Recent studies have explored optimizing several operational aspects of fronthaul networks for UDNs, specifically for RS-enabled CF-mMIMO systems. For instance, \cite{RadioStripes1} proposed an optimized sequential processing algorithm for RS-enabled CF-mMIMO, enhancing signal-to-interference-plus-noise ratio (SINR) and reducing latency under limited fronthaul capacity. In \cite{RadioStripes4}, a geometric programming approach was proposed for the strategic placement and grouping of APs in RS deployments, emphasizing the importance of effective network planning. In contrast, \cite{mmwave_source} explored mmWave fronthaul for RS-enabled CF-mMIMO, demonstrating its potential capacity for UDN deployments. Nevertheless, these studies predominantly focus on isolated performance dimensions, neglecting fronthaul deployment cost, scalability, and infrastructural constraints, which are critical considerations towards realizing cost-effective and resilient UDNs \cite{UDN, MyPaper}.

Motivated by these limitations, we aim to answer the following question: \textit{\textbf{How can we architect a scalable, cost-efficient, resilient, and high-capacity fronthaul infrastructure that supports practical CF-mMIMO deployments in future UDNs?}} To address this question, we investigate the feasibility and economic viability of an optimized HS and RS-enabled hybrid fronthaul network that integrates both wired and wireless technologies, namely, fiber optics, mmWave and FSO. We show that exclusive reliance on single-technology deployment does not meet the \textit{cost-effectiveness} and \textit{performance} needed for UDNs. 

Our primary objective is to minimize the Total Cost of Ownership (TCO) of the fronthaul network while ensuring compliance with critical metrics. Our framework supports diverse UDN scenarios, including small cells \cite{MyPaper}, and CF-mMIMO systems with RS or HS topologies, adhering to contemporary RAN architectures. It accommodates various decentralized processing levels, FS options, and AP groupings, offering actionable insights for Service Providers (SPs) on cost-efficient, high-capacity and resilient UDNs deployment. The main contributions of this paper are outlined as follows:

\begin{itemize}
\item We propose a two-tiered hybrid fronthaul design for UDNs through \textbf{a)} developing a Near-Optimal Fronthaul Association and Configuration (NOFAC) algorithm for the first tier, and \textbf{b)} formulating an Integer Linear Program (ILP) for the second tier. The proposed framework accommodates various fronthaul network connection schemes such as P2P small cells, along with HS and RS-enabled CF-mMIMO within the O-RAN architecture.
\item We formulate and solve an ILP optimization framework that minimizes fronthaul TCO% the second tier
, while satisfying Quality of Service (QoS) metrics, including reliability, individual link and overall network capacities, along with fronthaul technology-specific component requirements.
\item We analyze key network parameters, including varying the number of deployed CPUs, APs groups, homogeneous and non-homogeneous FS fronthaul rates for FS option 7.2x (FS7.2x) and FS option 8 (FS8), capturing practical factors affecting cost and performance. These include association distances, FS capacity thresholds, and trade-offs between group sizes and TCO.
\item We evaluate the proposed framework against multiple planning benchmarks, including traditional all-fiber fronthaul. We also assess the deployment resilience of both HS and RS-enabled CF-mMIMO connection topologies.
\end{itemize}

This paper thus addresses a critical research gap by presenting a \textit{robust fronthaul planning} framework for UDN and CF-mMIMO deployments. The remainder of the paper is organized as follows: Section \ref{CH2} describes the system model, FS options fronthaul rates, and channel models for the candidate fronthaul technologies. Section \ref{CH3} details the fronthaul network design to construct a NOFAC HS and RS-enabled CF-mMIMO. Section \ref{CH4} presents the proposed two-tier fronthaul TCO optimization. Section \ref{CH5} discusses the numerical results, highlighting the cost and performance effectiveness of the proposed framework. Finally, Section \ref{CH6} concludes the paper.

\vspace{-6pt}

\section{System Model for Hybrid Fronthaul Networks}\label{CH2}
This section presents the system model and key network components involved in building a hybrid fronthaul network for RS- and HS-enabled CF-mMIMO systems. We determine fronthaul capacity requirements and the channel models used to evaluate the achievable capacities of the candidate fronthaul technologies. To align with modern RAN architectures, we adopt the O-RAN and C-RAN terminologies, where the CPU is referred to as the Distributed Unit (DU) henceforth~\cite{C-RAN, O-RAN}.

\subsection{Hybrid Fronthaul Network Architecture}
We consider a hybrid fronthaul network comprising APs, DUs, and fronthaul links that leverage either fiber, mmWave or FSO. Without loss of generality, we model the deployment area as a two-dimensional square region of size $R \times R \; \text{m}^2$, containing $L$ APs that are randomly distributed following a uniform spatial distribution, given by the coordinates $ (x_{\ell}, y_{\ell}) \sim \mathcal{U}(0, R) $ for $\ell \in \mathcal{L} = \{1, 2, \ldots, L\}$. To emulate actual deployment perturbations, we consider multiple spatial realizations of APs. Initially, we employ the K-Means Clustering (KMC) algorithm to group nearby APs, constructing ${G}$ preliminary groups. Grouping of APs is a critical step for both HS and RS-enabled CF-mMIMO topologies, where each group is denoted by $\mathcal{G}_i \subseteq \mathcal{L}$, for $i = 1, 2, \ldots, G$, and the number of all APs in a group $i$ is denoted by $L_{\mathcal{G}_i}$. Furthermore, each group $\mathcal{G}_i$ is associated with one of $W$ distributed DUs based on proximity. These DUs, indexed by $w \in \mathcal{W} = \{1, 2, \ldots, W\}$, are responsible for serving several APs groups.

\begin{figure*}[t]
    \subfloat[Small cells network.\label{SmallCells}]{
        \includegraphics[width=0.315\textwidth, height=1.6in]{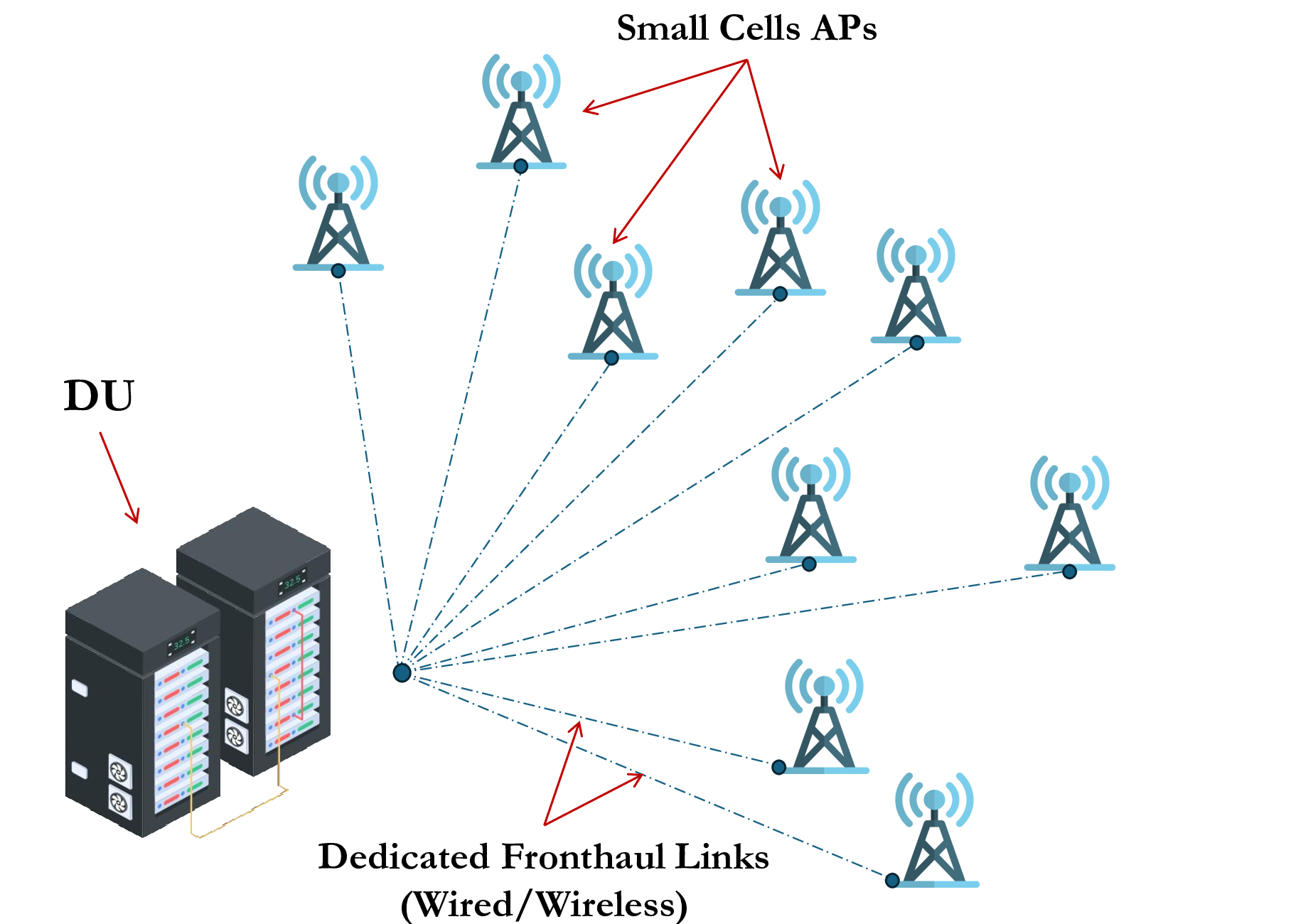}}
    \hfill    
    \subfloat[RS-enabled CF-mMIMO network.\label{RSFig}]{
        \includegraphics[width=0.315\textwidth, height=1.6in]{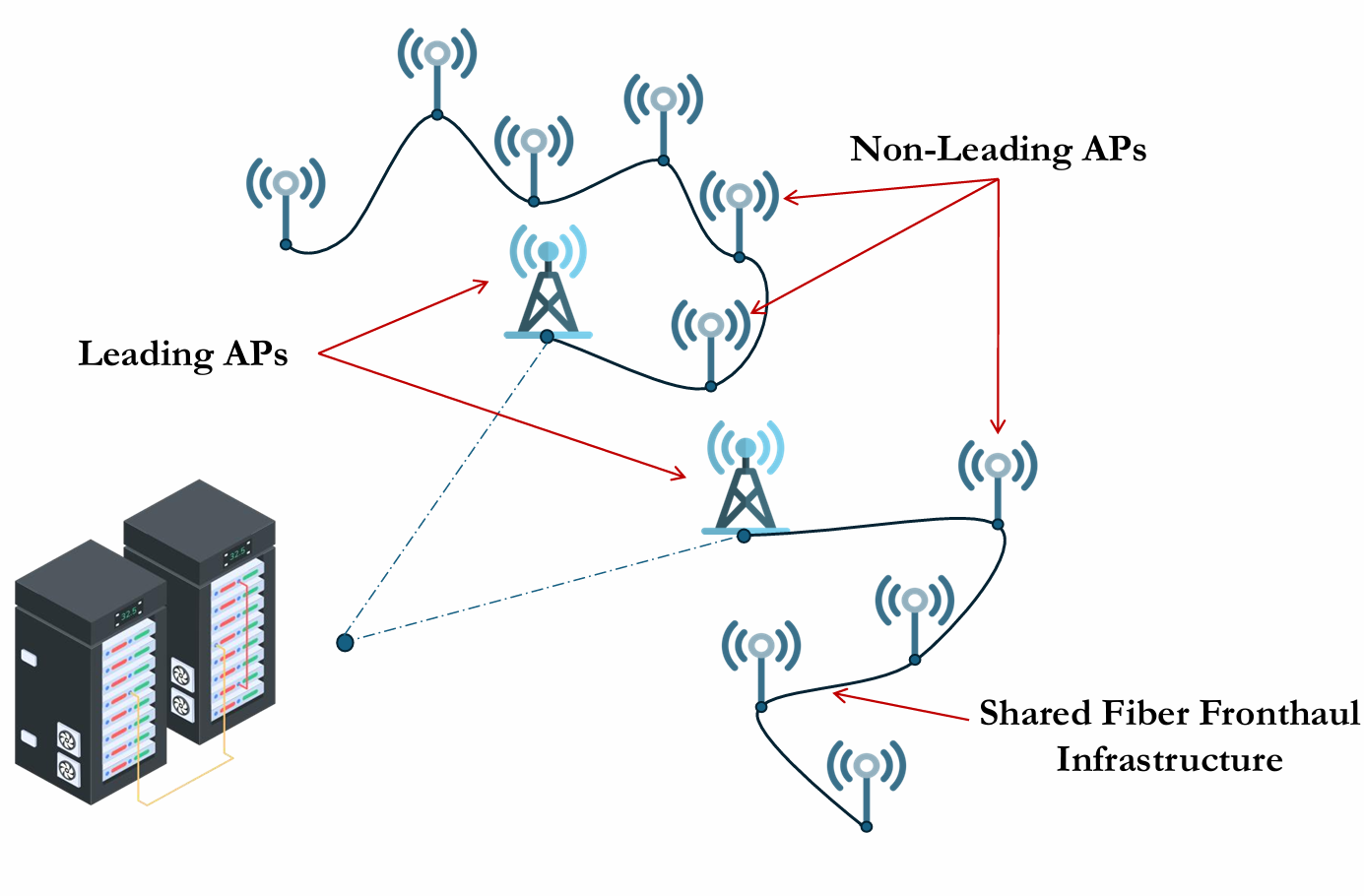}}
    \hfill
    \subfloat[HS-enabled CF-mMIMO network.\label{MSTFig}]{
        \includegraphics[width=0.315\textwidth, height=1.6in]{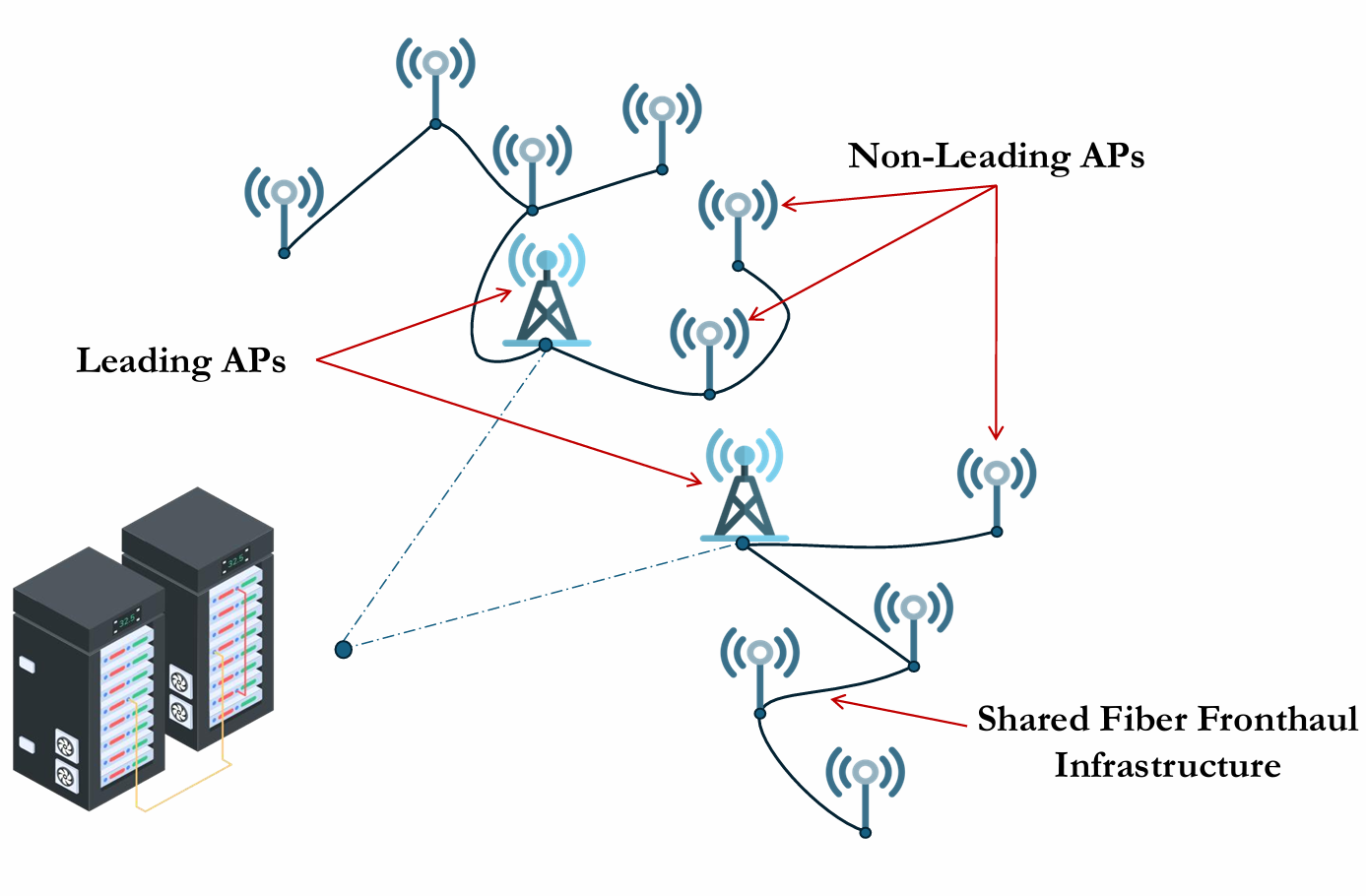}}
    \caption{Illustration of the different fronthaul topologies employed for UDNs schemes; (a) Conventional small-cell architecture with dedicated P2P fronthaul links, (b) RS topology and (c) HS topology.    }
    \label{fig1}
   \vspace{-1.2em}
\end{figure*}

In RS, and as described in the early RS antenna arrangement patent \cite{patentericsson}, APs within a group are connected serially via a shared wired fronthaul infrastructure. Specifically in this work, all APs within the same group $\mathcal{G}_i$ utilize a common intra-group fiber link to receive data streams from their serving DU, rather than each AP requiring a dedicated P2P link. This interpretation is strictly at the group level and does not imply that all APs in the network are connected through a single global fiber link. To reduce signaling overhead, a single AP is designated as the \textbf{leading AP} in each group. This AP, selected as one of the terminal points of the stripe, establishes a direct communication link with its serving DU via either a wired or wireless fronthaul connection \cite{RadioStripes2, mmwave_source}, as illustrated in Figure \ref{RSFig}. The leading AP processes and forwards fronthaul signals to other APs in the stripe, referred to as \textbf{non-leading APs}, through the shared fronthaul infrastructure \cite{mmwave_source, RadioStripes1, RadioStripes2}. This setup diverges from conventional small-cell architectures, which rely on dedicated P2P fronthaul links as depicted in Figure \ref{SmallCells}. We extend the concept of RS to the HS configuration shown in Figure \ref{MSTFig}, where the serial connection is replaced by a hierarchical topology. In this configuration, the leading AP is the one having the highest number of connection degrees, i.e., it has the largest number of dependent non-leading APs.

Efficient deployment of fronthaul technologies requires careful consideration of their unique components. For fiber-based fronthaul, the hardware needed for a typical Wavelength-Division Multiplexing Passive Optical Network (WDM-PON) includes fiber cables, Optical Add-Drop Multiplexers (OADMs) and Optical Network Units (ONUs) integrated with each AP $\ell$, along with the Optical Transport Network (OTN) colocated at each DU $w$. OTNs manage optical signals from multiple ONUs and employ components such as splitters, multiplexers (MUXs), and Optical Line Terminals (OLTs) to aggregate signals \cite{newfayad}. On the other hand, the antenna configuration in mmWave-based fronthaul influences the network model. For simplicity, we assume that all APs utilizing mmWave fronthaul have a single antenna (i.e., $N_{\text{AP}} = 1$), while DUs are equipped with $N_{\text{DU}}$ antennas. In contrast, FSO-based fronthaul uses P2P links, with each AP having a dedicated transceiver paired with its associated DU, ensuring single transmission and reception points.

We also assume that all APs within a group cooperate to provide spatial diversity for jointly-served users. That is for every group of APs $\mathcal{G}_i$, all APs need to receive a copy of the same message from their serving DU, hence,  eliminating the need for transmitting user-specific data to each AP individually. Instead, the same message received by leading APs is shared among all APs in the group, thereby simplifying fronthaul processing. %Additionally, we assume no compression techniques are employed to reduce data rates; thus, data is transmitted in its uncompressed, raw form. 
Given the ultra-dense nature of the network components deployment, line-of-sight (LoS) connectivity between APs and DUs in wireless fronthauling is assumed. Consequently, wireless relays and repeaters are excluded from consideration, as unobstructed connections are deemed achievable. 

We assume uncompressed fronthaul to isolate cost-performance tradeoffs between transmission technologies and avoid introducing codec-specific variability, FS options tradeoffs, technology-specific or transport medium limitations, as these could be additional metrics for comparison affecting the fronthaul rates (e.g., $\mu$-law and Block Floating Point (BFP) compression techniques, free space and fiber transport mediums, etc). Fronthaul compression is typically considered in specific scenarios rather than general infrastructure-level planning, and our analysis ensures a conservative estimate of fronthaul demands and preserves generality without biasing toward a specific compression method. 

\subsection{FS-Options Capacity Requirements}\label{FS-Traffic}
FS7.2x and FS8 are regarded as key candidates in B5G networks, due to their alignment with O-RAN architecture,  \cite{o-rancfmimo, O-RAN}. We introduce a fronthaul data rate threshold, denoted as $\psi$, to specify the minimum fronthaul capacity required for each AP, which will guide the optimization process and selection of appropriate fronthaul technologies.  TABLE \ref{T2:FS-options-Rates} lists the capacity requirements for FS8 and FS7.2x under standard 5G NR numerology 0 system configuration. This configuration is one of several possible numerologies defined by 5G NR, where numerology 0 and 1 correspond to a subcarrier spacing of 15 kHz and 30 kHz, respectively, with both commonly implemented in 5G systems. While these values may resemble LTE parameters, they are fully compliant with 5G NR as per 3GPP TS 38.211 (Release 18) \cite{3gpp38211}. We emphasize that our choice of numerology does not affect the solution design for fronthaul deployment, as subcarrier spacing $\Delta f$ only changes the frame structure and symbol timing, but not the overall data rate or required fronthaul capacity, which remains dependent on bandwidth and antenna configuration. Specifically, the system employs Orthogonal Frequency-Division Multiplexing (OFDM), a widely used modulation scheme in LTE and 5G systems, expected to be present in B5G networks \cite{o-rancfmimo}. Key parameters include OFDM symbol duration $T_{\text{symbol}}$, sampling frequency $f_s$, and the quantization bit-width $N_{\text{bits}}$, which denotes the number of bits per in-phase (I) and quadrature (Q) components. We assume that the access channel between users and APs is modeled as a block-fading channel, with each coherence block comprising $\tau_c$ time-frequency OFDM samples \cite{o-rancfmimo}. The available bandwidth $B$ is divided into $N_{\text{DFT}}$ subcarriers using the Discrete Fourier Transform (DFT), with $N_{\text{used}}$ effective subcarriers for data transmission and $N_{\text{null}}$ reserved for guard bands. The number of AP antennas on access channel, $N_{\text{AP}}^{\text{ac}}$, is distinct from those used for mmWave fronthaul, $N_{\text{AP}}$.
\begin{figure}[!t]
  \begin{center}
  \includegraphics[width=2.7in]{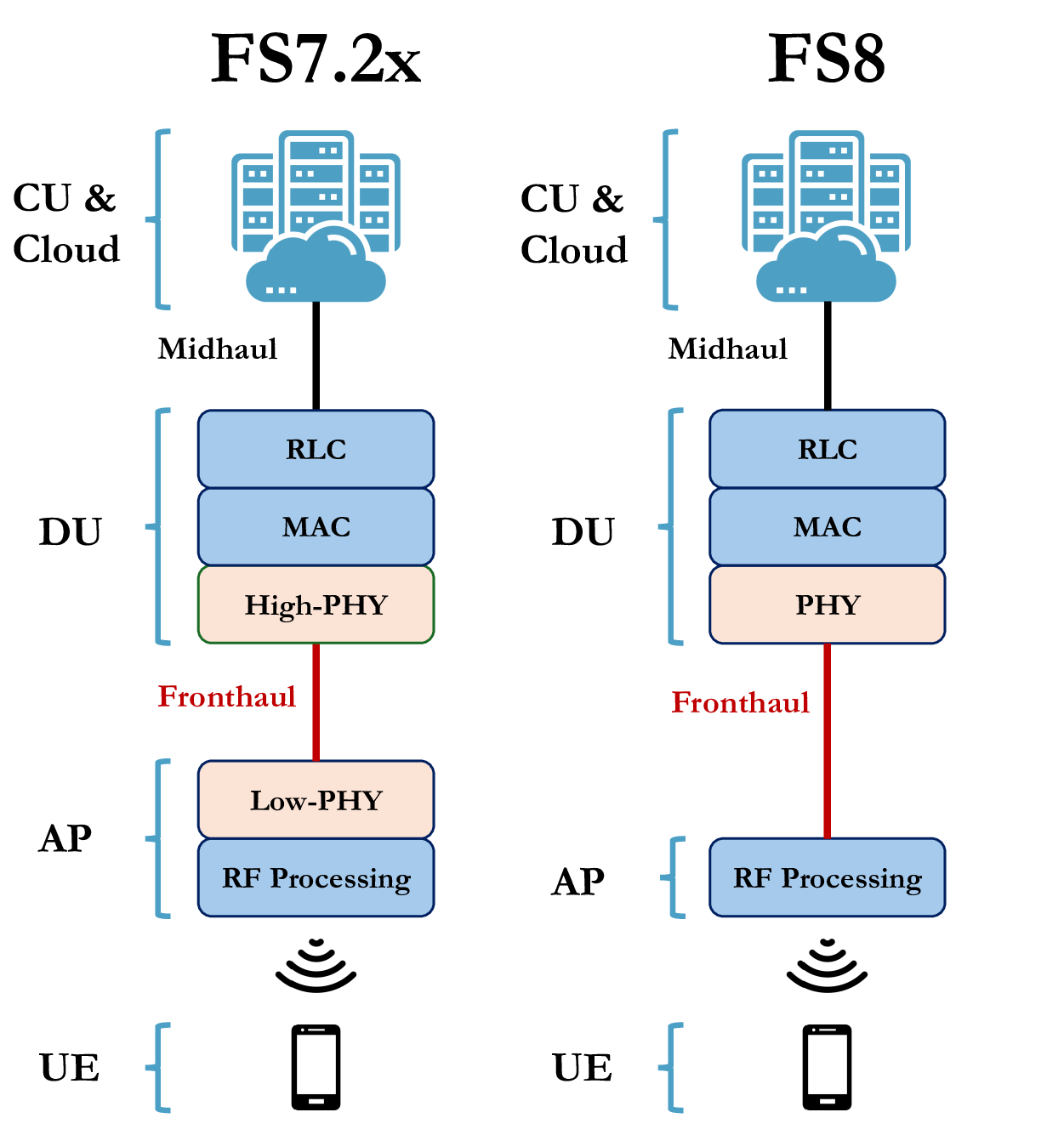}
  \vspace{-2pt}
\caption{Distribution of processing tasks in FS7.2x and FS8.}\label{newfs}
 \end{center}
     \vspace{-0.7em}
\end{figure}
\subsubsection{Functional Split 8 (FS8)}
The lowest layer split is defined as FS8, which is the PHY-RF split option aimed at fully benefiting from the efficient processing capabilities at the DUs, Centralized Units (CUs) and Cloud, while reducing APs complexity. As shown in Figure~\ref{newfs}, APs in FS8 only perform RF processing and receive raw, time-domain, quantized baseband signals from DUs through the fronthaul, leading to high fronthaul capacity requirements. For FS8, we assign each AP $\ell$ with the minimum required fronthaul capacity as follows \cite{o-rancfmimo}:
\begin{equation}
\label{eq:FS8}
    \psi^{\text{FS8}} = 2 \times N_{\text{bits}} f_s N_{\text{AP}}^{\text{ac}}.
\end{equation}

It is important to note that $N_{\text{bits}}$ in equation~\eqref{eq:FS8} refers to the bit-width per component (I or Q), and the multiplication factor of 2 accounts for both. Thus, the total bits per complex I/Q sample in our formulation equals $2 \times N_{\text{bits}}$.

\subsubsection{Functional Split  7.2x (FS7.2x)}
Similarly, FS7.2x assigns part of the intra-PHY functions to be processed at the APs, while the remaining high-PHY functions are shifted to the DUs, as shown in Figure \ref{newfs}. FS7.2x strikes a balance between AP complexity and fronthaul bandwidth, making it ideal for scalable O-RAN deployments in different UDNs schemes. Specifically, APs only receive the effective subcarriers ($N_{\text{used}}$) from DUs, leading to APs performing additional low-PHY functions, hence, lowering the fronthaul capacity requirements compared to FS8. The required fronthaul capacity in FS7.2x is \cite{o-rancfmimo}:
\begin{equation}
\label{eq:FS7}
    \psi^{\text{FS7.2x}} = \frac{2 \times N_{\text{bits}}  N_{\text{used}} N_{\text{AP}}^{\text{ac}}}{T_{\text{symbol}}}.
\end{equation}

Note that ~\eqref{eq:FS8} and~\eqref{eq:FS7} provide a baseline estimation of fronthaul data rate requirements under FS8 and FS7.2x options, following the widely used formulations in~\cite{o-rancfmimo}. These expressions primarily capture the user-plane data transfer. In practical systems, additional overhead from control signaling, CSI exchange, synchronization, and protocol encapsulation may further increase the fronthaul load, which we will account for next. The values listed in Table \ref{T2:FS-options-Rates} assume each AP operates at full load, serving the maximum expected user throughput. This worst-case planning approach, widely adopted in network design, ensures robustness under peak demand conditions. Consequently, our fronthaul deployment decisions are shaped by user traffic indirectly, through strict capacity constraints applied in the optimization framework. This approach is consistent with UDN/CF-mMIMO literature, where dense and uniform user demand is a standard modeling assumption~\cite{UDN, o-rancfmimo}.

% Although we do not explicitly model user locations or traffic distributions, user demand is implicitly embedded in our fronthaul capacity modeling. 

\subsubsection{Non-Homogeneous Traffic and the Control Plan}\label{new-section1}

Equations~\eqref{eq:FS8} and~\eqref{eq:FS7} define a minimum required fronthaul capacity for any AP assuming a homogeneous spatial traffic scenario, in which mobile operators assign this value for every AP in the network. In contrast, mobile operators can use a non-homogeneous spatial traffic scenario, where the expected traffic is determined based on a data traffic survey. In this scenario, each leading-AP~$\ell \in \mathcal{M}_w$ has a different minimum required fronthaul capacity that is defined as \cite{MyPaper}:
\begin{equation}
\label{eq:nonhomog}
    \psi^{\text{FS[8,7.2x]}}_\ell = f_{\rm traf.}(x_{\ell}, y_{\ell}),
\end{equation}

where $f_{\rm traf.}(\cdot)$ is the traffic-aware fronthaul capacity needed in the area centered on $(x_{\ell}, y_{\ell})$ which is the location of the leading-AP~$\ell \in \mathcal{M}_w$.

Adding control plane traffic for the equations in~\eqref{eq:FS8} and~\eqref{eq:FS7} can be a very involved task. Interestingly, we can add an overhead term that accounts for the control plane. According to ORAN technical report~\cite{ORAN_report}, there is a defined typical control plane that is sent on the fronthaul which includes PRACH channel I-Q data,  scheduling and beamforming commands, configuration parameters and request, ACK/NACK message, and other signals. We account for these messages that represent the control plane as a factor proportional to the data plane by defining the following:
\begin{align}
	\psi^{\text{FS8, CP}} &= \left(1 + \alpha^{\text{FS8, ovh}} \right) \psi^{\text{FS8}},
	\label{eq:FS8_CP2}
	\\
	\psi^{\text{FS7.2x, CP}} &= \left(1 + \alpha^{\text{FS7.2x, ovh}} \right)
	\psi^{\text{FS7.2x}},
	\label{eq:FS7_CP2}
\end{align}

where $\alpha^{\text{FS8, ovh}}, \alpha^{\text{FS7.2x, ovh}} \in [0, 1]$, and represent the amount of overhead introduced by the control plane. In our results, we will account for this control plane traffic using a safety margin that is calculated through surplus. We will illustrate this point in the results section.

\begin{table}[t]
\renewcommand{\arraystretch}{1.2}
\centering
\caption{OFDM-based standard parameters values for 5G NR numerology 0, and the required fronthaul capacity for FS8 and FS7.2x.}
\label{T2:FS-options-Rates}
\begin{small} 
\begin{tabular}{|c||c|}
\hline
\multicolumn{2}{|c|}{{\textbf{Configuration Parameters for 5G NR Numerology 0}}}  \\
\hline
\hline
{Bandwidth ($B$)} & $20$ MHz  \\
\hline
{Sampling Frequency ($f_s$)} & $30.72$ MHz \\
\hline
{Subcarrier Spacing ($\Delta f$)} & $15$ kHz \\
\hline
{Symbol Duration ($T_{\text{symbol}}$)} & $66.67\,\mu s$  \\
\hline
{Total Number of Subcarriers ($N_{\text{DFT}}$)} & $2048$  \\
\hline 
{Effective/Used Number of Subcarriers ($N_{\text{used}}$)} & $1200$  \\
\hline
{Quantization Bits ($N_{\text{bits}}$)} & $12$  \\
\hline
{Number of Access Antennas per AP} ($N_{\text{AP}}^{\text{ac}}$) & $4$ \\
\hline
\hline
\textbf{FS8 Required Capacity ($\bm{\psi^{\textbf{FS8}}}$)} & $2.95$ Gbps  \\
\hline
\hline
\textbf{FS7.2x Required Capacity ($\bm{\psi^{\textbf{FS7.2x}}}$)} & $1.73$ Gbps \\
\hline
\end{tabular}
\end{small}

\end{table}

\subsection{Fiber-Based Fronthaul Link Capacity}\label{sec35}
Due to the short distances between network elements in CF-mMIMO and the high efficiency of fiber optics, we assume lossless fronthaul links with a constant capacity denoted by $R_{w \ell}^{\text{Fiber}}$. We consider employing a 10 Gbps-capable symmetrical (XGS) WDM-PON, providing equal UL and DL data rates while accommodating evolving capacity demands. WDM-PON supports multiple wavelengths and is expected to remain a key solution for fronthaul/backhaul networks \cite{newfayad}. Consequently, each AP with a fiber-based fronthaul link has a constant capacity $ R_{w \ell}^{\text{Fiber}}$ of 10 Gbps.

\subsection{mmWave-based Fronthaul Link Capacity}
For mmWave, we adopt 3GPP 38.901 Urban Microcell street canyon (UMi-SC) model to characterize both LoS and Non-Line-of-Sight (NLoS) propagation, modeling the downlink (DL) performance from DUs to APs~\cite{3gpp}. SPs typically optimize terrestrial network fronthaul by leveraging the static positioning of APs and DUs to maintain LoS conditions during wireless fronthaul planning. Thus, LoS parameters are distance-based, while NLoS parameters, including the number of paths \(P \sim \mathcal{U}[1, 6]\) and angle-of-departure \(\theta_p \sim \mathcal{U}[-\frac{\pi}{2}, \frac{\pi}{2}]\), are randomly selected based on discrete uniform distribution to account for possible reflections \cite{mmwave_source}. Accordingly, the path loss for both LoS and NLoS scenarios is expressed respectively as follows:
\begin{equation}\label{equ3.18}
\text{PL}_{w\ell,\, \text{dB}}^{\text{LoS}} = 32.4 + 21 \log_{10}(d_{w \ell}) + 20 \log_{10}\left(f_{\text{c}}\right)  + \mathcal{S}_{\text{LoS}},
\end{equation}
\begin{equation}
\label{equ3.19}
\text{PL}_{w\ell,\, \text{dB}}^{\text{NLoS}} = 32.4 + 31.9 \log_{10}(d_{w\ell}) + 20 \log_{10}\left(f_{\text{c}}\right) + \mathcal{S}_{\text{NLoS}},
\end{equation}

where $f_{\text{c}}$ is the carrier frequency in GHz and $d_{w\ell}$ denotes the distance between AP $\ell$ and its serving DU $w$ in meters. While $\mathcal{S}_{\text{LOS}} \in \mathcal{N}(0, \sigma_{\text{LoS}}^2)$ and $\mathcal{S}_{\text{NLoS}} \in \mathcal{N}(0, \sigma_{\text{NLoS}}^2)$ are the shadowing terms modeled as Gaussian random variables with zero mean and standard deviation of 4 and 8.2 dB, respectively \cite{3gpp}. Moreover, let $\mathbf{h}_{w\boldsymbol{\ell}} \in \mathbb{C}^{N_{\text{DU}}}$ be the frequency-domain channel between the $\ell$-th AP and its serving $w$-\text{th} DU, and is expressed as the sum of LoS and NLoS components:
\begin{equation}
    \mathbf{h}_{w\ell} = \mathbf{h}_{w\ell}^{\text{LoS}} + \mathbf{h}_{w\ell}^{\text{NLoS}}.
\end{equation}

Before transmission to APs, DUs adopt analog beamforming with a network of quantized phase shifters to compute the beamforming vectors, approximating ZF beamforming to mitigate interference and focus the transmitted signal toward the intended APs \cite{mmwave_source}. The assumption of quantized phase shifters is for satisfying the practical constraints of mmWave, which means that the beamforming vector $\mathbf{f}_{w\ell}$ could only be selected from a certain set of vectors $\mathcal{F}$, thus, $\mathbf{f}_{w\ell} \in \mathcal{F}$. That is if every phase shifter has $q$ quantization bits, then we will have $2^q$ phase shift values defined by the discrete set $\mathcal{Q} = \{0, \frac{\pi}{2^q}, \ldots, \frac{(2^q - 1)\pi}{2^q}\}$. Hence, the set of all possible beamforming vectors for any DU is expressed as~\cite{mmwave_source}:
\begin{equation}
    \mathcal{F} = \left\{ \frac{\begin{bmatrix} e^{j\phi_1} ... e^{j\phi_{N_{\text{DU}}}}  \end{bmatrix}^T}{N_{\text{DU}}} ;  \phi_i \in \mathcal{Q}, \forall i \in \{1, .., N_{\text{DU}}\}
   \right\},
\end{equation}

and the fronthaul signal received at AP $\ell$ is given as:
\begin{equation}
\mathbf{r}_{\ell}^{\text{mmW}} =  \sqrt{p_{t}^{\text{mmW}}} \mathbf{h}^T_{w \ell}\mathbf{f}_{w \ell} \varsigma_{w\ell} + \!\! \underbrace{\sum_{\substack{w' = 1, \\ w' \neq w}}^{W} \sum_{\substack{i = 1, \\ i \neq \ell}}^{L}
\sqrt{p_{t}^{\text{mmW}}} \mathbf{h}_{w' i}^T \mathbf{f}_{w' i}\varsigma_{wi}}_{\text{interference}} + \mathbf{n}_{\ell}, 
\end{equation}

where $p_{t}^{\text{mmW}}$ is the normalized fronthaul transmission power, $\mathbf{n}_{\ell} \backsim \mathcal{N}_{\mathbb{C}}(0,\sigma^2)$ is the receiver (RX) noise at the $\ell$-\text{th} AP, and $\varsigma_{w\ell} \in \mathbb{C}$ is the signal transmitted from the $w$-\text{th} DU to AP $\ell$ with $\mathbb{E} \{|\varsigma_{w\ell}|^2\} = 1$. %Nevertheless, due to the static nature of our network components, we assume perfect channel estimation for fronthaul. Additionally, 
Given the relatively large distances between APs and non-serving DUs, combined with our use of beamforming, inter-DU interference is assumed negligible. Consequently, we focus on the Signal-to-Noise Ratio (SNR) to determine the actual rates of the mmWave link. Therefore, the SNR at AP $\ell$ and the corresponding mmWave fronthaul link capacity can expressed as follows:
\begin{align}
    \mathbf{SNR}_{\ell} &= \frac{p_{t}^{\text{mmW}} \left | \mathbf{h}^T_{w  \ell}  \mathbf{f}_{w  \ell} \right |^2 }{\sigma^2},
\\
    R_{w \ell}^{\text{mmW}} &= {\rm BW}^{\text{mmW}} \log_2(1 + \mathbf{SNR}_{\ell}).
\end{align}
\begin{remark}\label{remarkNew}
Although analog beamforming with quantized phase shifters may not exactly replicate ZF performance leading to an imperfection in canceling interference, it remains effective in focusing energy toward serving APs and attenuating unintended directions, particularly under LoS or dominant-path channel conditions, as assumed in our fronthaul model. In practice, DUs and APs are statically positioned as part of a planned deployment, enabling precomputed beam directions and stable LoS-dominant channels. Additionally, the usage of mmWave frequencies limits the range of communication and hence interference. These factors make the assumption of negligible inter-DU interference stronger. Motivated by this, we use SNR for our performance evaluation for the mathematical tractability advantage. In an actual deployment with static fronthaul scenario, mobile operators can relax our assumption by setting a target network surplus capacity (Figure \ref{SurplusFS}). 
\end{remark}

\subsection{FSO-Based Fronthaul Link Capacity}\label{sec33}

FSO has been proposed as a viable communication solution for fronthaul and backhaul, demonstrating sufficient availability and data rates over long distances \cite{FSO-Main, FSOandFiberCFMIMO, Turbulence2}. Nevertheless, FSO technology remains severely affected by environmental conditions, where losses such as scattering, turbulence, scintillation, geometrical spreading, and optical inefficiencies can jeopardize its reliability. Scattering occurs when the FSO beam interacts with atmospheric particles, and expressed as \cite{FSO-Main}:

\begin{equation}
    L_{\text{sca}} = 4.34 \left(\frac{3.91}{V} \left(\frac{\lambda}{550}\right)^{- \delta}\right) \frac{d_{w\ell}}{1000},
\end{equation}

where $L_{\text{sca}}$ is the scattering loss in dB, $V$ is the visibility range in km,  $\lambda$ is the signal wavelength in nm, and $\delta$ is a visibility-dependent constant given as $\delta =  
0.585 V^{1/3}, \text{for} \; V < 6 \text{ km}$ \cite{FSO-Main}. Additionally, atmospheric turbulence losses are caused by variations in the refractive index structure parameter $ C_n^2(h)$ of the atmosphere at altitude $h$, measured in $\text{m}^{-2/3}$, and transiently affected by wind speeds \cite{Turbulence2}. During the planning stage, an average value for the refractive index structure parameter for moderate turbulence is $\left[ 10^{-15} \text{m}^{-2/3} \right]$, as considered in \cite{Turbulence2}. Scintillation loss, denoted as $ L_{\text{sci}} $, is caused by the rapid fluctuations in light intensity when it propagates through a turbulent atmosphere, and the resultant attenuation in dB is given by:
    \begin{equation}\label{scint}
    L_{\text{sci}} = 2 \sqrt{23.17 \left( \frac{2\pi}{\lambda} 10^9 \right)^{\frac{7}{6}} C_n^2(h) d_{w\ell}^{\frac{11}{6}}}.
\end{equation}

Finally, geometrical losses $ L_{\text{geo}}$ result from light spreading over a larger area. Assuming circular mirrors at both TX and RX, the geometrical loss in linear scale is \cite{FSO-Main}:
\begin{equation}
    L_{\text{geo}} \approx \left( \frac{r_r}{ (\frac{\theta_t d_{w\ell}}{2})}\right)^2,
\end{equation}

where $r_r$ is the RX aperture radii in meters, and $\theta_t$ is the divergence angle of the transmitter (TX) in radians. Several assumptions are made before deriving the achievable data rate for FSO links ($R_{w\ell}^{\text{FSO}}$). Due to the short APs-DUs distances, an average visibility range of \(V = 0.4\) km is assumed. A study in Berlin, Germany \cite{FSO-Availability} shows that visibility drops below 0.4 km only 0.25\% of the year, resulting in an average FSO link availability of 99.75\% (\(\zeta^{\text{FSO}} = 0.9975\)). Consistent with the assumptions made in \cite{FSO-Main}, we consider LoS FSO transmission. Under moderate fog conditions, fog loss is calculated as \(L_{\text{fog}} = 20.99 \, d_{w\ell}\) dB, where \(d_{w\ell}\) is the distance in km between the AP and its associated DU \cite{FSO-Main}. Lastly, a 10 dB loss (\(L_{\text{rain}}\)) is included to account for rain and other worst-case scenarios. Therefore, total atmospheric losses are expressed as:
 \begin{equation}
    L_{\text{dB}}^{\text{atm}} = \left( L_{\text{sca}} + L_{\text{rain}} + L_{\text{fog}} + L_{\text{sci}} \right)_{\text{dB}}.
\end{equation}

The rate for the FSO link between DU $w$ and AP $\ell$ is~\cite{FSO-Main}:
\begin{equation}\label{rateeee}
R_{w\ell}^{\text{FSO}} =  \frac{p_t^{\text{FSO}} \eta_t \eta_r  r_r^2}{L_{}^{\text{atm}} E_p N_b (\frac{\theta_t d_{w\ell}}{2})^2},
\end{equation}

where $p_t^{\text{FSO}}$ is the transmit power, $L_{}^{\text{atm}}$ is the atmospheric losses in linear scale, $\eta_t$ and $\eta_r$ are the optical losses caused by imperfections in optical TX and RX, respectively. $E_p$ is the photon energy expressed as $E_p = \frac{h_p c}{\lambda}$, where $h_p = 6.625 \times 10^{-34}$ J-s is Max Planck's constant, and $N_b$ is the RX sensitivity in photons/bit.

\section{Hierarchical and Radio-Stripes-Enabled CF-mMIMO Network Configuration}\label{CH3}
To simplify the problem formulation for selecting optimal fronthaul technologies, and without any loss of practicality, a \textit{brownfield deployment scenario} is assumed. In this scenario, the locations of APs are assumed to be predetermined by SPs according to the specific coverage and economic requirements. In this section, we introduce NOFAC, an iterative algorithm that minimizes inter-APs distances, balances cluster formation, efficiently deploys DUs, and achieves a \textbf{Near-Optimal Fronthaul Association and Configuration (NOFAC)}. 

\begin{figure}[t]
\centering
\subfloat[Initial RS network deployment.\label{RSDUsa}]{
\includegraphics[width=0.485\columnwidth]{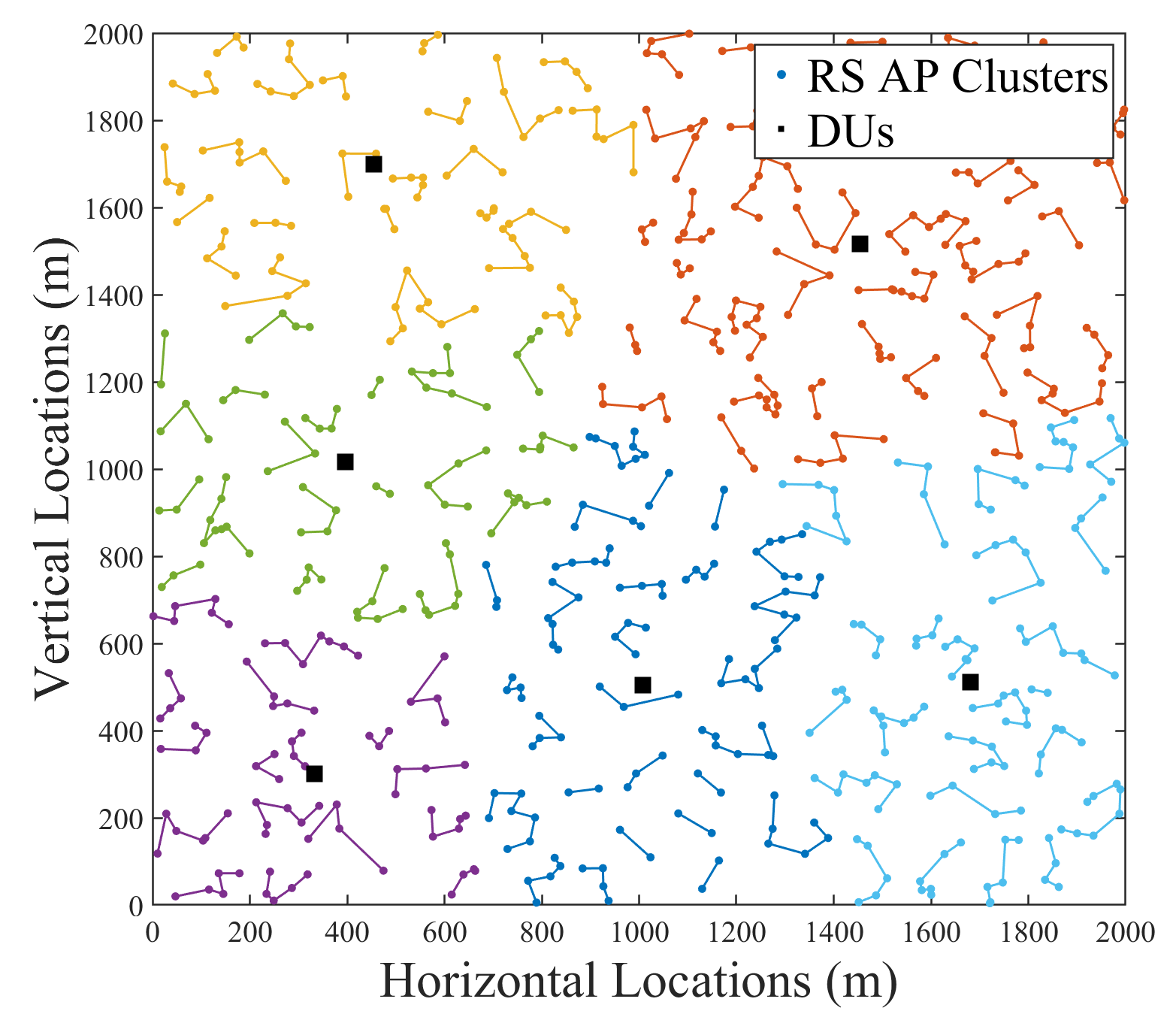}}
\hfill    
    \subfloat[Final optimized associations.\label{RSDUsb}]{
\includegraphics[width=0.485\columnwidth]{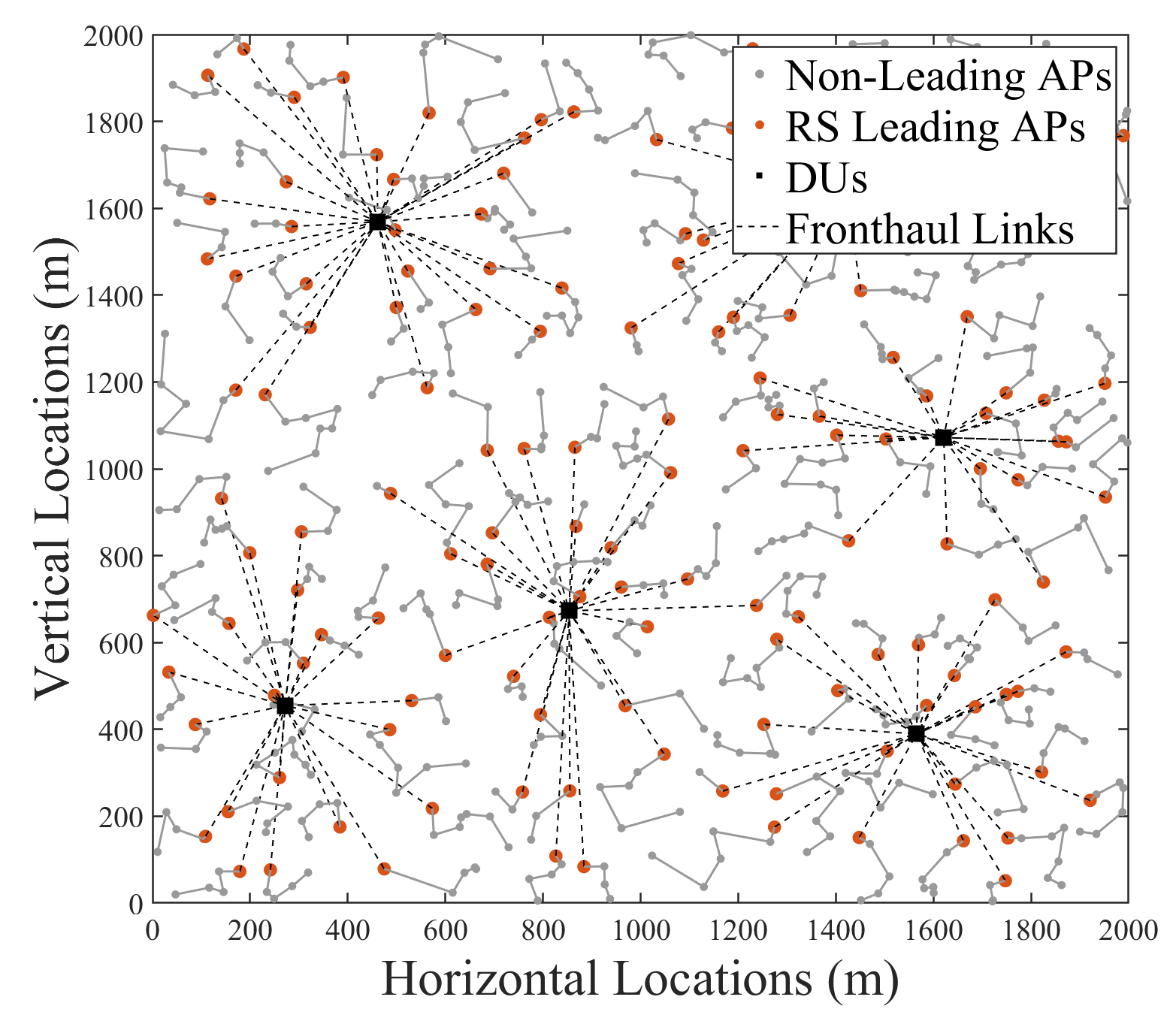}}

\caption{Sample realization of the optimized RS network.}
\label{RSDUs}

\end{figure}
\begin{figure}[t]
\centering
\subfloat[Initial HS network deployment.\label{HSDUsa}]{
    \includegraphics[width=0.485\columnwidth]{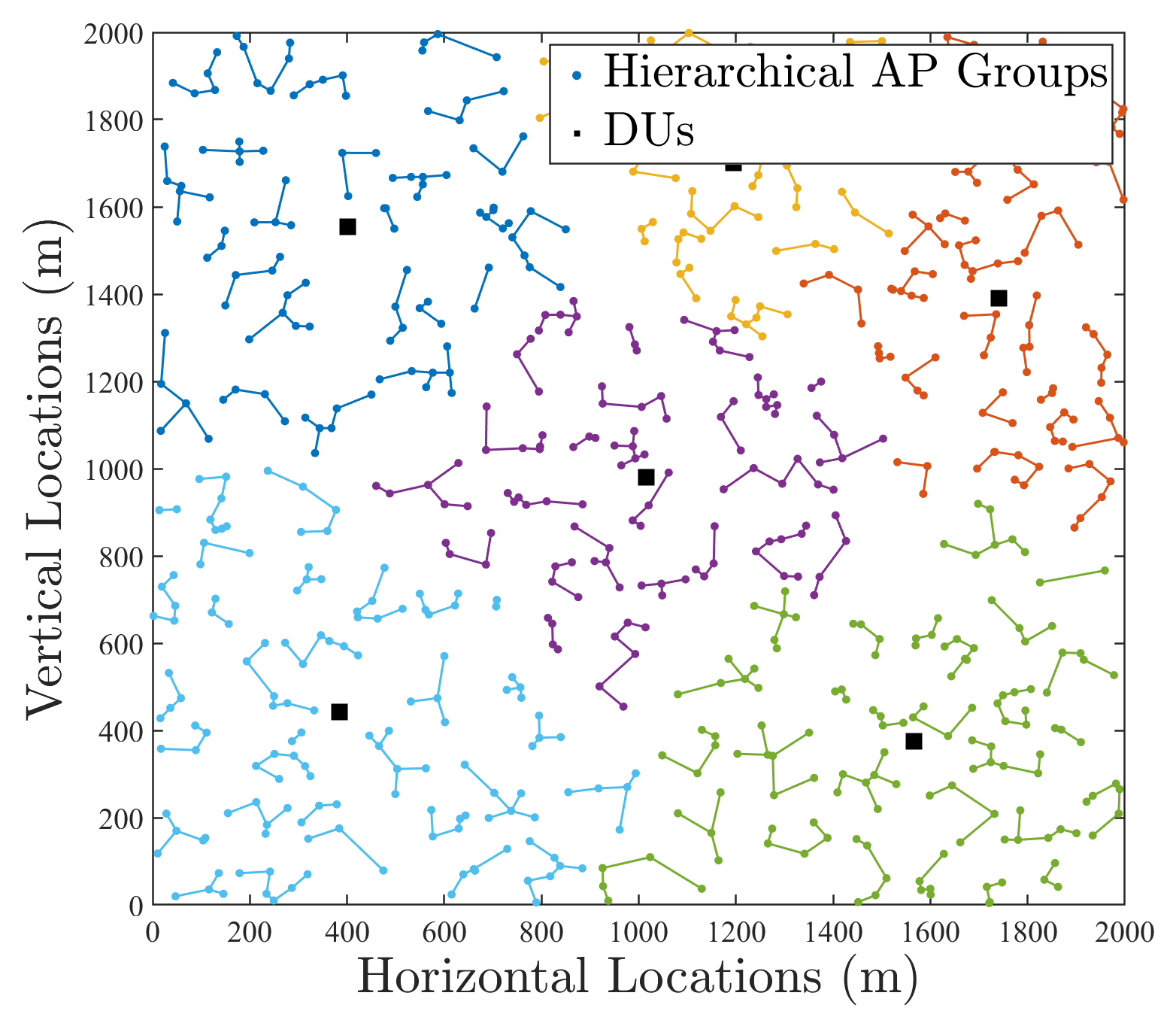}}
\hfill    
\subfloat[Final optimized associations.\label{HSDUsb}]{
    \includegraphics[width=0.485\columnwidth]{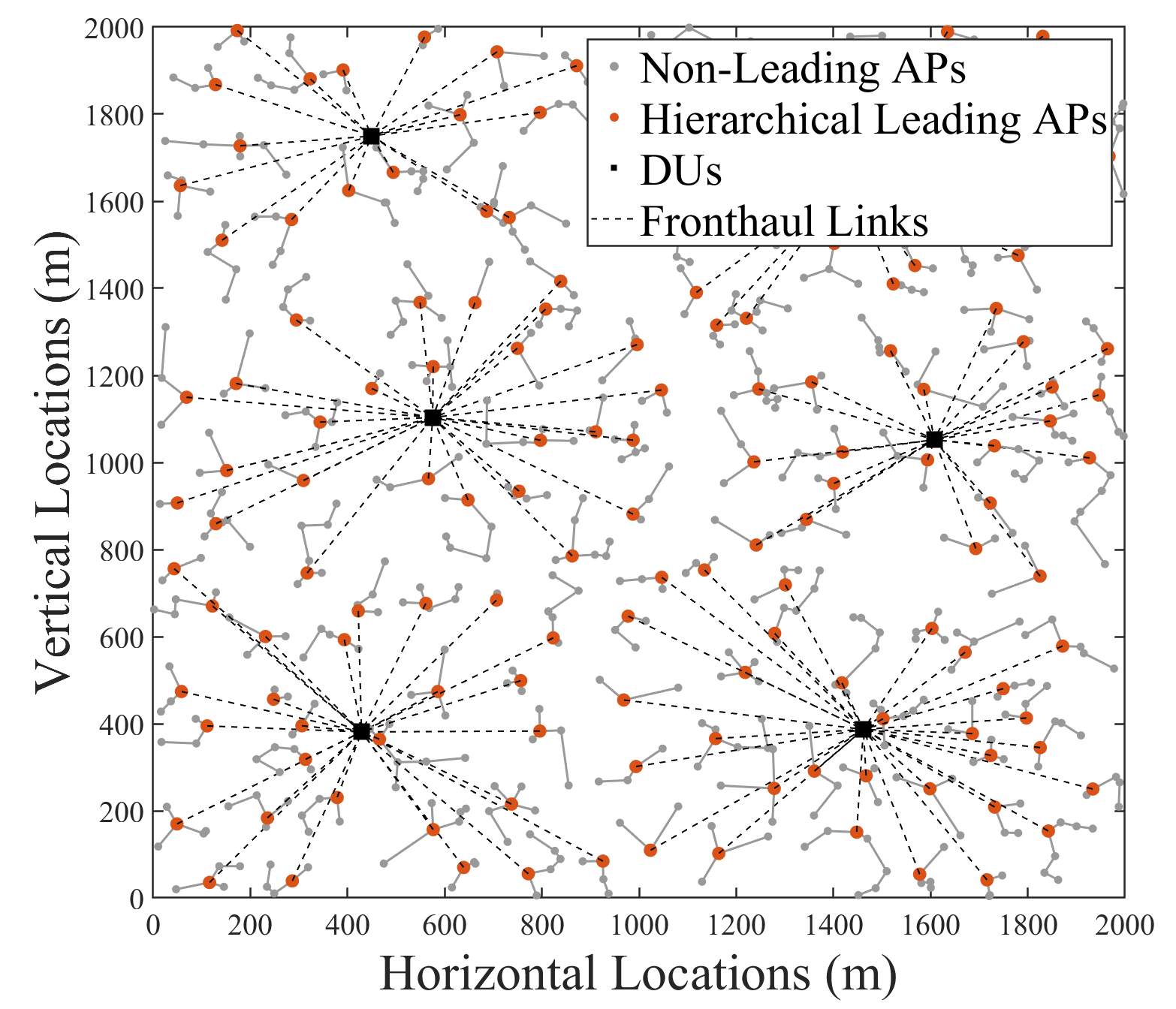}}
    
\caption{Sample realization of the optimized HS network.}
\label{HSDUs}

\end{figure}

\subsection{Preliminary Network Construction Methodology}\label{sec:preliminary}

The spatial density and geographic positioning of APs influence the grouping process, ensuring that APs in close proximity are grouped together to minimize inter-AP distances, fronthaul costs, and latency \cite{mmwave_source, RadioStripes3}. To mitigate limitations of basic KMC, which generates $G$ preliminary AP groups of irregular sizes, initial clusters are refined by applying Split-Merge Rules (SMR) based on predefined system parameters, namely: split ($g_s$) and merge ($g_m$) thresholds, defining the maximum and minimum permissible number of APs per group, respectively. 

Following the initial clustering, groups exceeding the maximum permissible size ($L_{\mathcal{G}_i} > g_s$) are split into smaller clusters. Conversely, groups smaller than the minimum allowable size ($L_{\mathcal{G}_i} < g_m$) are merged with adjacent groups, provided the resulting group size does not exceed $g_s$ (i.e., $g_s < L_{\mathcal{G}_i} < g_m$). These refinement steps ensure that all groups maintain a balanced number of APs, balanced groups sizes, eliminating oversized and undersized clusters of both HS and RS-enabled CF-mMIMO networks construction. 
The sets of all APs and groups associated with DU $ w $ are represented by $\mathbf{L}_w \subseteq \mathcal{L}$, and $ \mathbf{G}_w \subseteq \mathbf{L}_w$, respectively.

\subsection{Constructing Radio-Stripes-Based CF-mMIMO Network}\label{sec:RadioStripesConstruction}
After refining the AP groups, APs within each group $\mathcal{G}_i$ for in RS are connected in a serial topology to minimize the total inter-AP distances. This involves identifying a sequence of APs that forms the lowest total traversal path, which is analogous to the Traveling Salesmen Problem (TSP). For smaller groups with $L_{\mathcal{G}_i} \leq L^{\text{max}}$, where $L^{\text{max}}$ is a system parameter ensuring that the factorial complexity ($L_{\mathcal{G}_i}!$) remains computationally tractable, the TSP is solved optimally using brute-force search across all APs permutations. Nevertheless, when $L_{\mathcal{G}_i} > L^{\text{max}}$, we use the alternative Nearest Neighbor (NN) algorithm as a computationally efficient approximation. Thus, a practical approach to minimize the overall serial connection distance between the same-group APs is done through the mix use of TSP and NN algorithms. The process for groups with $L_{\mathcal{G}_i} \leq L^{\text{max}}$ begins by computing the pairwise distances between all APs in a group $\mathcal{G}_i$. The path with the minimal total distance is then selected as follows
\begin{equation}
    d_{\mathcal{G}_i}^{\text{RS}} = \arg \min_{\pi \in \mathcal{P}} L_{\pi}, \; \; \; \forall i = 1, 2, \ldots, G,
\end{equation}

where $L_{\pi}$ represents the total path length of a permutation $\pi$, and $\mathcal{P}$ is the set of all possible paths. If the number of APs in a cluster exceeds $L^{\text{max}}$, NN algorithm is applied with multiple starting points to improve the chances of approximating the optimal solution. Following the construction of RS, we deploy $W$ DUs with their placement initially optimized using KMC, ensuring proper association planning between DUs  $ w = 1, 2, \ldots, W $ and APs groups $ \mathcal{G}_i = 1, 2, \ldots, G $ based on proximity to groups centroids. This approach ensures that each group is fully assigned to a single DU, as shown in Figure \ref{RSDUsa}. From each group $\mathcal{G}_i$, a leading AP $\ell_i$ is selected from the stripe endpoints, prioritizing the AP closest to the DU to minimize the total association distance. Since DU positions and leading AP selections are interdependent, Algorithm \ref{ALG1} implements NOFAC, which attempts to minimize the total association distance while maintaining effective AP-DU associations through updating DU positions, re-assigning leading APs, and re-iterating over the previous steps. Convergence is achieved when the maximum movement of any DU $w$ between iterations falls below a predefined threshold $\epsilon$:
\begin{equation}\label{OptimizedDUs}
    \max_{w} \|{\mu}_w^{\text{new}} - {\mu}_w^{\text{old}}\| < \epsilon,  \;\;\; \forall w \in \mathcal{W},
\end{equation}

where ${\mu}_w^{\text{new}}$ and ${\mu}_w^{\text{old}}$ represent the updated and previous positions of DU $w$, and $\epsilon$ is a very small positive number. If the convergence condition is not met, the process restarts. The final positions of the DUs and the selection of leading APs forming the final RS-based CF-mMIMO network are depicted in Figure \ref{RSDUsb}. After convergence, each DU $w$ is assigned a set of leading APs $\mathcal{M}_w$ and non-leading APs $\mathcal{A}_w$, with corresponding sizes $M_w$ and $A_w$, such that $\mathcal{M}_w, \mathcal{A}_w \subseteq \mathbf{L}_w$.
Although Algorithm 1 is a heuristic procedure, it includes a well-defined convergence condition, i.e.~\eqref{OptimizedDUs} based on the stabilization of DU placements. Given the finite number of APs and the deterministic refinement rules applied to the groupings, the algorithm consistently converges within a small number of iterations.
\begin{algorithm}[h]
\caption{NOFAC: For a near-optimal HS and RS-based CF-mMIMO fronthaul network association and configuration.}\label{ALG1}
\begin{algorithmic}[1]
\State \textbf{Input:} $W$, \,$L$, \,$G$, \,$\epsilon$, \,$L^{\text{max}}$, \,$g_s$ and $g_m$.
\State \textbf{Initialization:}
\State {Uniformly distribute $L$ APs.}
\State {Construct $G$ initial AP groups $(\mathcal{G}_{\text{initial}}  = \{ \mathcal{G}_1, \mathcal{G}_2, \dots, \mathcal{G}_{G}\})$ using KMC.}
\State {Deploy $W$ DUs using KMC ($\boldsymbol{\mu}^{\text{initial}}$).}
\State  $\boldsymbol{\mu}_w^{\text{old}} \gets \boldsymbol{\mu}_w^{\text{initial}} \; \;\;\; \forall w \in \mathcal{W}$; 
\While{true}
\For {$\bm{\forall w \in \mathcal{W}}$}
\For {$\bm{\forall i \in \mathbf{G}_w}$}
\If{$L_{\mathcal{G}_i} > g_s$}
\State Split group $\mathcal{G}_i$.
\ElsIf{$L_{\mathcal{G}_i} < g_m$}
\State Merge group $\mathcal{G}_i$ with the nearest group.
\Else 
\State Skip $\mathcal{G}_i$.
\EndIf
\If{RS-enabled CF-mMIMO network,}
\If{$L_{\mathcal{G}_i} \leq L^{\text{max}}$}
\State Apply TSP.
\Else
\State Apply NN.
\EndIf
\State Compute $d_{\mathcal{G}_i}^{\text{RS}} = \arg \min_{\pi \in \mathcal{P}} L_{\pi}.$
\Else \Comment{HS-enabled CF-mMIMO network}
\State Construct MST for $\mathcal{G}_i$.
\State Compute $d_{\mathcal{G}_i}^{\text{MST}} = \min_{\mathcal{T}_i \subseteq \mathcal{G}_i} \sum_{e_{jk} \in \mathcal{T}_i} w_{jk}.$
\EndIf
\EndFor      
\EndFor
\State {Assign Leading APs ($\mathcal{M}_w$).}
\State {Refine DUs positions using KMC ($\boldsymbol{\mu}_w^{\text{new}}$).}
\If {$\max_{w} \|\boldsymbol{\mu}_w^{\text{new}} - \boldsymbol{\mu}_w^{\text{old}}\| < \epsilon,  \;\; \forall w \in \mathcal{W} \;$} 
\State \textbf{Break}
\Else
\State $\boldsymbol{\mu}_w^{\text{old}} \gets \boldsymbol{\mu}_w^{\text{new}} \; \;\;\; \forall w \in \mathcal{W}$; 
\EndIf
\EndWhile
\State \textbf{Output:} 
$\boldsymbol{\mu}_w^{\text{new}}, \, \mathbf{L}_w, \, \mathbf{G}_w, \, \mathcal{M}_w, \, \mathcal{A}_w \;  \forall w \in \mathcal{W}$, \; and \; $\mathcal{G}_i, \, L_{\mathcal{G}_i},$  \, $d_{\mathcal{G}_i}^{\text{RS}} \;\; \text{or} \;\; d_{\mathcal{G}_i}^{\text{MST}} \;\; \forall i = 1, 2, \ldots, G$.
\end{algorithmic}
\end{algorithm}

\subsection{Constructing Hierarchical-Based CF-mMIMO Network}\label{sec:MSTConstruction}
The HS-enabled CF-mMIMO network construction builds on the principles employed in the RS network described in Section \ref{sec:RadioStripesConstruction}. However, instead of using serial connections, the APs within each group $\mathcal{G}_i$ are connected in a hierarchical topology to minimize inter-AP distances using Minimum Spanning Tree (MST) algorithm. This graph-theoretic approach ensures cost-effective connectivity between APs \cite{MSTISAC}. Each group $\mathcal{G}_i$ is represented as a complete weighted graph $\mathcal{G}_i = (\mathcal{V}_i, \mathcal{E}_i)$, where $\mathcal{V}_i$ is the set of APs (vertices) in group $\mathcal{G}_i$, and $\mathcal{E}_i$ represents the set of links (edges) connecting every pair of APs in a group. Thus, the weight of each link $e_{jk} \in \mathcal{E}_i$ is calculated as the Euclidean distance between APs $j$ and $k$. The MST for each group $\mathcal{G}_i$ is constructed by minimizing the total edge weight of the tree as follows:
\begin{equation}\label{MSTeq}
d_{\mathcal{G}_i}^{\text{MST}} = \min_{\mathcal{T}_i \subseteq \mathcal{G}_i} \sum_{e_{jk} \in \mathcal{T}_i} w_{jk}, \quad \forall i = 1, 2, \ldots, G,
\end{equation}

where $d_{\mathcal{G}_i}^{\text{MST}}$ is the total weight (i.e., total distance) of the MST for group $\mathcal{G}_i$, and $\mathcal{T}_i$ is the MST of group $\mathcal{G}_i$, which is a subgraph of $\mathcal{G}_i$ that connects all APs with the minimum possible total link distances. The MST problem in \eqref{MSTeq} can be optimally computed using Prim's or Kruskal's algorithm \cite{MSTISAC}, with the resulting HS illustrated in Figure \ref{HSDUsa}. Then, a similar process to the RS construction steps in Section \ref{sec:RadioStripesConstruction} is followed to deploy $W$ DUs. The leading AP from each group $\mathcal{G}_i$ is chosen as the node with the highest number of edges connected to it, which reflects the number of neighboring APs in the tree. In cases of tie, the AP nearest to the DU is selected to minimize association distance, with remaining APs classified as non-leading. The positions of DUs and selection of leading APs are iteratively optimized following the NOFAC approach in Algorithm \ref{ALG1}, and the process
repeats until convergence is achieved as per equation \eqref{OptimizedDUs}.

\section{Fronthaul TCO Optimization Formulation}\label{CH4}
This section presents the planning and cost optimization of the fronthaul network for HS and RS-based CF-mMIMO schemes. The objective is to minimize the TCO while ensuring optimal fronthaul technology selection. Our formulation starts by incorporating critical information, including the locations of APs and DUs, APs-DUs clusters ($\mathbf{L}_w$) and distances ($d_{w\ell}$), the effective capacities of leading APs for all fronthaul technologies ($R_{w\ell}^{\text{Fiber}}$, $R_{w\ell}^{\text{mmW}}$, and $R_{w\ell}^{\text{FSO}}$), alongside the outputs of NOFAC in Algorithm \ref{ALG1}. We consider LTE-based functional split options 8 and 7.2x for capacity requirements as defined in Section \ref{FS-Traffic}, collectively represented as $\psi^{\text{FSX}}$, where $X$ denotes either 7.2x or 8. This optimization model employs a two-tiered approach to minimize TCO, encompassing both operational (OPEX) and capital expenditures (CAPEX), with OPEX averaged over a fixed deployment period $N_{\text{period}}$.

\subsection{Tier 1 Optimization}
The first tier of optimization focuses on minimizing the deployment cost the shared fiber fronthaul infrastructure that consists of ONUs/OADMs and fiber cables per meter for the set of non-leading APs ($\mathcal{A}_w$), while masking leading APs ($\mathcal{M}_w$) in every group. This has been addressed by NOFAC in Algorithm \ref{ALG1}, by optimizing the grouping of APs and minimizing the pairwise connection distances $d_{\mathcal{G}_i}^{\text{RS}}$ and  $d_{\mathcal{G}_i}^{\text{MST}}$ between all APs in a group $\mathcal{G}_i$. Following the traditional construction of RS \cite{mmwave_source, patentericsson}, we assume that all APs in a group are connected via fiber cables, with each AP equipped with an ONU and an OADM to facilitate signal transmission and conversion from electrical to optical over a stripe. At each AP in a group, one signal is dropped using OADMs and inserted to a receiving ONU. For all the set of non-leading APs ($\mathcal{A}_w$) associated with DU $w$, the incurred costs comprise the ONU cost ($C_{}^{\text{ONU}}$), which incorporates the OADM cost, installation cost, O\&M costs over $N_{\text{period}}$ years, collectively represented as $C_{\ell}^{\text{Fiber}}$. Considering that both the HS and RS systems were optimized in Sections \ref{sec:preliminary}, \ref{sec:RadioStripesConstruction} and \ref{sec:MSTConstruction}, the TCO for Tier 1 remains constant at this stage, where X refers to either RS or MST, and it can be expressed as:

\begin{equation}\label{Tier2TCO}
    C_{\text{T}_1}^{\mathcal{A}} = \sum_{w=1}^{W} A_w C_{}^{\text{ONU}} + \sum_{i=1}^{G} \eta^{\text{Fiber}} d_{\mathcal{G}_i}^{\text{X}}.
\end{equation}

\subsection{Tier 2 optimization}
The second tier of optimization focuses exclusively on optimizing the fronthaul links between leading APs ($\mathcal{M}_w$) and their associated DUs. We formulate a unique objective function for each candidate fronthaul technology to minimize the TCO of the fronthaul network by accounting for both OPEX and CAPEX of each technology, where CAPEX is further divided into deployment costs per AP ($C_{\ell}$) and DU ($C_{w}$).

\textbf{Fiber-based Fronthaul:}
For every leading AP $\ell$ utilizing fiber, the incurred costs comprise the ONU cost ($C_{}^{\text{ONU}}$), which incorporates the OADM cost, installation cost, O\&M costs over $N_{\text{period}}$ years, collectively represented as $C_{\ell}^{\text{Fiber}}$. Additionally, the cost of trenching and burial of fiber cables per meter is given by $\eta^{\text{Fiber}}$. At the DU side, the OTN cost associated with DU $w$ serving fiber-connected leading APs is denoted by $C_{w}^{\text{Fiber}}$. This includes the necessary colocated infrastructure at the DU to facilitate fiber-based fronthaul, such as MUXs, OLTs, etc. Therefore, the TCO for fiber-based fronthaul is expressed as:
\begin{equation}\label{pre-Fiber}
\begin{aligned}
C^{\text{Fiber}} = &  \sum_{w=1}^{W} \sum_{\ell=1}^{M_w}  \left( \underbrace{C_{}^{\text{ONU}} + N_{\text{period}}  C_{\text{O\&M}}^{\text{Fiber}}}_{C_{\ell}^{\text{Fiber}}} + \eta^{\text{Fiber}}   d_{w \ell}\right) \\
& + \sum_{w=1}^{W} \left( \underbrace{C^{\text{OLT}} + C^{\text{OTN}} + C^{\text{Other}}}_{C_{w}^{\text{Fiber}}} \right).
\end{aligned}
\end{equation}

\textbf{mmWave-based Fronthaul:} For mmWave-based fronthaul, the TCO per leading AP connection, including annual power consumption, installation, and O\&M costs, is denoted by $C_{\ell}^{\text{mmW}}$. On the other hand, each DU that serves mmWave-based leading APs will have a massive MIMO antenna device with $N_{\text{DU}}$ elements, with its cost denoted by $C_{w}^{\text{mmW}}$. Hence, the TCO for the mmWave-based fronthaul is given by:
\begin{equation}
\begin{aligned} 
C^{\text{mmW}} = & \sum_{w=1}^{W} \sum_{\ell=1}^{M_w}  \left( \underbrace{C_{\ell}^{\text{mmW-RX}} + N_{\text{period}} C_{\text{O\&M}}^{\text{mmW}}}_{C_{\ell}^{\text{mmW}}} \right) + \sum_{w=1}^{W} C_{w}^{\text{mmW}}. \nonumber
\end{aligned}
\end{equation}
\begin{remark}
The DU cost $C_{\mathrm{DU}}^{\mathrm{pool}}$ used in this study reflects a pooled compute infrastructure model rather than standalone edge units. This cost is fixed at \$91,035 based on FCC estimates~\cite{FCC-UScosts}, and includes shared resources such as cooling, power, and rack space. While our model does not explicitly scale cost with the number of connected APs or implement load-aware pooling, it captures the trade-off between centralized and distributed computing by simulating scenarios with different numbers of DU pools (ranging from 2 to 12). This setup enables analysis of centralization impact on cost and fronthaul deployment. Future work may incorporate dynamic, load-dependent pooling models to more precisely capture cloudification efficiency.    
\end{remark}

\textbf{FSO-based Fronthaul Network TCO:} In FSO-based fronthaul, we focus on P2P FSO transmission with dedicated TXs and RXs. Simply, for every leading AP that uses FSO, we have a reserved exclusive link and equipment, and we could aggregate these costs into a single term $C_{}^{\text{FSO}}$, representing the average cost of deployment for every group $\mathcal{G}_i$ with its leading AP utilizing FSO for fronthauling. Hence, the TCO for FSO-based fronthaul links can be expressed as follows:
\begin{equation}\label{FSO}
\begin{aligned} 
C^{\text{FSO}} = & \sum_{w=1}^{W} \sum_{\ell=1}^{M_w} \left( \underbrace{C_{\ell}^{\text{FSO}} + C_{\text{install}}^{\text{FSO}} + N_{\text{period}} C_{\text{O\& M}}^{\text{FSO}} + C_{w}^{\text{FSO}} }_{C_{}^{\text{FSO}}} \right). 
\end{aligned}
\end{equation}

\textbf{Final Objective Function:} By combining the TCO of all fronthaul technologies (Fiber, mmWave, and FSO), the final joint cost objective function can be expressed as follows:
\begin{equation}\label{finalobjectiveRS}
\begin{aligned}
g_{\rm o}({\mathbf{x}}, {\mathbf{z}}, {\bf u}, {\bf v}, {\bm \kappa}) & = \sum_{w=1}^{W} \sum_{\ell=1}^{M_w}   x_{w \ell}  \left( C_{\ell}^{\text{Fiber}} + \eta^{\text{Fiber}}   d_{w \ell} \right) \\ 
& +  \sum_{w=1}^{W} \sum_{\ell=1}^{M_w} \left[  z_{w \ell} C_{\ell}^{\text{mmW}} +  u_{w \ell}  C_{}^{\text{FSO}} \right] \\
& + \sum_{w=1}^{W} \left[ v_w  C_{w}^{\text{mmW}} + \kappa_w  C_{w}^{\text{Fiber}} \right],  
\end{aligned}
\end{equation}

where $x_{w \ell}, z_{w \ell}, u_{w \ell} \in \{0, 1\}$ are the binary controlling variables for leading APs selecting fiber, FSO or mmWave, respectively. These vectors will have their $\ell$-th entry being 1 to indicate if the $\ell$-th AP is using the corresponding technology for fronthauling, and 0 otherwise.  The integer $\kappa_w \in \mathbb{Z}$ is a variable indicating the number of fiber OTNs required at the DU side, based on the number of leading APs using fiber associated with the $w$-th DU, and the capacity of OTNs ($\Theta$) in terms of how many fiber-based leading APs they can support. Lastly, $v_w \in \{0, 1\}$ is a binary variable indicating if a DU $w$ has any connected mmWave leading APs if 1, and 0 otherwise.

Effective fronthaul planning and joint optimization of multiple technologies necessitate the incorporation of practical and realistic constraints guiding the selection process of the most cost-effective technologies. These constraints can be divided into three main categories, namely: General architectural constraints, technology-specific constraints, QoS metrics constraints, and are as follows: 

\textbf{General Architectural Constraints:} This category ensures robust formulation and network components association, and the constraints in this category include the following:
       \begin{equation}\label{first-constraint}
\sum_{w=1}^{W} \left( x_{w \ell} +  z_{w \ell} + u_{w \ell} \right) = 1 , \;\;\; \forall \ell \in \mathcal{M}_w, \;\; \forall w \in \mathcal{W},
        \end{equation}
        \begin{equation}\label{second-constraint}
        x_{w \ell}, \, z_{w \ell}, \, u_{w \ell}\in \{0, 1\} , \;\; \forall \ell \in \mathcal{M}_w,
    \end{equation}
\begin{equation}\label{thirdthird-constraint}
    v_w \in \{0, 1\}, \;\; \forall w \in \mathcal{W},
\end{equation}
\begin{equation}\label{third-constraint}
    \kappa_w \in \mathbb{Z},  \;\;\; \forall w \in \mathcal{W},
\end{equation} 

where \eqref{first-constraint} ensures that each group $\mathcal{G}_i$ is only associated with a single DU $w$ and selecting one fronthaul technology for every leading AP $\ell \in \mathcal{M}_w$, while both \eqref{second-constraint} and \eqref{thirdthird-constraint} define the binary controlling variables, and \eqref{third-constraint} defines the fiber-associated controlling variable $\kappa_w$ as an integer.

\textbf{QoS Metrics Constraints:} This category ensures that individual leading APs and the entire network meet performance standards, and it includes:
    \begin{equation}\label{sixth-constraint}
x_{w \ell}  R_{w \ell}^{\text{Fiber}}  + z_{w \ell}  R_{w \ell}^{\text{mmW}} +  u_{w \ell}  R_{w \ell}^{\text{FSO}}   \geq  \psi^{\text{FSX}}, \, \forall \ell \in \mathcal{M}_w ,  \forall w \in \mathcal{W},
    \end{equation}
    \begin{equation}\label{Availability-constraint}
    \sum_{\ell=1}^{M_w}  \left(x_{w \ell}  \zeta_{}^{\text{Fiber}} + z_{w \ell} \zeta_{}^{\text{mmW}} + u_{w \ell} \zeta_{}^{\text{FSO}}  \right)  \geq M_w  \zeta_{}^{\text{SLA}}, \; \forall w \in \mathcal{W},
\end{equation}

where constraint \eqref{sixth-constraint} guarantees that the selected fronthaul technology minimizing the TCO for each leading AP $\ell \in \mathcal{M}_w$ is also meeting a fixed fronthaul capacity threshold imposed by the FS option requirements $\psi^{\text{FSX}}$, where FSX refers to options FS7.2x or FS8. While the availability constraint \eqref{Availability-constraint} ensures that for the entire $N_{\text{period}}$, the Service Level Agreement (SLA) of SPs is not breached, by capturing the average uptime for a network. Availability means that all leading APs in the network are ready for immediate use, by ensuring that the number of fully operational leading APs associated with each DU $w$ are always larger than or equal to $M_w\zeta_{}^{\text{SLA}}$.

\textbf{Technology-Specific Constraints:} This category ensures that the unique characteristics of each technology are accurately represented, and it includes:

\textit{Fiber Constraint:} In fiber-based fronthaul, the maximum number of groups with their leading APs choosing fiber for fronthauling connected to a single DU depends on the capacity ($\Theta$) of the OTN deployed at the DU side. For simplicity, we are taking the ratio option of splitters to determine the bottleneck capacity of OTNs. Assuming a single splitter supports $\Theta$ number of links (i.e., 1:$\Theta$ PON), then if more than $\Theta$ APs associated with $w$-th DU are using fiber, and additional OTN must be deployed, effectively doubling the cost of $C_{w}^{\text{Fiber}}$. 
\begin{equation}\label{fourth-constraint}
\sum_{\ell=1}^{M_w} \frac{x_{w \ell}}{\Theta} \leq \; \kappa_w \; \leq \sum_{\ell=1}^{M_w} \frac{x_{w \ell}}{\Theta} + 1 - \epsilon,  \;\;\; \forall w \in \mathcal{W}.
\end{equation}

\textit{mmWave Constraint:} Similar to fiber, this constraint guarantees that the $w$-th DU will not be equipped with a mmWave antenna device unless there is at least one leading AP employs mmWave technology for fronthauling. As a result, since $v_w$ is a binary variable, $v_w$ will be equal to 1 if and only if $ \sum_{\ell=1}^{L_w} z_{w \ell} \neq 0$.
    \begin{equation}\label{FinalConstraint}
\sum_{\ell=1}^{M_w} \frac{z_{w \ell}}{M_w} \leq \; v_w \; \leq \sum_{\ell=1}^{M_w} z_{w \ell},  \;\;\; \forall w \in \mathcal{W}.
    \end{equation}

Constraints \eqref{fourth-constraint} and \eqref{FinalConstraint}, combined with the definition of their decision variables in constraints \eqref{thirdthird-constraint} and \eqref{third-constraint} resemble a linearized form of a ceiling function applied to the decision variables $\kappa_w$ and $v_w$.

\subsection{Final Formulation and Proposed Algorithm}
By combining the TCO of both tiers, the final optimization problem for both the studied HS and RS-enabled CF-mMIMO networks that aims to minimize the fronthaul network TCO and ensures effective performance through the selection of fronthaul technologies is presented as follows:
\begin{subequations}\label{Final-Formulation}
\begin{align}
\min_{\substack{
x_{w \ell}, \, z_{w \ell}, \, u_{w \ell} \\
v_{w}, \kappa_{w} 
}}  
\;\;  g_{\rm o}({\mathbf{x}},  & {\mathbf{z}}, {\bf u}, {\bf v}, {\bm \kappa}) +  C_{\text{T}_1}^{\mathcal{A}},  \\
\text{subject to} \quad &
\begin{aligned}
&\eqref{first-constraint} - 
\eqref{FinalConstraint}.
\end{aligned}
\end{align}
\end{subequations}

The above formulation is classified as an Integer Linear Program (ILP), which is a combinatorial optimization problem characterized by the combination of binary ($x_{w \ell}, z_{w \ell}, u_{w \ell}, v_{w}$) and integer ($\kappa_w$) variables, alongside the linear nature of the objective function and constraints. To achieve the optimal solution of equation \eqref{Final-Formulation}, specifically for the second tier, we employ the branch-and-bound method \cite{gurobi}, and the steps are outlined in Algorithm \ref{ALGFinal}. The algorithm returns the fronthaul technologies selection ($\mathbf{x}, \mathbf{z}, \mathbf{u}, \mathbf{v},$ and $\bm{\kappa}$), cost values of each technology ($C^{\text{Fiber}},$ $C^{\text{mmW}},$ and $C^{\text{FSO}}$), in addition to the TCO of both tiers of the optimized network. Furthermore, while Algorithm 1 provides the topology-level input, the core optimization problem presented in Section IV is an ILP, which is solved using a branch-and-bound method. This guarantees convergence to a globally optimal solution due to the discrete and bounded nature of the solution space. Since this framework is intended for offline planning rather than real-time deployment, convergence speed is not a limiting factor. Nonetheless, we observe stable and efficient convergence behavior in all tested configurations.

\begin{algorithm}[t!]
\caption{Proposed algorithm for solving the hierarchical and radio-stripes-based CF-mMIMO fronthaul planning and cost minimization ILP in eq.~\eqref{Final-Formulation}.}\label{ALGFinal}
\begin{algorithmic}[1]
\State \textbf{Input:} Set up the given technology-specific and general parameters values from TABLE \ref{Table5:input-parametersRS} as input to the system.
\For{realization $r = 0$}
\State {$r \gets r + 1$;}
\State \textbf{Initialization:}
\State {Obtain NOFAC Algorithm \ref{ALG1} outputs as input.}
\For {$\bm{\forall w \in \mathcal{W}}$}
\State \text{Initialize $\mathbf{v}_{}^{(r)}$ and $\bm{\kappa}_{}^{(r)}$ $\gets$ $0$;}
\For {$\bm{\forall \ell \in \mathbf{M}_w}$}
\State \text{Compute $d_{w\ell}$, $R_{w\ell}^{\text{Fiber}}$, $R_{w\ell}^{\text{mmW}}$, and  $R_{w\ell}^{\text{FSO}}$. }
 \State \text{Initialize $\mathbf{x}^{(r)}$, $\mathbf{z}^{(r)}$ and $\mathbf{u}^{(r)}$ $\gets$ $0$; }
\EndFor
\EndFor
\State \text{Compute $C_{\text{T}_1}^{\mathcal{A}}$ from eq.~\eqref{Tier2TCO}}.
\State \text{Solve eq.~\eqref{Final-Formulation} using branch and bound to obtain:}
\State   $\mathbf{x}_*^{(r)}$, $\mathbf{z}_*^{(r)},  \mathbf{u}_*^{(r)}, \mathbf{v}_*^{(r)}, \text{and }  \bm{\kappa}_*^{(r)}.$
    \EndFor
\State \textbf{Output:} $\mathbf{x}, \mathbf{z}, \mathbf{u}, \mathbf{v},$ $\bm{\kappa},$ $C^{\text{Fiber}},$ $C^{\text{mmW}},$ $C^{\text{FSO}},$ $C_{\text{T}_1}^{\mathcal{A}},$ and $g_{\rm o}({\mathbf{x}}, {\mathbf{z}}, {\bf u}, {\bf v}, {\bm \kappa})$.
\end{algorithmic}
\end{algorithm}

\begin{table*}[t]
\renewcommand{\arraystretch}{1.25}
\centering
\caption{Input system parameters values used in simulation for the HS and RS-based CF-mMIMO system.}
\label{Table5:input-parametersRS}
\begin{small} 
\begin{tabular}{|c|c||c|c||c|c||c|c|}
\hline
\multicolumn{2}{|c||}{\textbf{Fiber}} &
\multicolumn{2}{|c||}{\textbf{mmWave}} & \multicolumn{2}{c|}{\textbf{FSO}} & \multicolumn{2}{c|}{\textbf{General}} \\
\hline
\hline
{$\eta^{\text{Fiber}}$} & \$26 & {$C_{w}^{\text{mmW}}$} & \$34,500 & {$C^{\text{FSO}}$} & \$15,000 \cite{FSO-Main} &  \textbf{Coverage area (A)} & {2 km $\times$ 2 km} \\
\hline
{$C_{}^{\text{ONU}}$} & \$6,502 & {$C_{\ell}^{\text{mmW-RX}}$} & \$6,000 & {$C_{\text{O\&M}}^{\text{FSO}}$} & \$13,000 \cite{FSO-Main} &{${C}_{{DU}}^{{pool}}$} & \$91,035 \\
\hline
{$C^{\text{OTN}}$} & \$61,727 & {$C_{\text{O\&M}}^{\text{mmW}}$} & \$13,000 & {$r_r$, $V$} & {0.05 m, 400 m} & $\bm{\psi^{\text{FS7.2x}}}$ & {1.73} Gbps \\
\hline
{$C^{\text{OLT}}$} & \$20,100  & {$N_{\text{AP}}$}, {$N_{\text{DU}}$} & {1}, {256} & {$\eta_t$}, {$\eta_r$} & {0.5, 0.5} &  {$\bm{\psi^{\text{FS8}}}$} & {2.95} Gbps\\
\hline 
 {$C_w^{\text{Fiber}}$} & \$81,827 & {$p_{t}^{\text{mmW}}$} & {120 W} & {$p_t^{\text{FSO}}$}, {$\theta_t$} & {0.5 W}, 10 mrad  & {${L}$} & {1000} \\
\hline
{$C_{\text{O\&M}}^{\text{Fiber}}$} & \$2,285 & $q$ & $6$  & $C_n^2(h)$ & $10^{-15} \text{m}^{-2/3}$~\cite{Turbulence2} & ${L^{\text{max}}}$ & 9  \\
\hline
{$R_{}^{\text{Fiber}}$} & {10 Gbps} & {$\text{BW}^{\text{mmW}}$} & {2.5 GHz} & {$E_p$} & $1.2823 \times 10^{-19}$ &  {$N_{\text{period}}$} & {1 year} \\
\hline
{$\zeta_{}^{\text{Fiber}}$} & {1.00} & {$\zeta_{}^{\text{mmW}}$} & {0.99999} & {$\zeta_{}^{\text{FSO}}$} & {0.9975}   & {${\zeta}_{}^{{SLA}}$} & {0.9999} \\
\hline
{$\Theta$} & {16} & {$f_{c}$} & 80 GHz & {$\lambda$}, {$N_b$} & 1550 nm, {$100$} & $\mathbf{g_s}$, $\mathbf{g_m}$ & {15, 3}  \\
\hline
\hline
\multicolumn{2}{|c||}{\multirow{1}{*}{\textbf{Variables:}}} & \multicolumn{2}{c||}{No. of groups (${G}$)} & \multicolumn{2}{c||}{No. of DUs (${W}$)} & \multicolumn{2}{c|}{No. of APs per group ($L_{\mathcal{G}_i}$)}  \\
\hline
\end{tabular}

\end{small}
\end{table*}

\section{Numerical Results}\label{CH5}
We evaluate the effectiveness of the proposed framework to understand the critical role of fronthaul network planning in CF-mMIMO and, more broadly, in UDNs. In particular, we focus on the use of either HS or RS connection schemes within the O-RAN paradigm. The selection of fronthaul technologies, optimization of TCO and network capacity are examined with respect to several key factors, mainly: \textbf{(1) Varying the number of DUs} (${W}$), impacting distances between DUs and APs. \textbf{(2) Varying the number of AP groups} ($G$) for the same total number of deployed APs $L$, which affects the number of leading APs and non-leading APs, eventually impacting the fronthaul infrastructural units needed. \textbf{(3) Fronthaul capacity thresholds}, namely $\psi^{\text{FS7.2x}}$ and $\psi^{\text{FS8}}$, where $\psi^{\text{FS8}}$ makes constraint \eqref{sixth-constraint} more stringent. Unless otherwise specified, the cost estimates for all equipment in TABLE \ref{Table5:input-parametersRS} are primarily derived from average values reported by the US FCC \cite{FCC-UScosts}.

It is important to note that the simulation results presented in this section are averaged over hundreds of spatially randomized network realizations, reflecting diverse AP distributions, group topologies, and DU placements. This extensive sampling implicitly captures the performance variability induced by different deployment topologies, including effects stemming from radio-stripe interconnections. Although our framework does not explicitly model physical layer processing strategies such as sequential decoding or cooperative fusion, the averaged performance over varied topologies ensures robustness and generalizability of the results. Moreover, the proposed planning strategy is modular and can be adapted to operator-specified topology constraints or processing architectures as needed.

\begin{figure}[t]
    \centering
    \subfloat[Sample of 4 DUs and 150 stripes.\label{Sample1}]{
        \includegraphics[width=0.485\columnwidth]{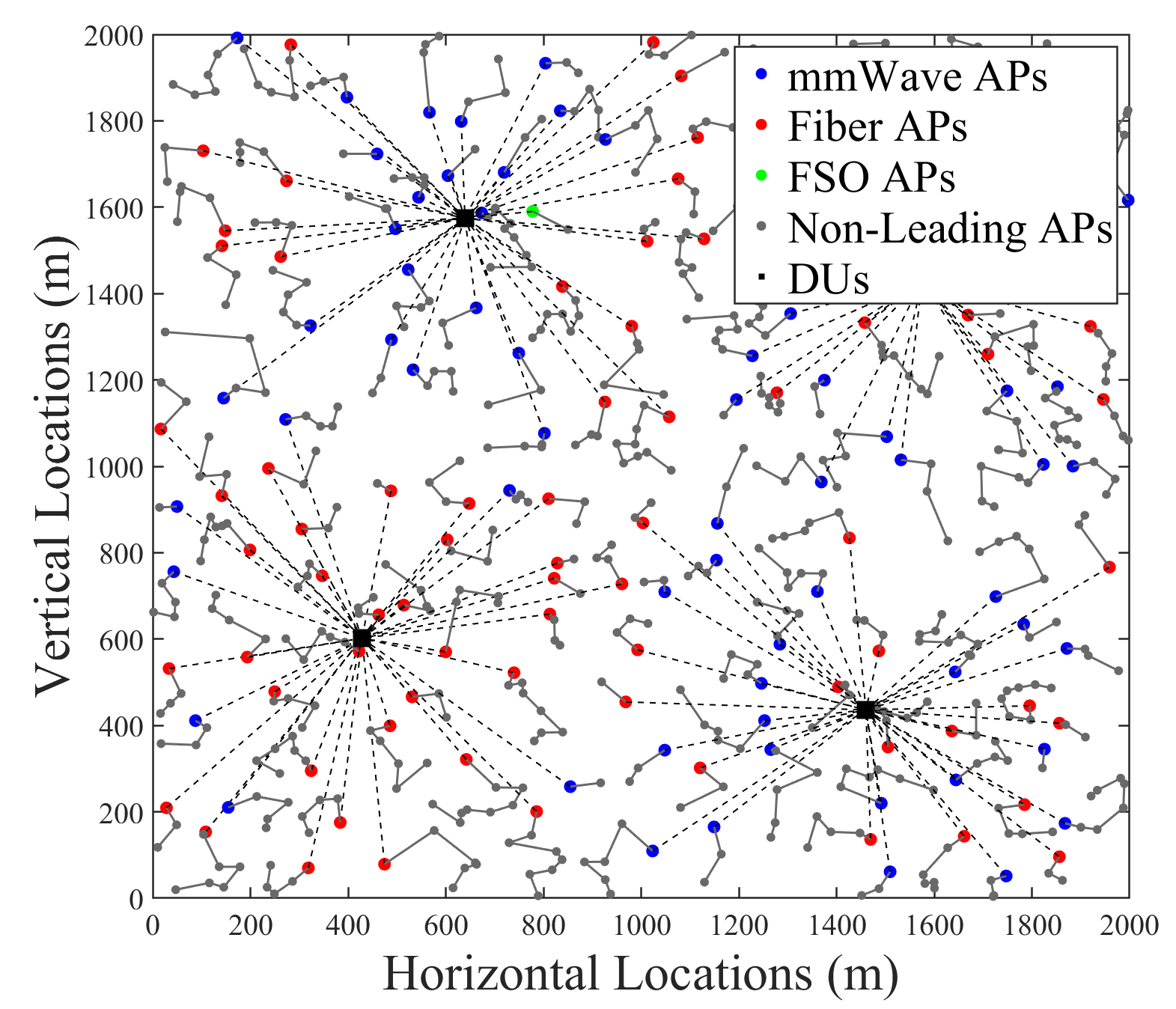}}
    \hfill    
    \subfloat[Sample of 8 DUs and 150 stripes.\label{Sample2}]{
        \includegraphics[width=0.485\columnwidth]{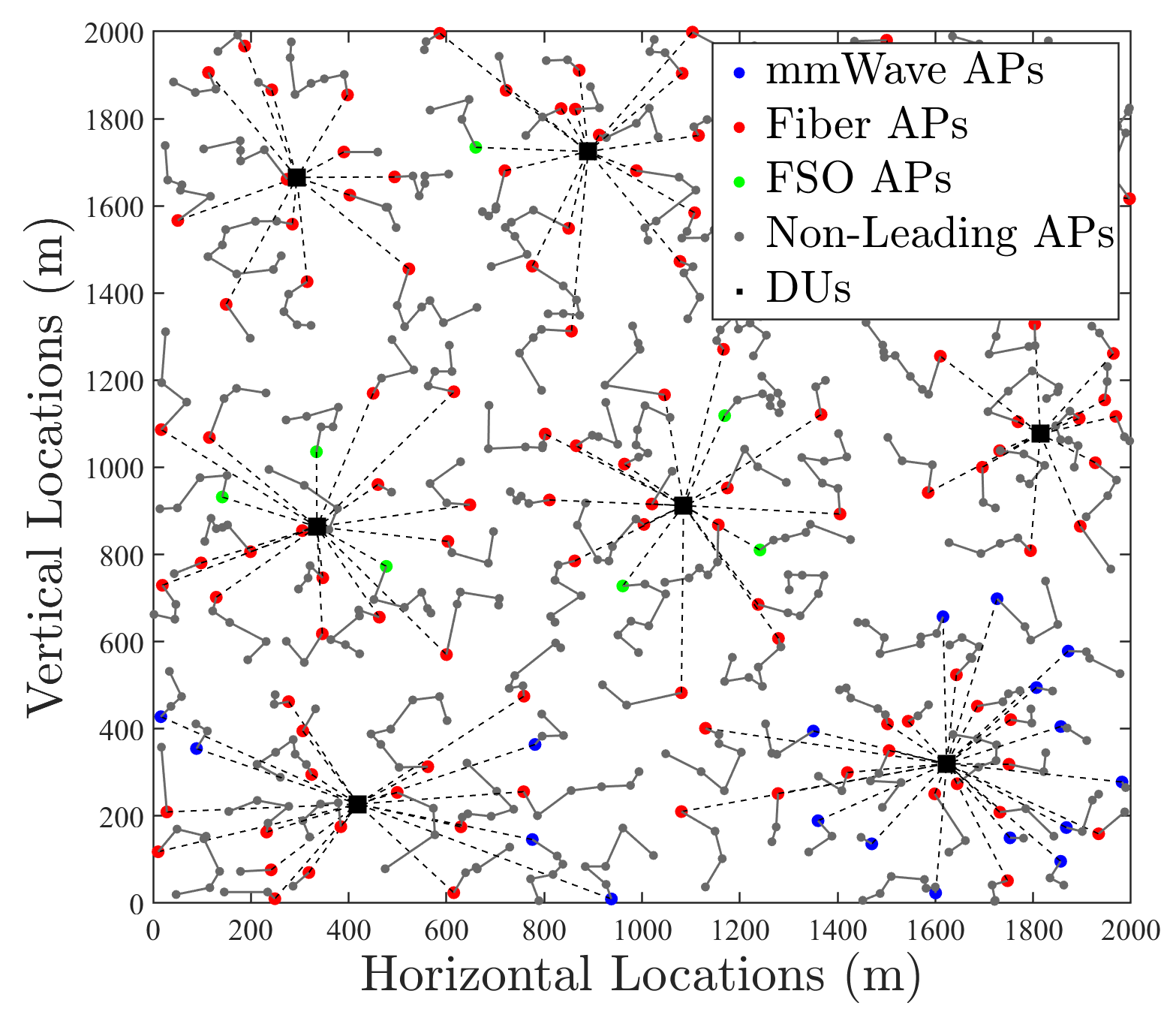}}
    \caption{Optimized fronthaul mixed-technology selection in RS with FS7.2x under different levels of decentralized processing.}
    \label{RSSelection}
\end{figure}
\begin{figure}[t]
    \centering
    \subfloat[Sample of 4 DUs and 150 groups.\label{Sample3}]{
        \includegraphics[width=0.485\columnwidth]{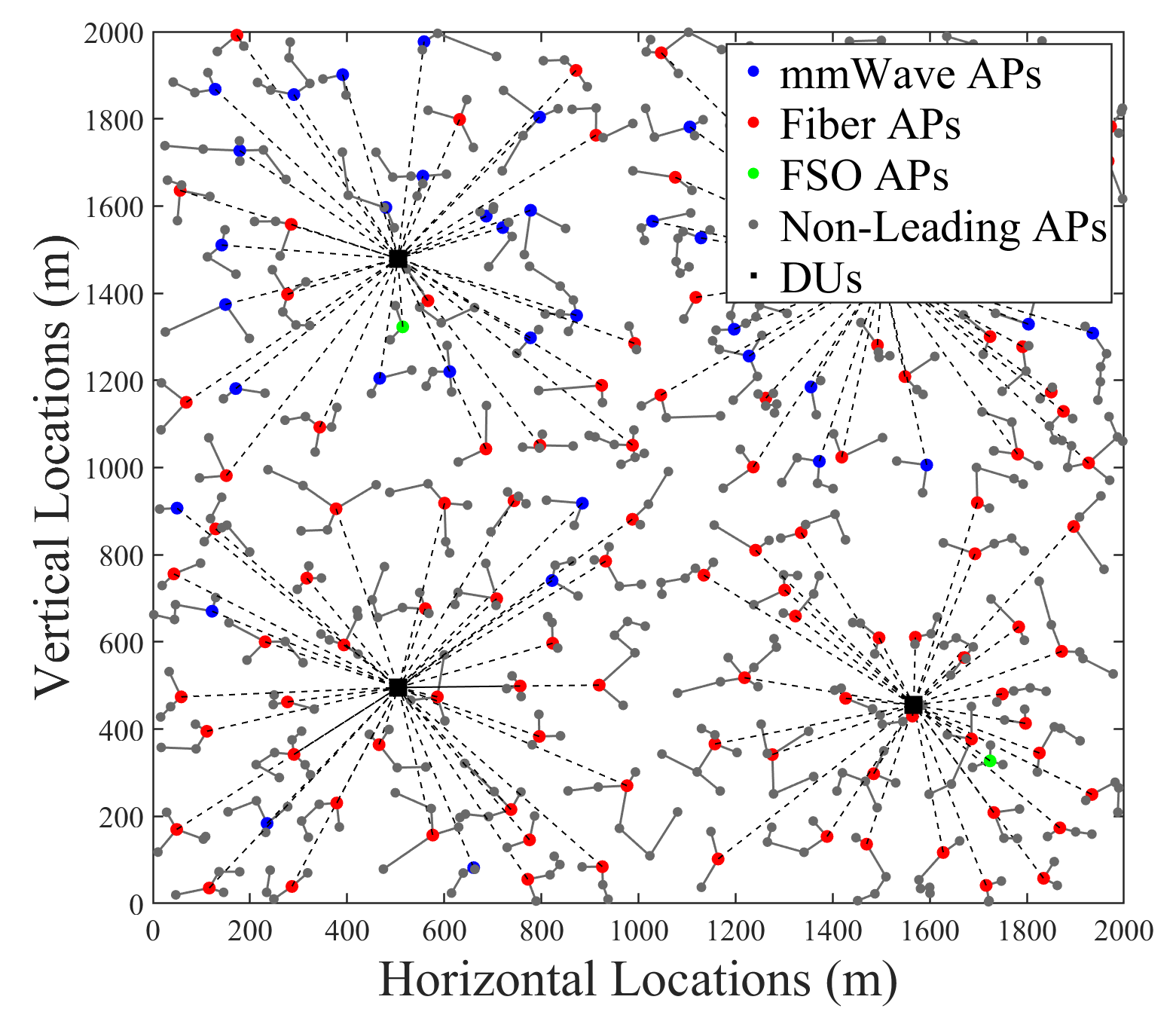}}
    \hfill    
    \subfloat[Sample of 8 DUs and 150 groups.\label{Sample4}]{
        \includegraphics[width=0.485\columnwidth]{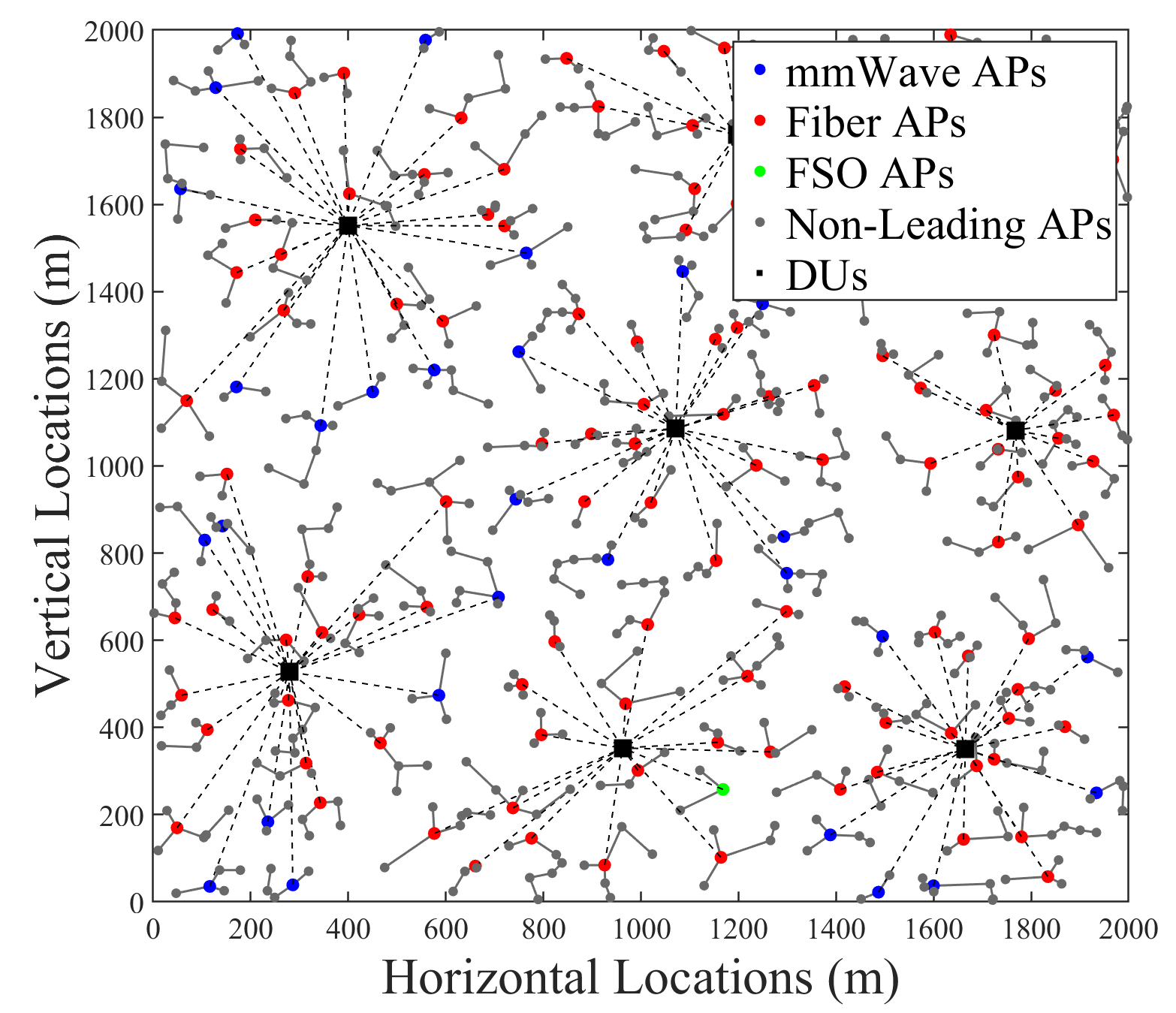}}
    \caption{Optimized fronthaul mixed-technology selection in HS with FS7.2x under different levels of decentralized processing.}
    \label{HSSelection}
\end{figure}

The performance of the optimization framework is evaluated by comparing it against three benchmarks, each focusing on different fronthaul technology strategies for leading APs in the second tier of the network, and these are: \textbf{(a)} \textbf{Standard all-Fiber fronthaul network}, serving as a benchmark for the most reliable performance, satisfying all constraints \eqref{first-constraint} - 
\eqref{FinalConstraint}. \textbf{(b)} \textbf{Suboptimal all-mmWave fronthaul network}, representing an economically appealing benchmark, but falls short of meeting the capacity $\psi^{\text{FSX}}$ constraint \eqref{sixth-constraint} for all leading APs. Lastly, \textbf{(c)} \textbf{Heuristic method for hybrid deployment}, balancing cost and capacity without necessarily achieving optimal TCO. In this method, mmWave is initially assigned to all APs, and if the capacity threshold ($\psi^{\text{FSX}}$) for a leading AP $\ell$ exceeds its mmWave capacity $R_{w\ell}^{\text{mmW}}$, it is switched to fiber, while ensuring compliance with all the remaining constraints.

In our solution, we do not focus on enhancing the bandwidth capabilities of fiber optics or any other fronthaul technology. Instead, we adopt a planning perspective that works within the constraints of existing technologies. While our study considers standard fronthaul capacity requirements based on uncompressed data for FS7.2x and FS8, it is agnostic to specific fronthaul enhancements (e.g., compression, protocol efficiency improvements). Our optimization framework remains applicable and flexible to adapt to future technology upgrades, as it fundamentally operates on cost, capacity, and reliability parameters, regardless of underlying physical layer innovations.

\begin{figure*}[t]
    \centering
    \subfloat[TCO per AP for $W$ = 4.\label{cost1}]{
\includegraphics[width=0.315\textwidth]{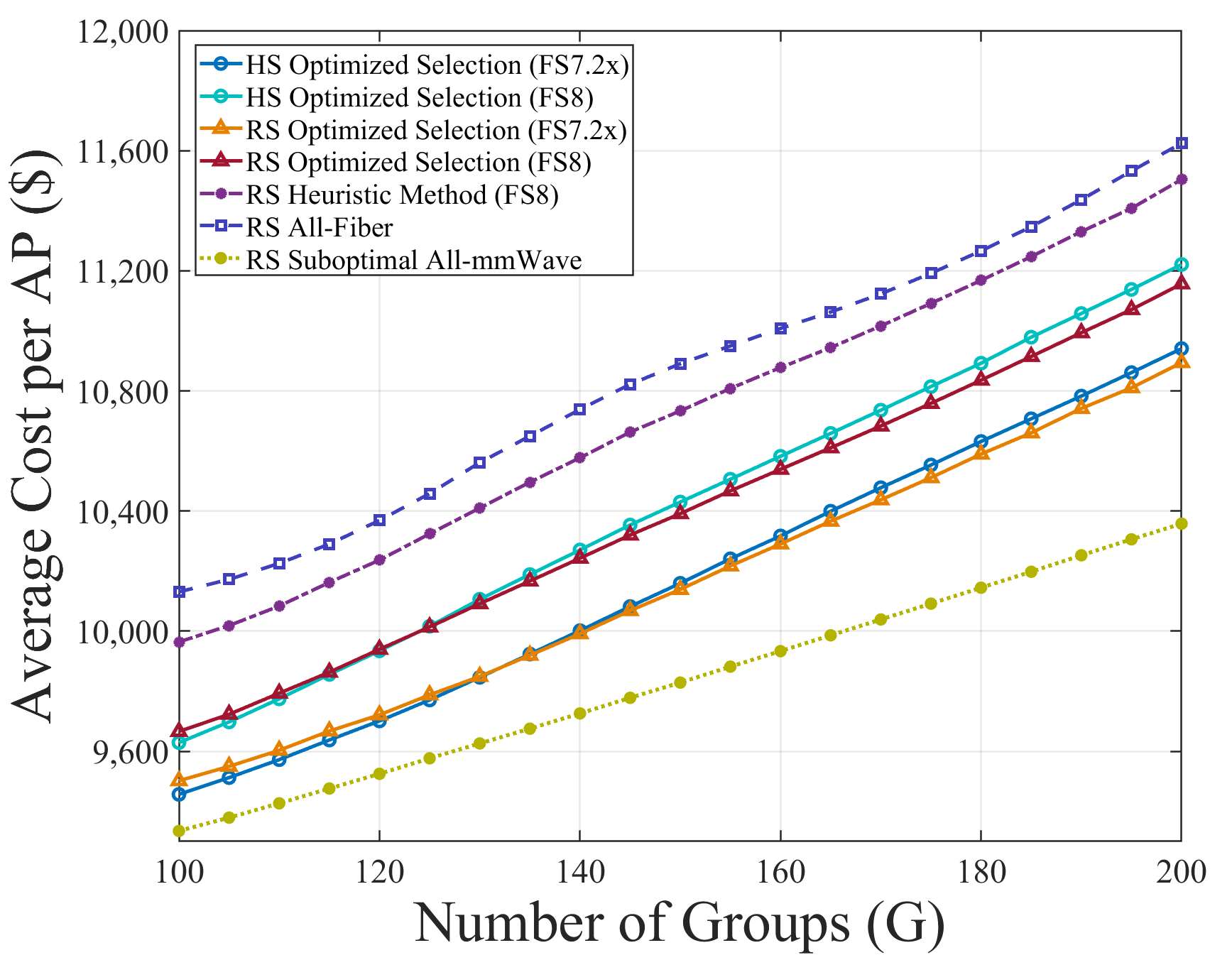}}
    \hfill
    \raisebox{0pt}[\height][\depth]{\hspace{0.1pt}  % Move cost2 two points to the left
    \subfloat[TCO per AP for $W$ = 6.\label{cost2}]{
    \includegraphics[width=0.315\textwidth]{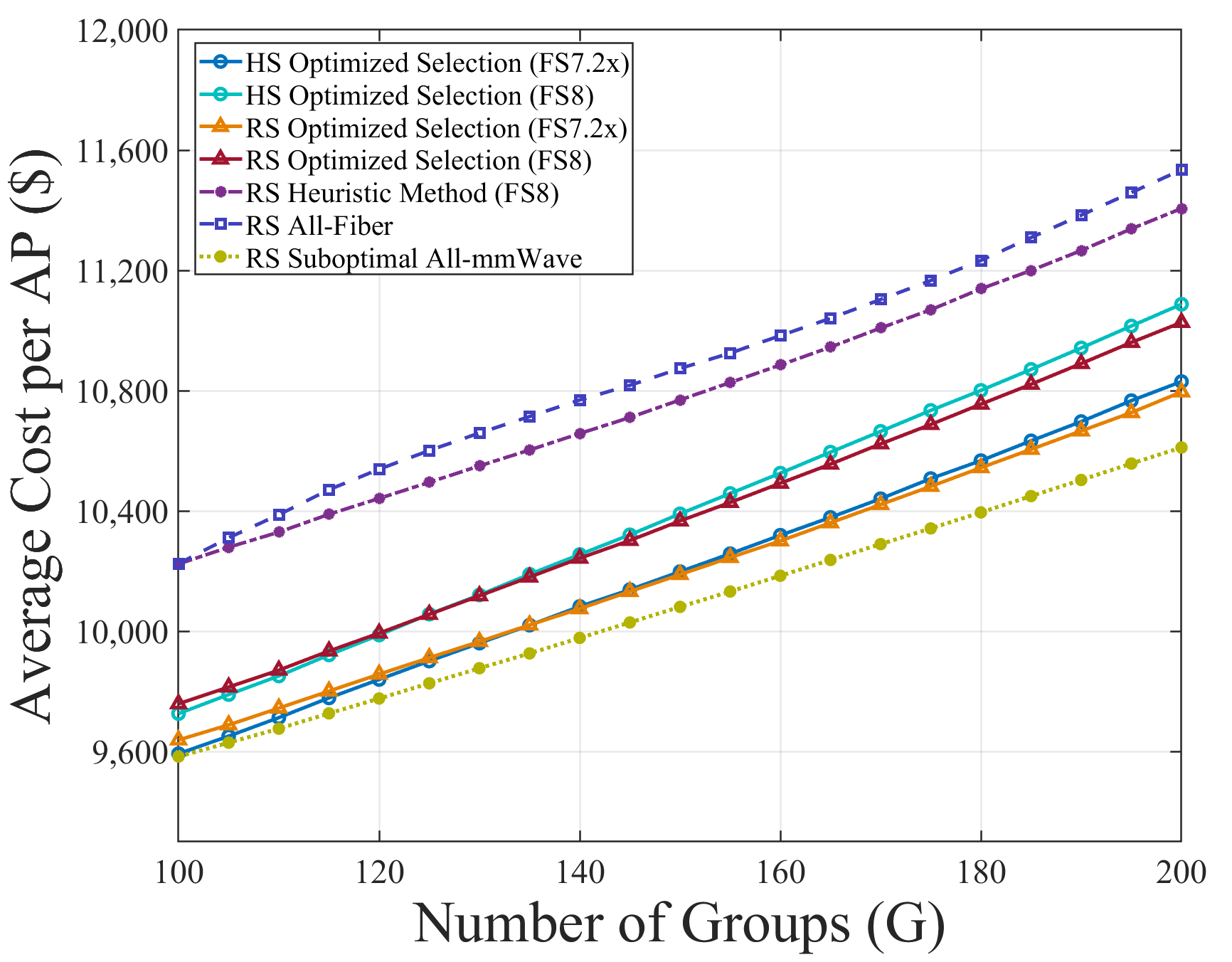}}}
    \hfill
    \raisebox{0pt}[\height][\depth]{\hspace{0.1pt}  % Move cost3 two points to the left
    \subfloat[TCO per AP for $W$ = 8.\label{cost3}]{
\includegraphics[width=0.315\textwidth]{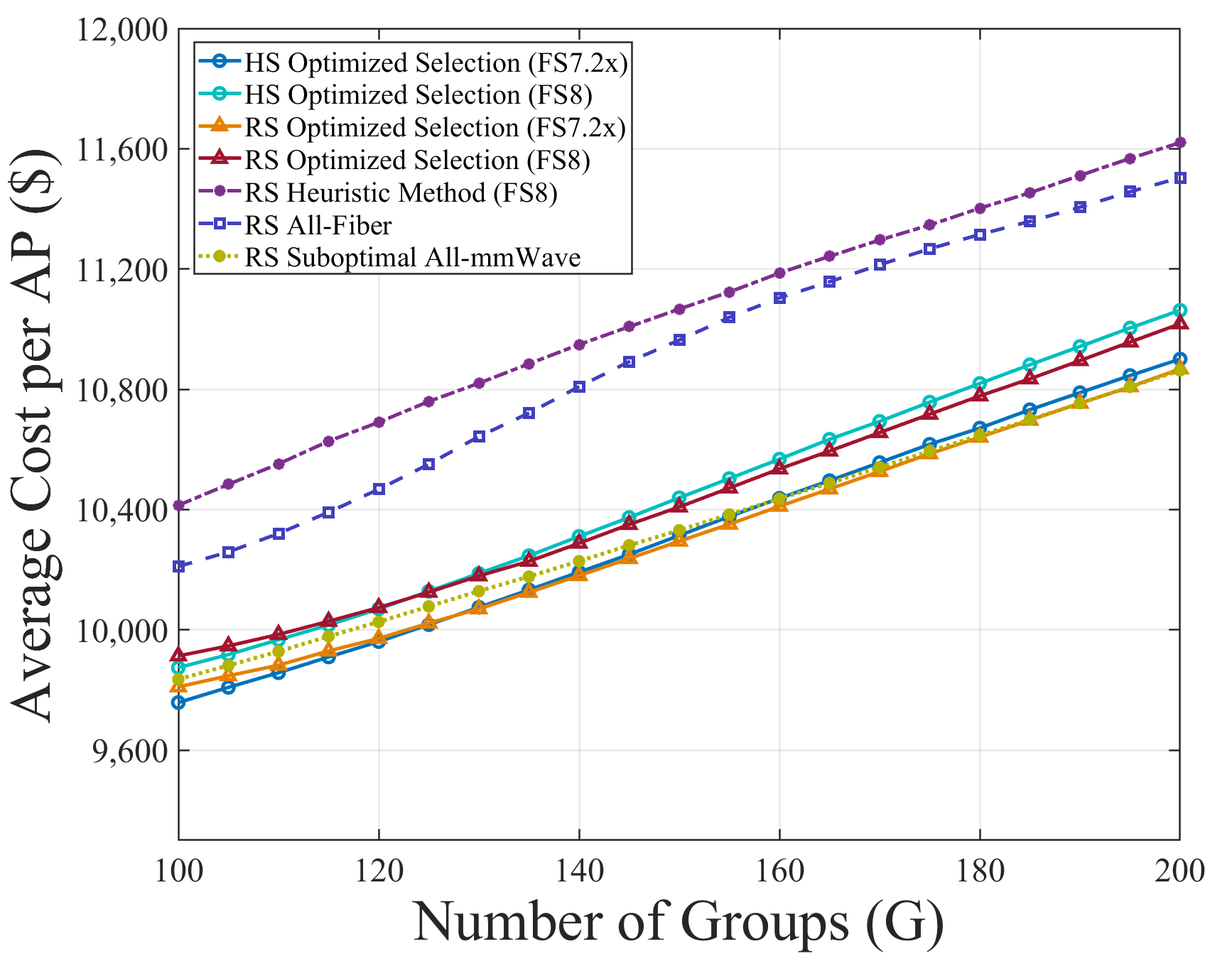}}}
    \caption{Average TCO per AP vs. number of groups ($G$) across different levels of decentralized processing and benchmarks.}
    \label{fig5}
\end{figure*}
\begin{figure*}[t]
    \centering
    \subfloat[Network surplus capacity for $W$ = 4.\label{surplus1}]{
\includegraphics[width=0.315\textwidth]{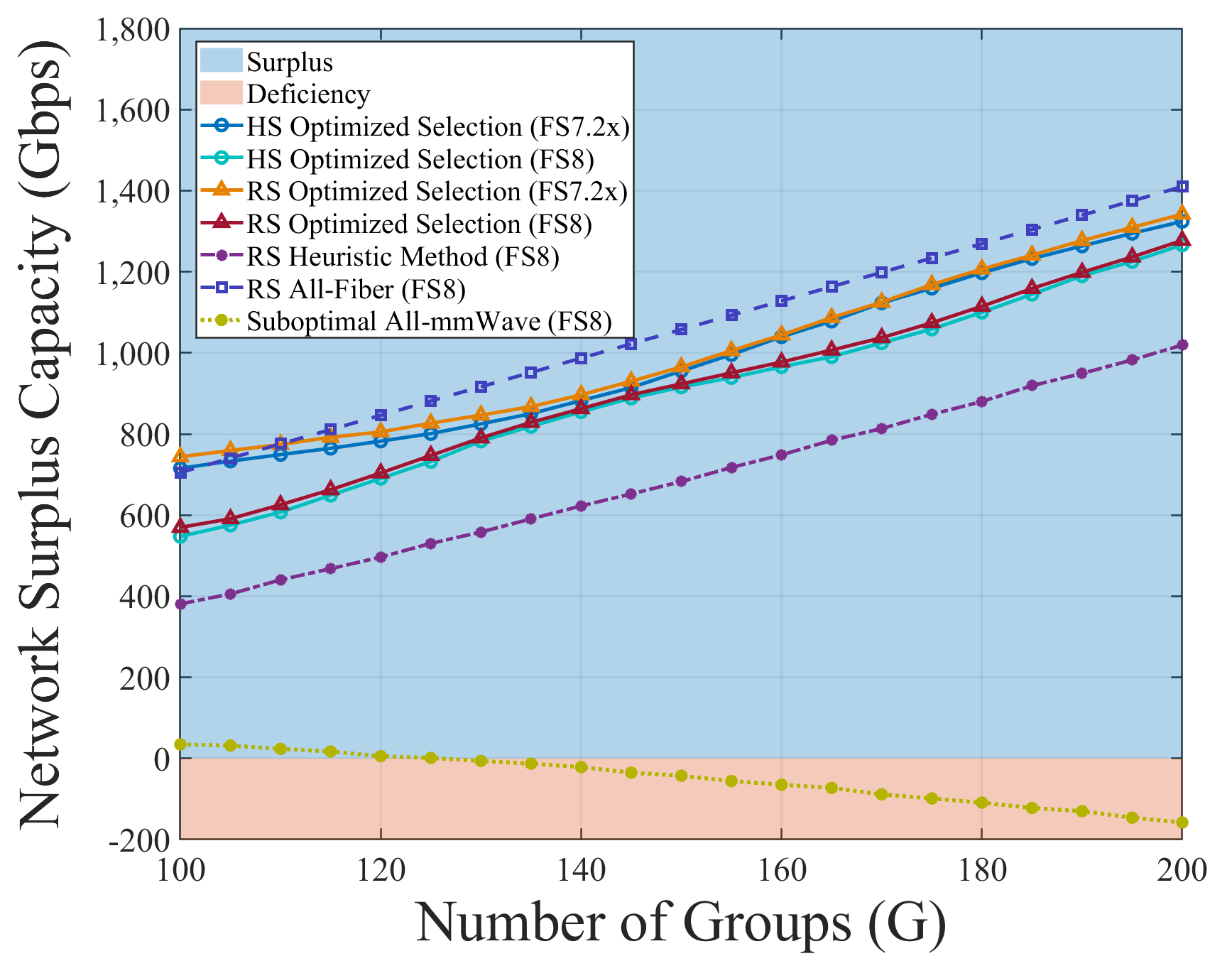}}
    \hfill    
    \subfloat[Network surplus capacity for $W$ = 6.\label{surplus2}]{
\includegraphics[width=0.315\textwidth]{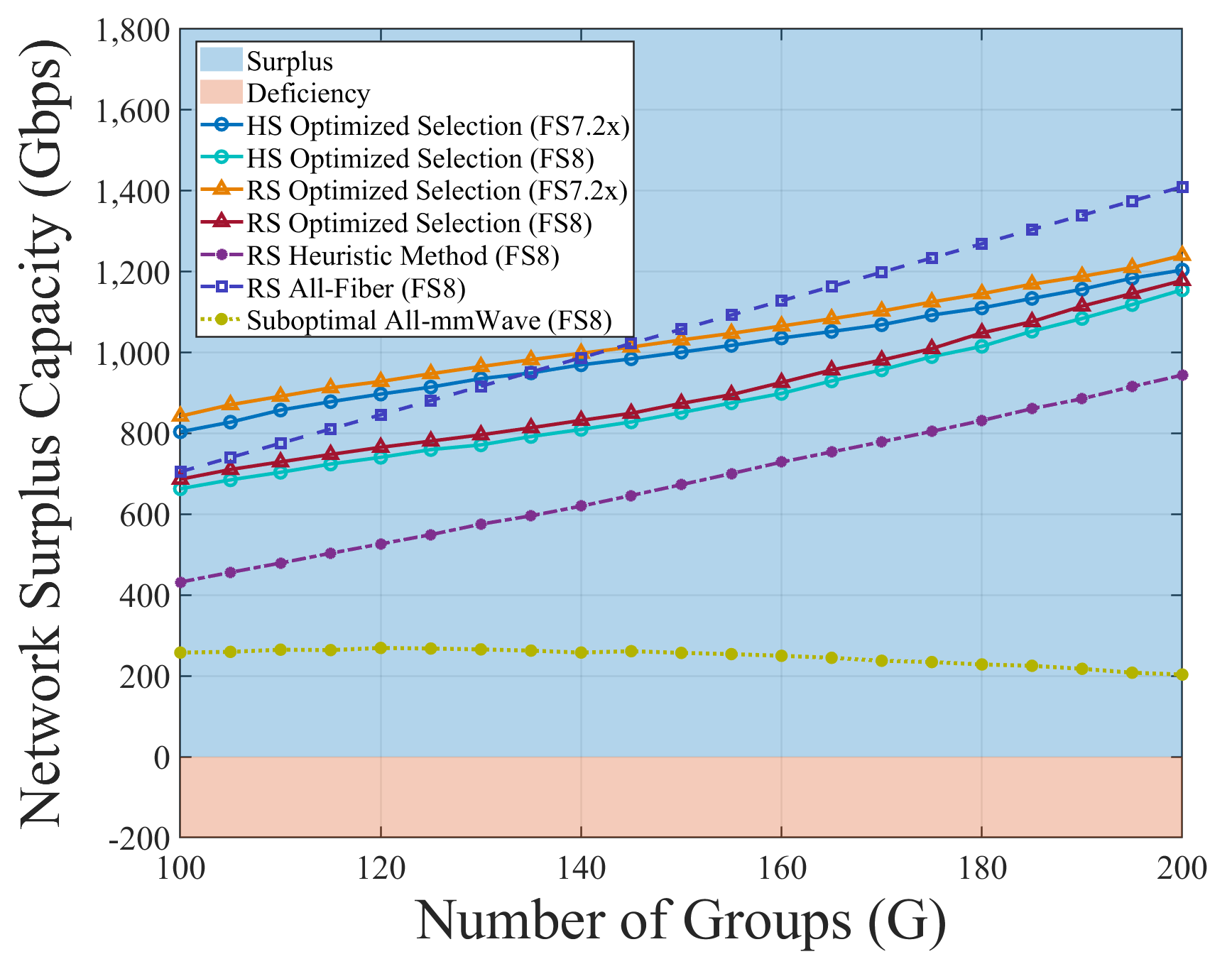}}
    \hfill
    \subfloat[Network surplus capacity for $W$ = 8.\label{surplus3}]{
\includegraphics[width=0.315\textwidth]{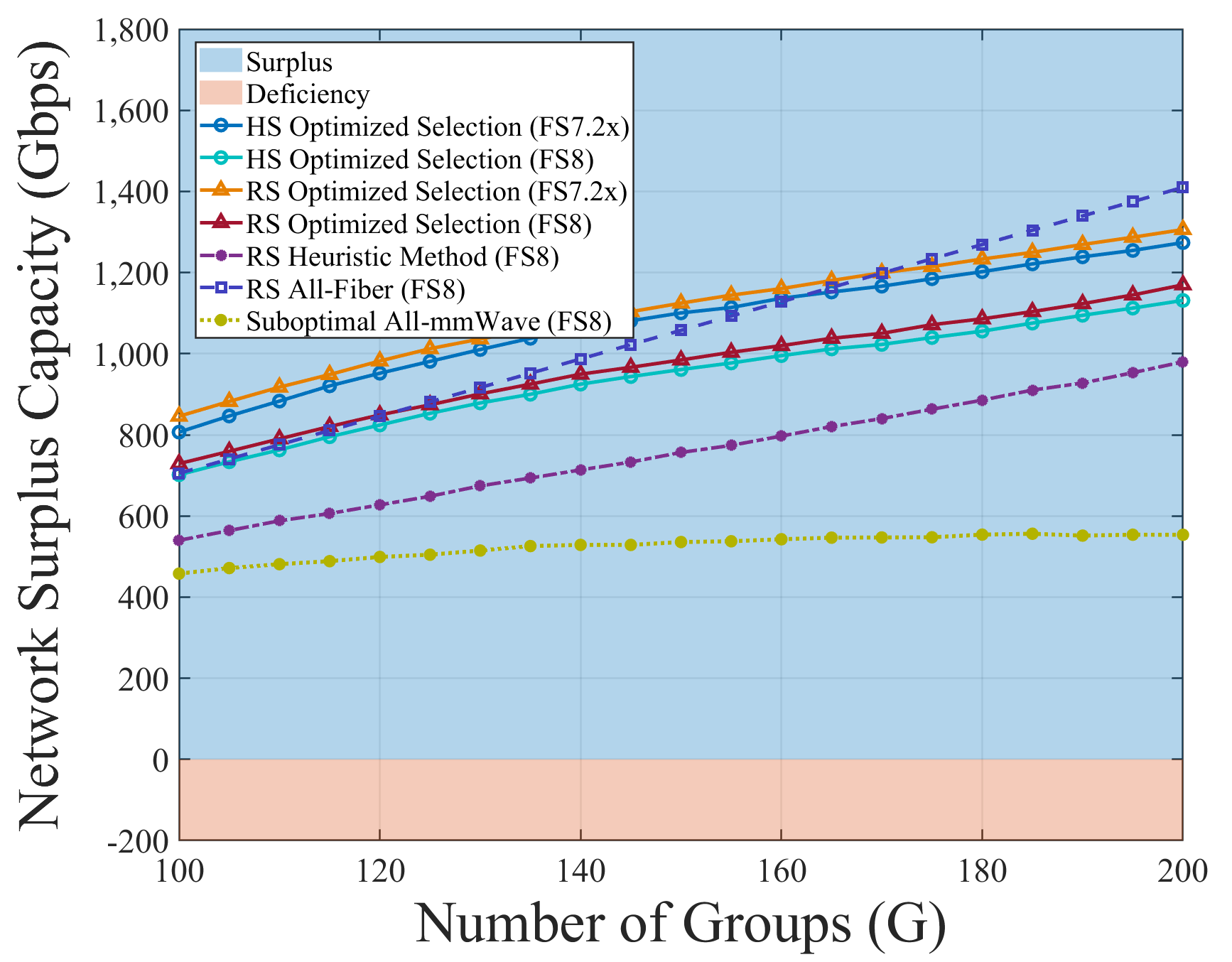}}
    \caption{\centering Surplus capacity vs. number of groups ($G$) across different levels of decentralized processing and benchmarks.}
    \label{SurplusFS}
\end{figure*}

\subsection{Fronthaul Technologies Selection}
 
Figures \ref{RSSelection} and \ref{HSSelection} illustrate fronthaul technology choices for RS and HS topologies under FS7.2x, considering different numbers of deployed DUs (${W}$). Although the results reflect specific realizations, the insights extend to other configurations and traffic scenarios \cite{MyPaper}. For both HS- and RS-enabled CF-mMIMO, APs near DUs typically use fiber due to lower cost, while distant APs rely on mmWave when capacity and reliability allow. FSO adoption remains limited, given its higher cost and lower reliability. Interestingly, and contrary to widely held assumptions, as processing becomes more decentralized and AP–DU distances shrink, fiber deployment increases. Additionally, the optimization framework prioritizes efficient utilization of DU-associated infrastructure, by maximizing connections using the same technology, hence minimizing infrastructural redundancy across technologies.

\subsection{Network Optimized TCO and Number of Groups}  

Figure \ref{fig5} shows the average TCO per AP for different numbers of DUs ($W$) and groups ($G$). Although an all-mmWave deployment often appears to have the lowest cost, it consistently violates the capacity constraint \eqref{sixth-constraint} and is therefore infeasible. Our optimized design achieves the most cost-efficient deployment across both connection schemes and FS options. Increasing group sizes ($L_{\mathcal{G}_i}$) lowers fronthaul costs but adds processing burden on each DU, while deploying more DUs distributes the load more evenly and reduces overhead.

\begin{figure}[t]
    \centering
    \subfloat[FS7.2x average network TCO and Tier 2 cost breakdown.\label{AvgPercentaa}]
        {\includegraphics[width=\linewidth]{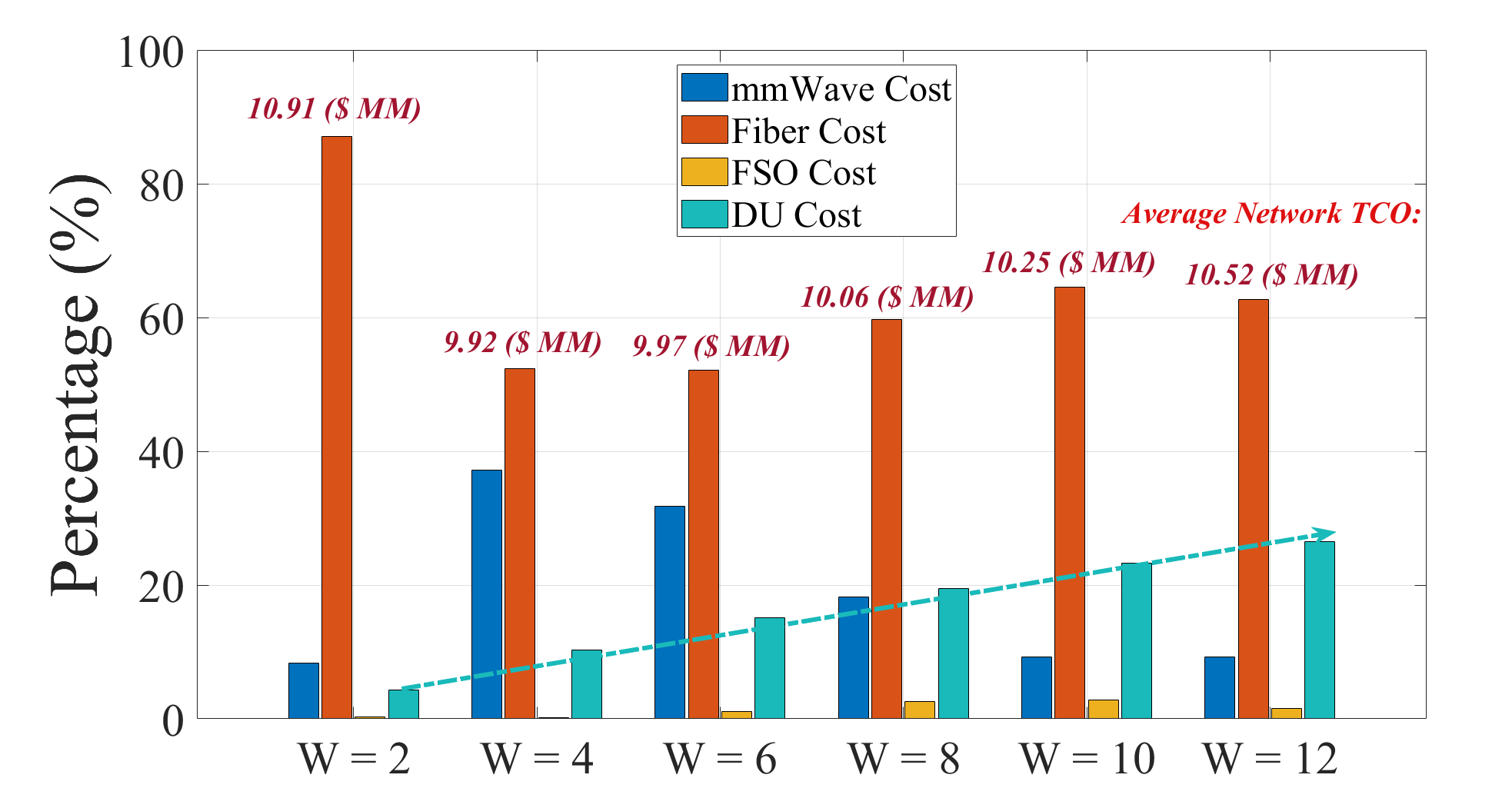} }
\hfill
    \subfloat[FS8 average network TCO and Tier 2 cost breakdown.\label{AvgPercentbb}]
{\includegraphics[width=\linewidth]{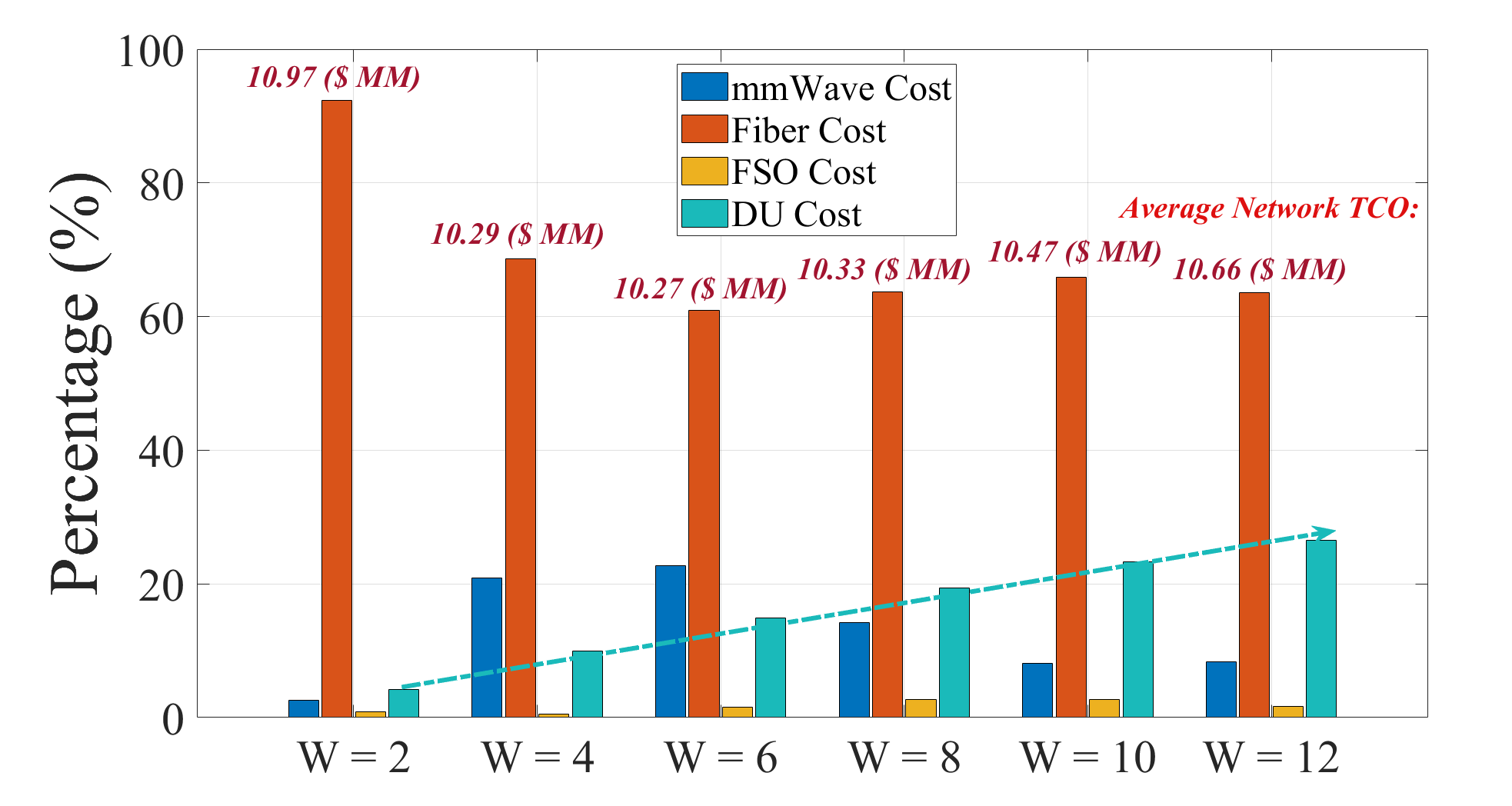}}
\caption{Average network TCO and technology cost distribution for different DU counts in RS-based CF-mMIMO.}
    \label{fig20:AvgTCO}
\end{figure}

Moreover, at lower decentralization levels (e.g., Figures \ref{cost1} and \ref{cost2}), fiber-only deployments incur the highest TCO due to the long distances between APs and DUs, whereas the suboptimal all-mmWave case typically shows the lowest cost despite constraint violations. On the other hand, at higher decentralization levels (e.g., Figure \ref{cost3}), the cost-effectiveness of the suboptimal all-mmWave scheme diminishes. Unlike common belief, this shift occurs due to the reduced AP-DU distances, which sometimes render fiber to be a more economically appealing option. The average TCO per AP is nearly identical for RS and HS-enabled CF-mMIMO under the same FS option. However, FS8 incurs slightly higher costs than FS7.2x due to stricter fronthaul capacity requirements. This cost difference becomes negligible at DU densities (e.g., $W = 8$), where fiber dominates for both FS options. Nevertheless, the selection of fronthaul technologies is contingent on the network configuration, FS option, connection scheme employed, and the number of deployed groups. 

To generalize these findings, Figure \ref{fig20:AvgTCO} highlights the average network TCO, summing both tiers, in Millions of US dollars (\$ MM), along with a breakdown of Tier 2 cost contributions of each technology and their respective percentages for RS-enabled CF-mMIMO. The results are averaged over numerous Monte Carlo simulations with group counts $G$ ranging from 100 to 200, for both FS7.2x and FS8 options. Figure \ref{fig20:AvgTCO} results are also applicable to the HS topology, providing valuable insights into the economic considerations associated with different deployment strategies and O-RAN FS options. With sparse DU deployment ($W = 2$), long AP-DU distances result in higher attenuation for wireless links, making fiber the preferred choice despite its high TCO. As $W$ increases (e.g., $W = 4$-$8$), AP-DU distances become more manageable, and wireless fronthaul technologies become more viable options. At higher levels of decentralized processing ($W>8$), fiber deployment costs drop significantly, further solidifying its role as an effective fronthaul solution in these scenarios. Finally, Figure \ref{AvgPercentbb} reveals that the stringent fronthaul capacity requirement of FS8 ($\psi^{\text{FS8}}$) leads to a higher percentage of fiber deployment compared to FS7.2x (Figure \ref{AvgPercentaa}), across all DU densities and connection schemes.

%\vspace{-0.3cm}

\subsection{Network Surplus Capacity and Number of Groups} 
The efficiency of deployment strategies can be assessed through fronthaul surplus capacity,  which measures the difference between the total capacity provided by the deployed fronthaul network and the actual demand imposed by various FS options. A positive surplus indicates that the network has excess capacity beyond the current demand, offering room for future traffic growth or accommodating unexpected surges. Conversely, a deficiency signifies that the network is operating below the traffic demand, compromising its ability to maintain QoS. Figure \ref{SurplusFS} shows that while the all-fiber benchmark offers the highest capacity, our hybrid optimized schemes consistently achieve substantial surplus at comparatively lower cost, demonstrating the effectiveness of the proposed framework. Additionally, the suboptimal all-mmWave frequently fails to meet the minimum required rates ($\psi^{\text{FSX}}$), resulting in the lowest surplus capacity, or even capacity deficiencies. While the current optimization does not enforce any surplus constraint, network operators may extend the model by imposing minimum surplus thresholds to ensure future-proofing in static deployments.

These results emphasize that low-cost strategies must not compromise QoS requirements to ensure reliable network performance. Moreover, the findings presented in Figure \ref{SurplusFS} consistently demonstrate the superiority of FS7.2x, offering greater surplus capacity compared to FS8 in both schemes. When combined with the cost-effectiveness insights from Figure \ref{fig5}, this reinforces FS7.2x as the preferred O-RAN functional split for future mobile networks \cite{O-RAN, o-rancfmimo}. Furthermore, FS7.2x aligns well with UDNs requirements, where the O-RAN architecture plays a role in complementing the vision of future UDNs deployment. Across all presented results, the heuristic method consistently yields lower surplus capacity and higher TCO compared to the optimized network. Therefore, we conclude that the best strategy in UDNs involves a well-planned and diversified mix of fronthaul technologies, as exemplified by the superior performance of our optimized network, in both FS options. The results highlights that cost considerations should not overshadow QoS requirements to reliably meet performance targets for SPs.

\begin{remark}
    Surplus fronthaul capacity is not used as a measure of algorithm performance but rather as an indicator of network resilience and headroom. It quantifies how much additional traffic the fronthaul infrastructure could support beyond the minimum required by the selected functional split. Most importantly, the surplus plots can be used to account for the additional control plane traffic as discussed in~\eqref{eq:FS8_CP2} and~\eqref{eq:FS7_CP2}. This metric helps assess the network’s future readiness under rising throughput demands, sudden traffic surges, and evolving service requirements. It is further important to note that any observed surplus fronthaul capacity is not the result of intentional over-dimensioning, but rather an artefact of discrete fronthaul technology selection under strict cost minimization and feasibility constraints, and is therefore interpreted solely as a resilience and headroom indicator.
\end{remark}

\begin{figure}[t]
  \begin{center}
  \includegraphics[width=0.8\linewidth]{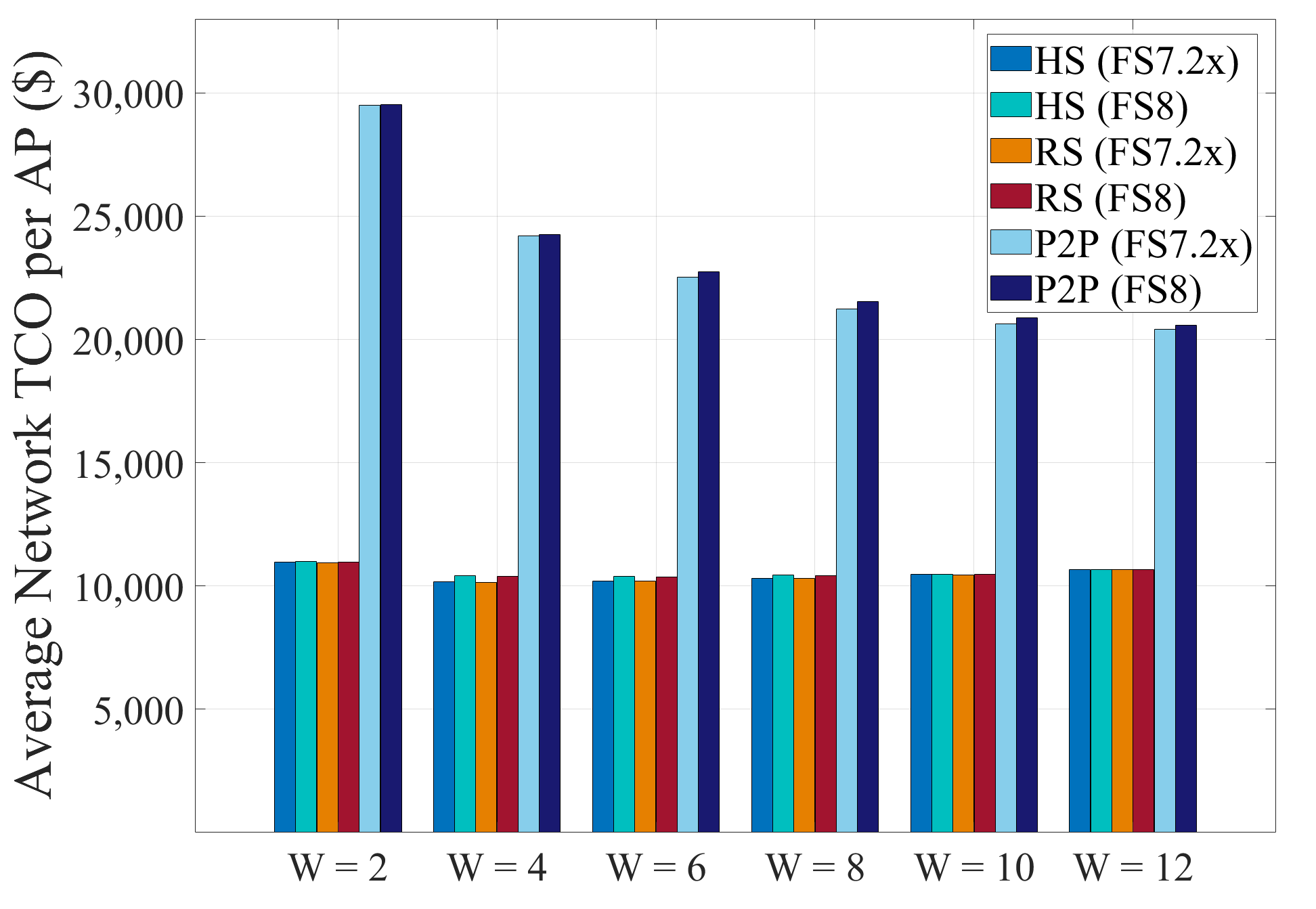}
  \vspace{-1pt}
\caption{Comparison of average TCO per AP among small cells (P2P fronthaul), RS, and HS topologies under varying FS options and different levels of decentralized processing.}\label{Comparison}
 \end{center}
   \vspace{-1.2em}
\end{figure}

\begin{figure}[t]
\centering
\subfloat[Traffic distribution mesh plot.\label{RS-6DUs-MidTraffMesh}]{
\includegraphics[width=0.485\columnwidth]{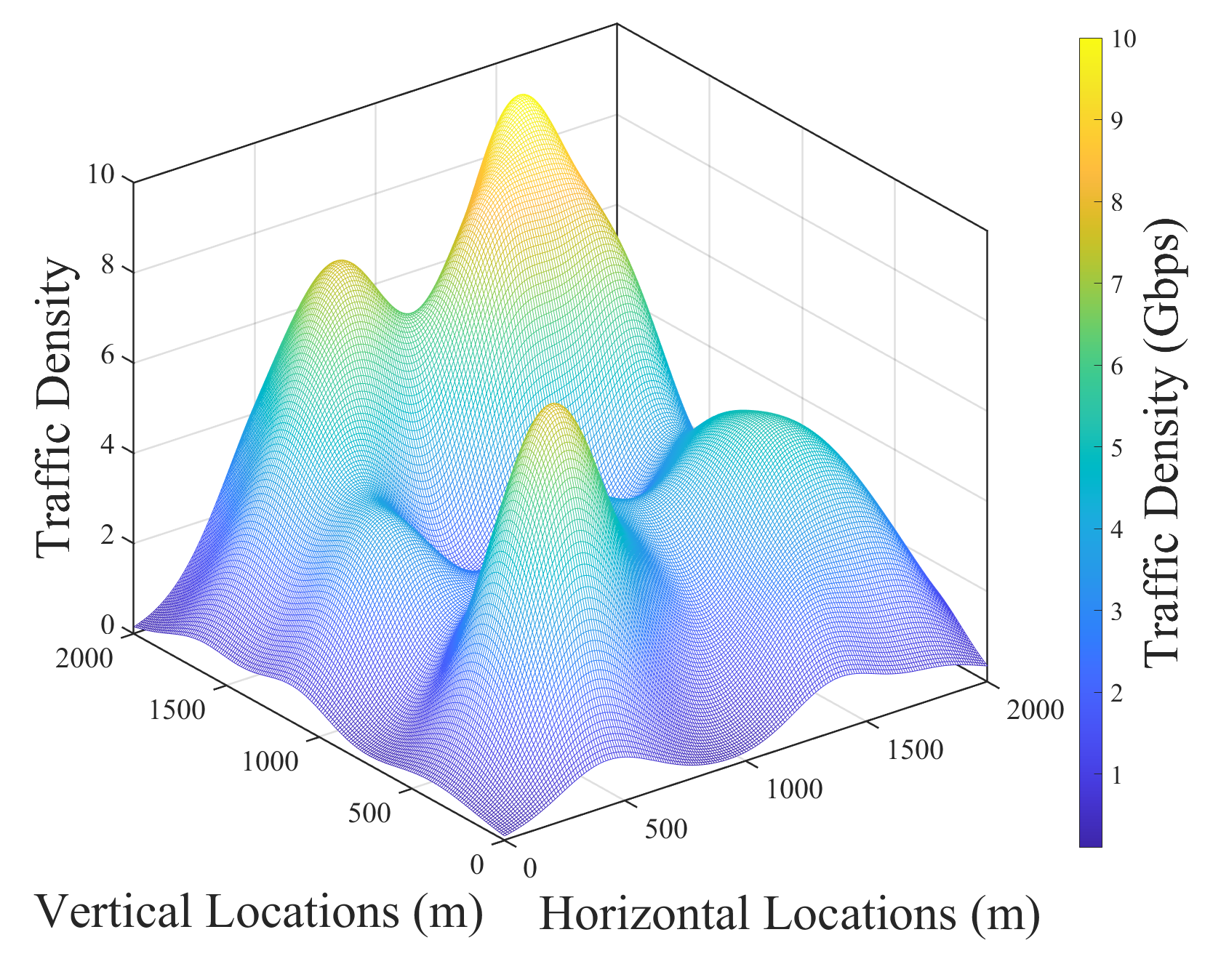}}
\hfill    
    \subfloat[Optimized fronthaul selection.\label{RS-6DUs-MidTraff}]{
\includegraphics[width=0.485\columnwidth]{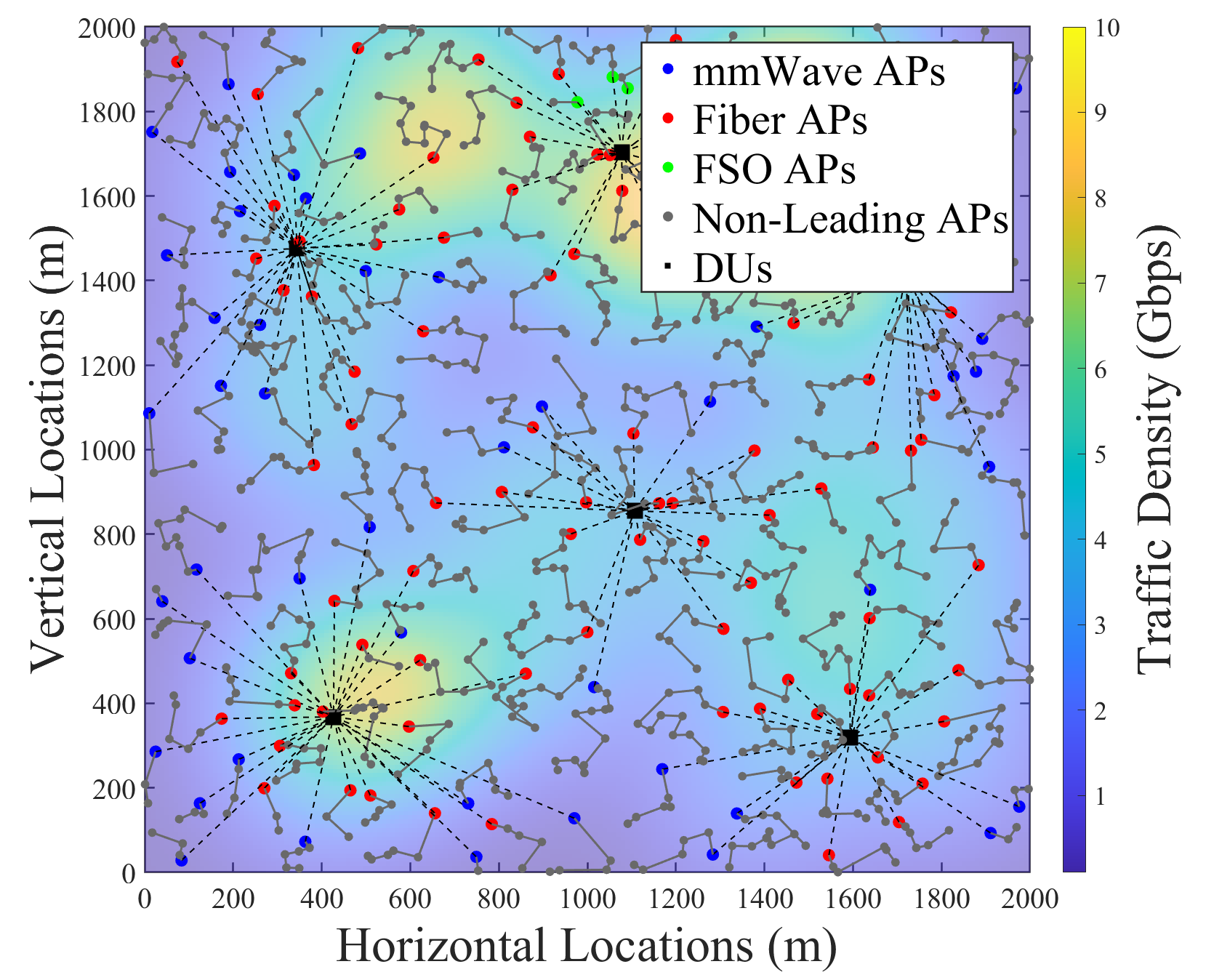}}
\caption{Sample realization of combined RS-enabled CF-mMIMO network and traffic distribution with optimized fronthaul technology selection for $G = 150$ and $W = 6$.}
\label{Tech-Selection-TrafficAware}
\end{figure}
%\vspace{-0.2cm}
\subsection{Radio-Stripes, Hierarchical and Small Cells TCO}

To underscore the cost-effectiveness of the optimized fronthaul connection schemes in supporting ultra-dense CF-mMIMO deployments, Figure \ref{Comparison} compares the average TCO per AP in RS, HS, and the traditional P2P fronthaul connection scheme for small cells networks  \cite{MyPaper}. To ensure comparison fairness, each of the aforementioned systems deploys an identical configuration of $L$ APs within the same coverage area and satisfies the same capacity requirements ($\psi^{\text{FSX}}$) in constraint \eqref{sixth-constraint}, for both FS7.2x and FS8. Moreover, both RS and HS setups are configured with a fixed number of groups ($G=200$). Figure \ref{Comparison} reveals that both RS and HS setups significantly reduce the average TCO per AP by sharing fronthaul resources among AP within the same group, particularly pronounced at lower levels of decentralized processing (e.g., $W < 4$). These schemes provide a practical and efficient alternative to conventional P2P connections for UDNs. When combined with the proposed optimization framework, RS and HS topologies further enhance the economic viability and scalability of UDN architectures for next-generation mobile networks.

\subsection{Non-Homogeneous Traffic Analysis} \label{subsec:stress_test}

To further assess the robustness and scalability of the proposed framework beyond standardized functional split operating conditions, we conduct a systematic stress test by increasing the minimum fronthaul bandwidth requirement using a traffic-aware methodology, similar to what is described in Section \ref{new-section1}. Importantly, this analysis does not involve redesigning fronthaul technologies or modifying their underlying models. Motivated by our previous work \cite{MyPaper}, we replace the fixed functional split requirement $\psi^{\text{FSX}}$ in \eqref{sixth-constraint} with a spatially varying threshold $\psi_{\ell}^{\text{het.}}$ assigned to each leading AP $\ell \in \mathcal{M}_w$. This modification preserves the original optimization structure and cost formulation, replacing only the static traffic requirement with a traffic-aware threshold that captures operator-style spatial demand field. Specifically, the minimum fronthaul requirement for each leading AP is defined as a function of its spatial coordinates:

\begin{equation}\label{traffic-aware-threshold}
\psi_{\ell}^{\text{het.}} = f_{\text{traffic}}(x_{\ell}, y_{\ell}), \qquad \forall \ell \in \mathcal{M}_w .
\end{equation}

\begin{figure}[t]
\centering
\subfloat[TCO per AP for $W$ = 2.\label{2DUs-New-Cost-Trend}]{
\includegraphics[width=0.485\columnwidth]{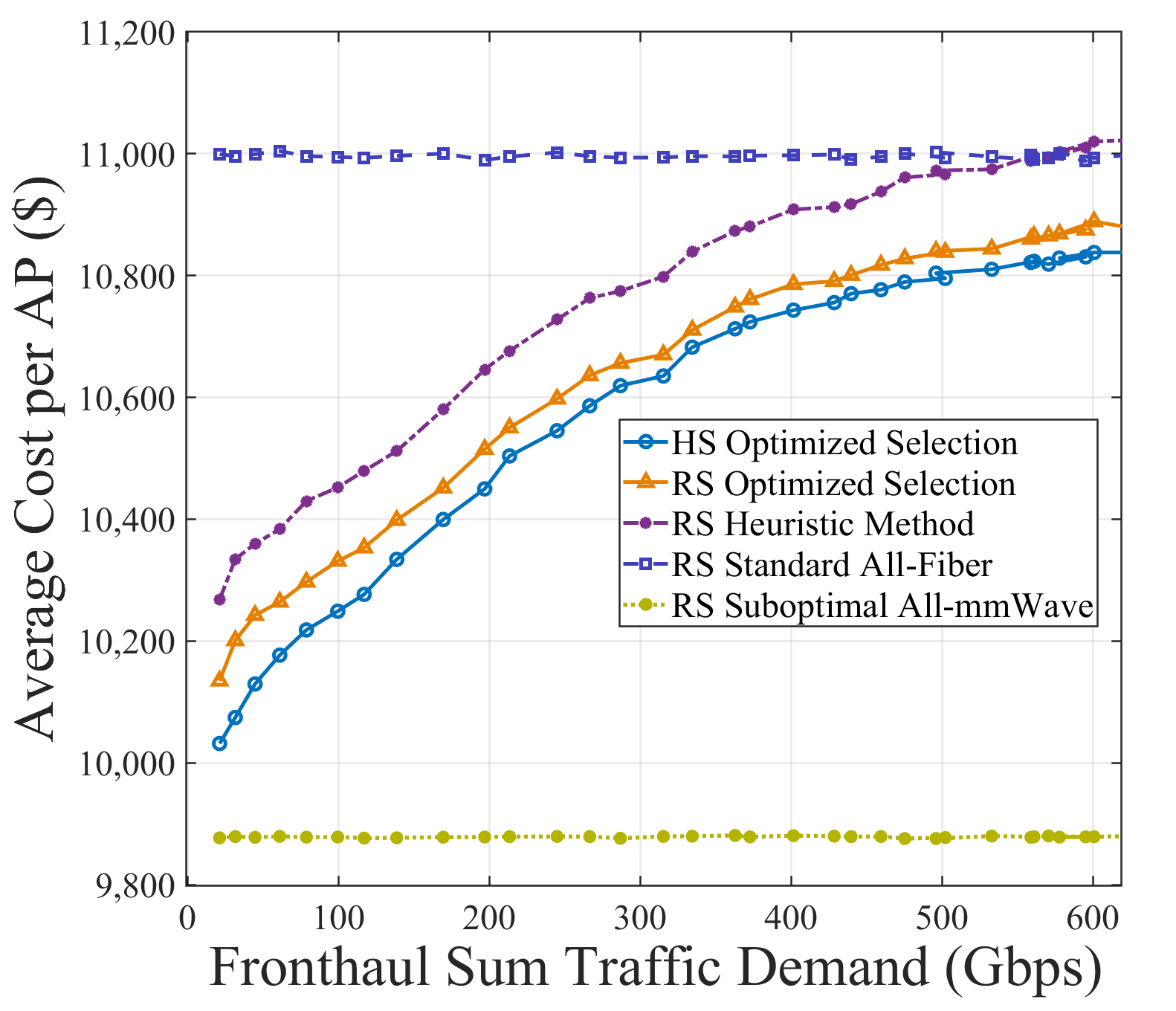}}
\hfill    
    \subfloat[TCO per AP for $W$ = 6.\label{6DUs-New-Cost-Trend}]{
\includegraphics[width=0.485\columnwidth]{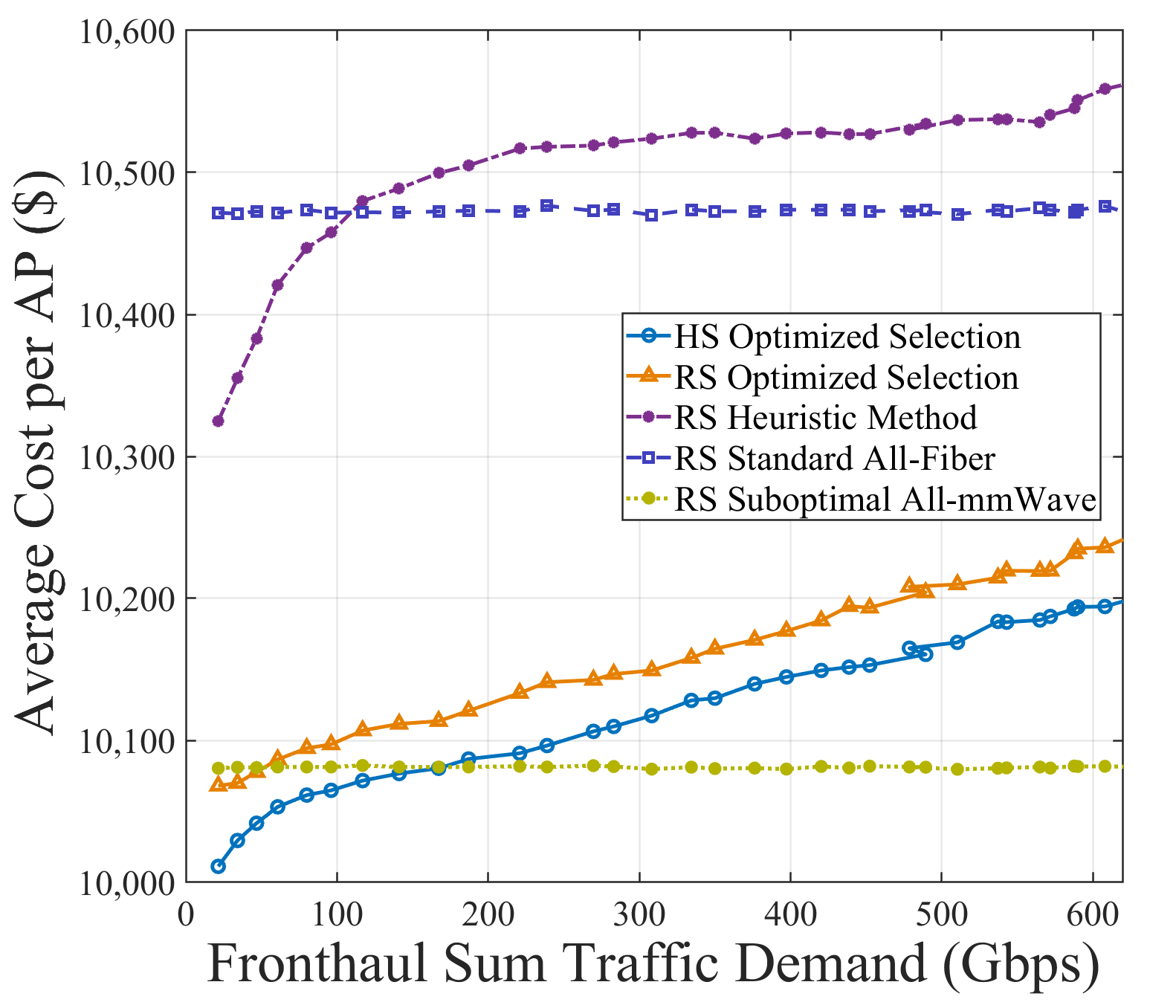}}
\caption{Average TCO per AP vs. fronthaul sum traffic thresholds for different DU counts and $G = 150$.}\label{New-Cost-Trend}
\end{figure}
\begin{figure}[t]
\centering
\subfloat[Surplus capacity for $W$ = 2.\label{2DUs-New-capacity-Trend}]{
\includegraphics[width=0.485\columnwidth]{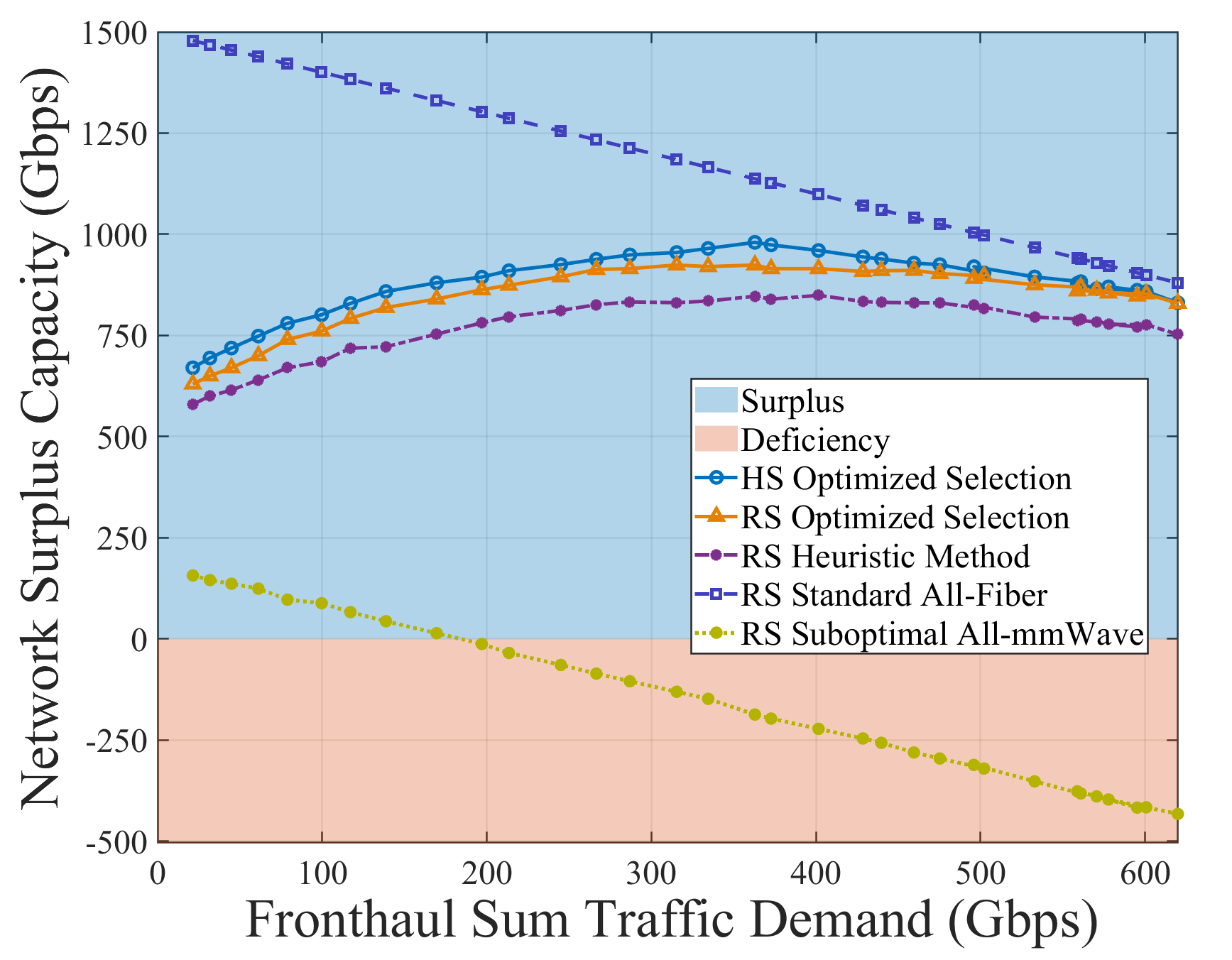}}
\hfill    
    \subfloat[Surplus capacity for $W$ = 6.\label{6DUs-New-capacity-Trend}]{
\includegraphics[width=0.485\columnwidth]{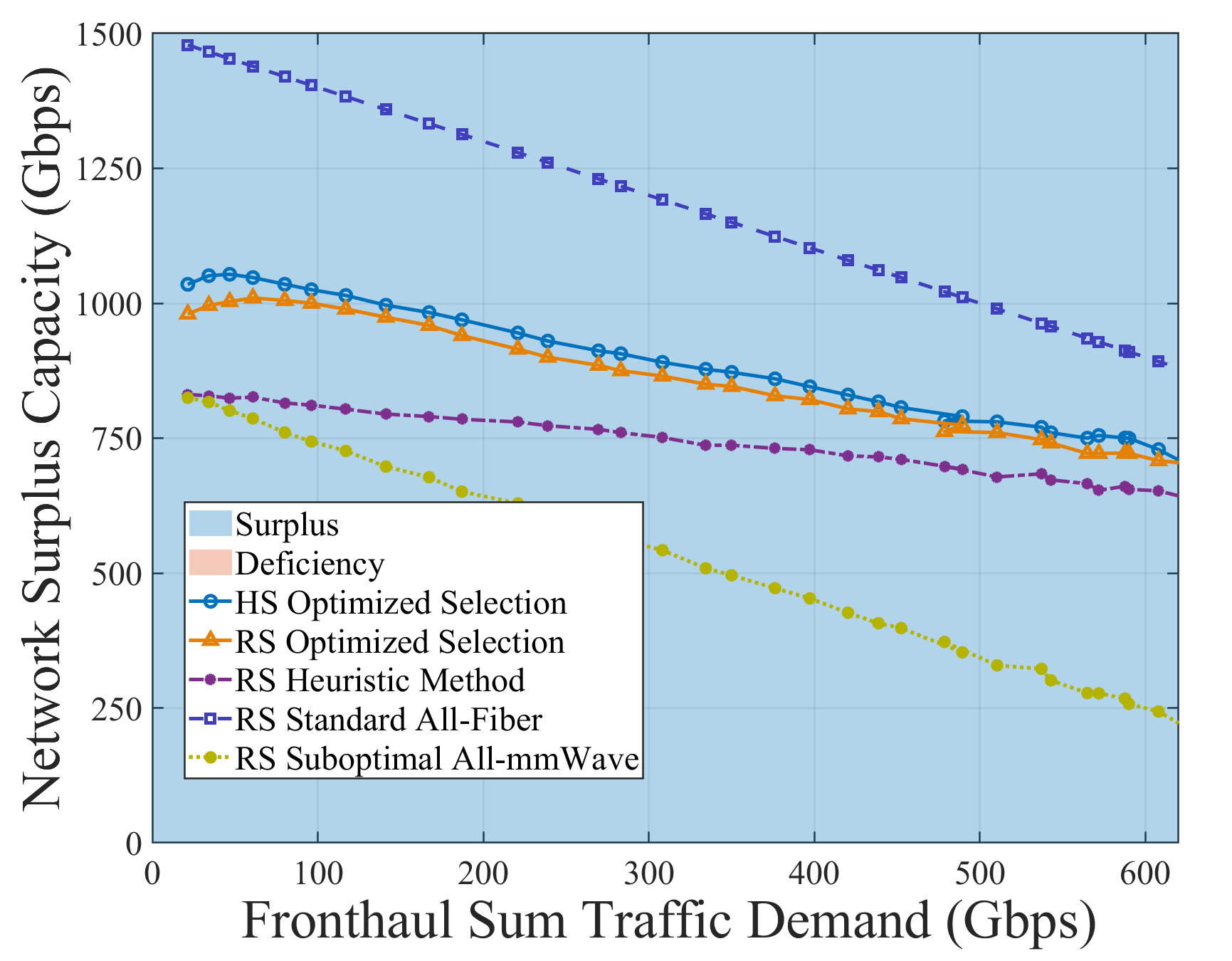}}
\caption{Network surplus capacity vs. fronthaul sum traffic thresholds for different DU counts and $G = 150$.}\label{New-Capacity-Trend}
\end{figure}

The spatial traffic field is generated using a hotspot-based Gaussian mixture model that emulates operator-style demand distributions \cite{MyPaper}. A total of $N_s$ traffic hotspots are randomly placed over the coverage area, as shown in Figure \ref{Tech-Selection-TrafficAware}. Moreover, the peak demand is limited by $\mathbf{X}^{\text{max}}$, which has to be less than the highest physical capacity of the used technologies (Fiber, $10$ Gbps in this study). Accordingly, the fronthaul capacity constraint in \eqref{sixth-constraint} becomes as follows:

\begin{equation}\label{sixth-constraint-stress}
x_{w \ell} R_{w \ell}^{\text{Fiber}} + z_{w \ell} R_{w \ell}^{\text{mmW}} + u_{w \ell} R_{w \ell}^{\text{FSO}}
\geq
\psi_{\ell}^{\text{het.}}, \, \forall \ell \in \mathcal{M}_w ,  \forall w \in \mathcal{W}.
\end{equation}

Finally, each leading AP $\ell \in \mathcal{M}_w$ is assigned a minimum capacity threshold according to its location through \eqref{traffic-aware-threshold}, which directly governs the feasibility of \eqref{sixth-constraint-stress} and, consequently, the fronthaul technology selection. As $\psi_{\ell}^{\text{het.}}$ increases, the optimization progressively prioritizes higher-capacity and more reliable fronthaul technologies for these APs. This trend clearly leads to an increase in the optimized TCO and network capacity, along with a gradual convergence toward more fiber-dominated deployments, especially under highly centralized deployments (i.e., $W = 2$), as backed by previous results in \cite{MyPaper}, and demonstrated in Figures \ref{New-Cost-Trend} and \ref{New-Capacity-Trend}. This behavior reflects that by increasing fronthaul traffic demand, under which low- and medium-capacity links become infeasible, high-capacity fronthaul solutions dominate the feasible design space. The resulting technology selection patterns shown in Figure \ref{Tech-Selection-TrafficAware} is consistent with the core objective of the framework. It trades cost for capacity and reliability only when needed, favoring fiber in high-traffic regions while preserving feasibility within the physical capabilities of the available fronthaul technologies.

\begin{remark}
Infeasible links under extreme demand is a direct consequence of physical fronthaul capacity limits rather than a limitation of the proposed optimization framework. If $\psi_{\ell}^{\text{het.}}$ at a given leading AP exceeds the maximum achievable rate supported by all available technologies ($R^{\text{Fiber}}$, $R^{\text{mmW}}$, or $R^{\text{FSO}}$), no feasible fronthaul assignment exists by construction. In such cases, the framework correctly identifies these locations instead of forcing invalid assignments. From a planning perspective, this outcome signals that the deployment has reached a technology-imposed boundary and motivates actionable remedies such as network densification, topology reconfiguration, or upgrading the fronthaul technology set. Therefore, this analysis serves as a diagnostic scalability tool that clearly distinguishes algorithmic behavior from fundamental physical capacity ceilings.
\end{remark}

\subsection{Network Resilience Analysis}
\begin{figure}[t]
    \centering
    \subfloat[Sample of main failures in RS.\label{Resilience1RS}]{
        \includegraphics[width=0.48\columnwidth]{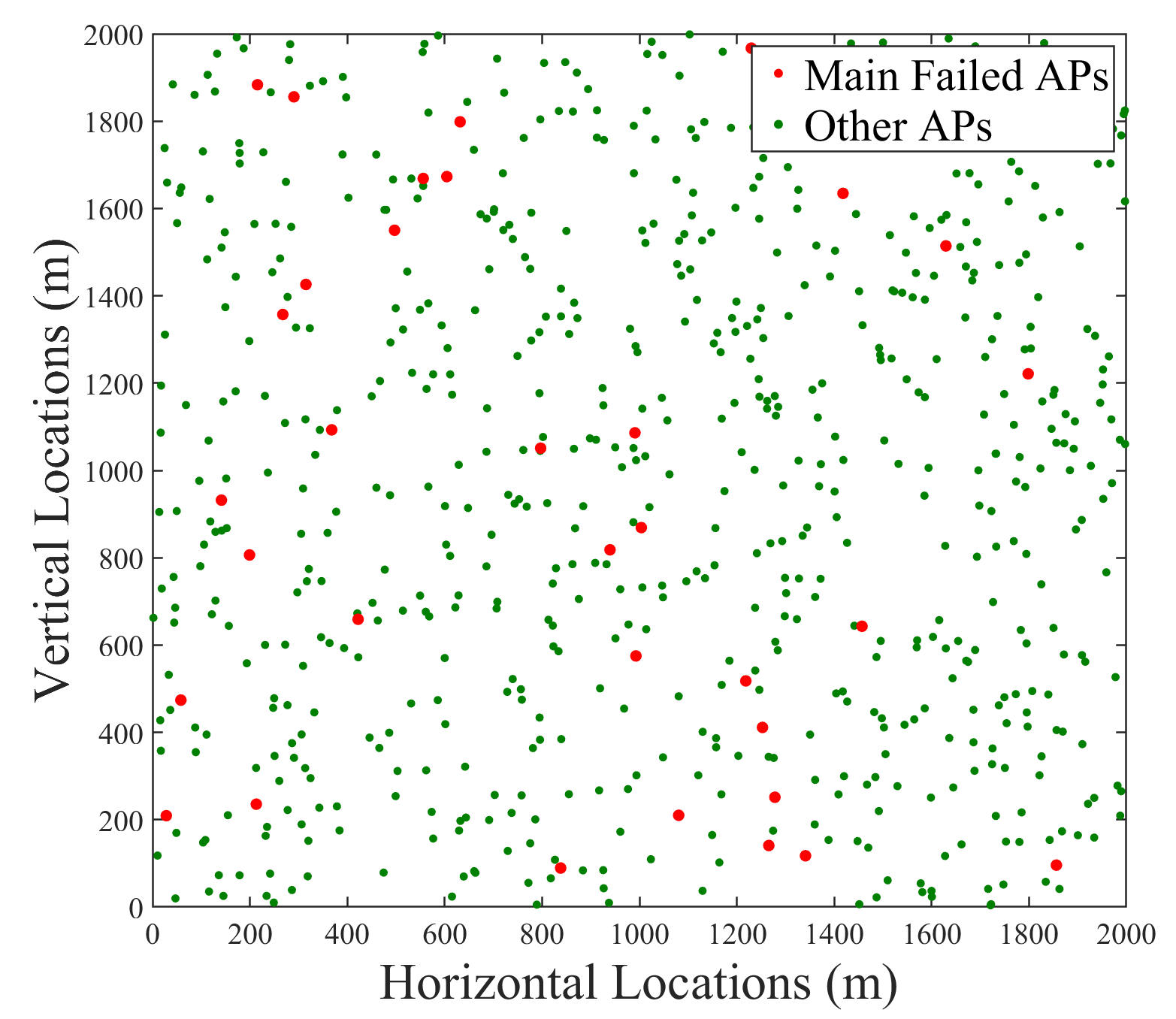}}
    \hfill    
    \subfloat[Sample of cascaded failures in RS.\label{Resilience2RS}]{
        \includegraphics[width=0.49\columnwidth]{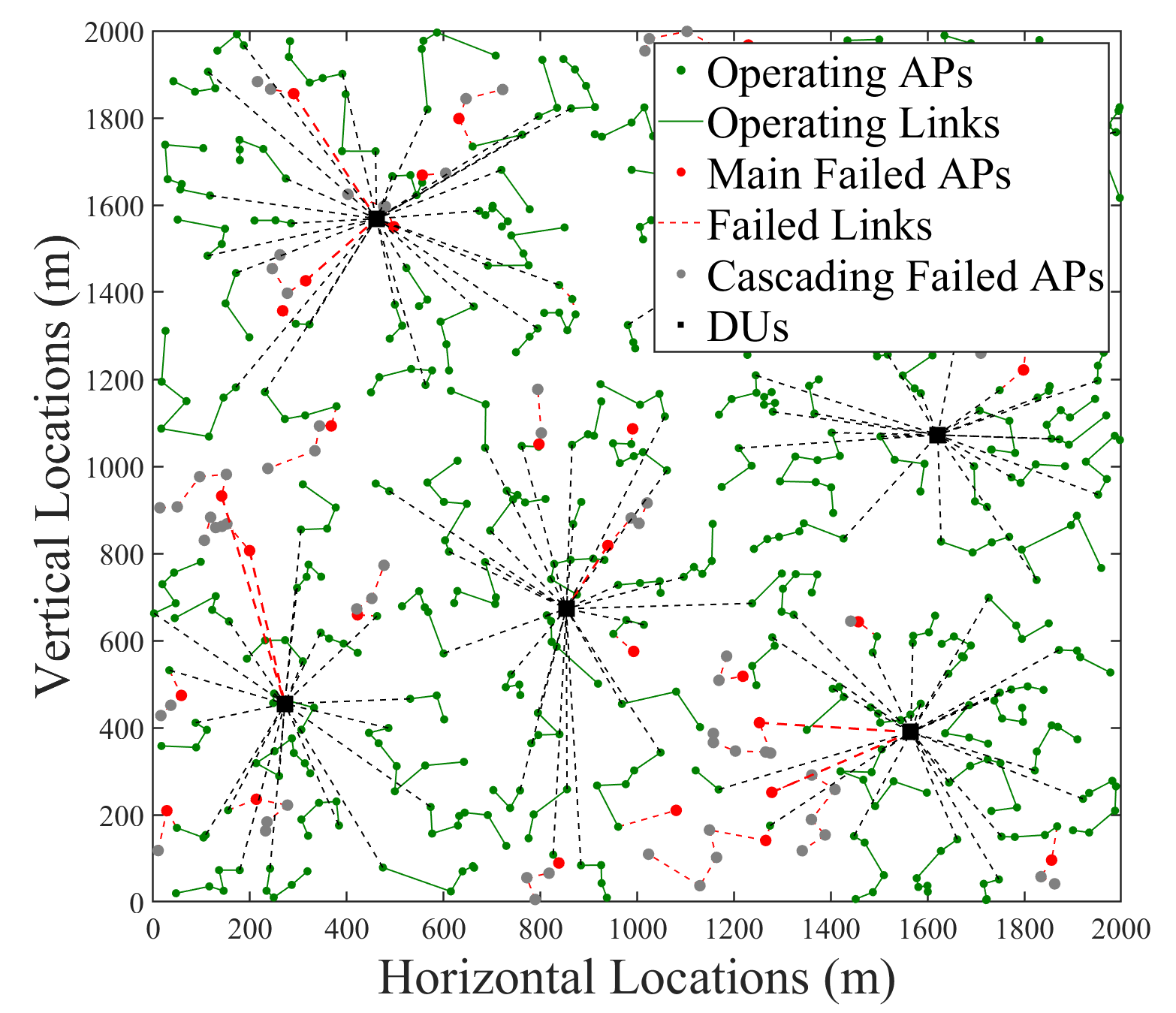}}
    
    \caption{RS-enabled CF-mMIMO network resilience sample realization for $G = 150$ and $p = 5\%$.}
    \label{CascadedFailuresRS}
    
\end{figure}

\begin{figure}[t]
    \centering
    \subfloat[Sample of main failures in HS.\label{Resilience1HS}]{
        \includegraphics[width=0.485\columnwidth]{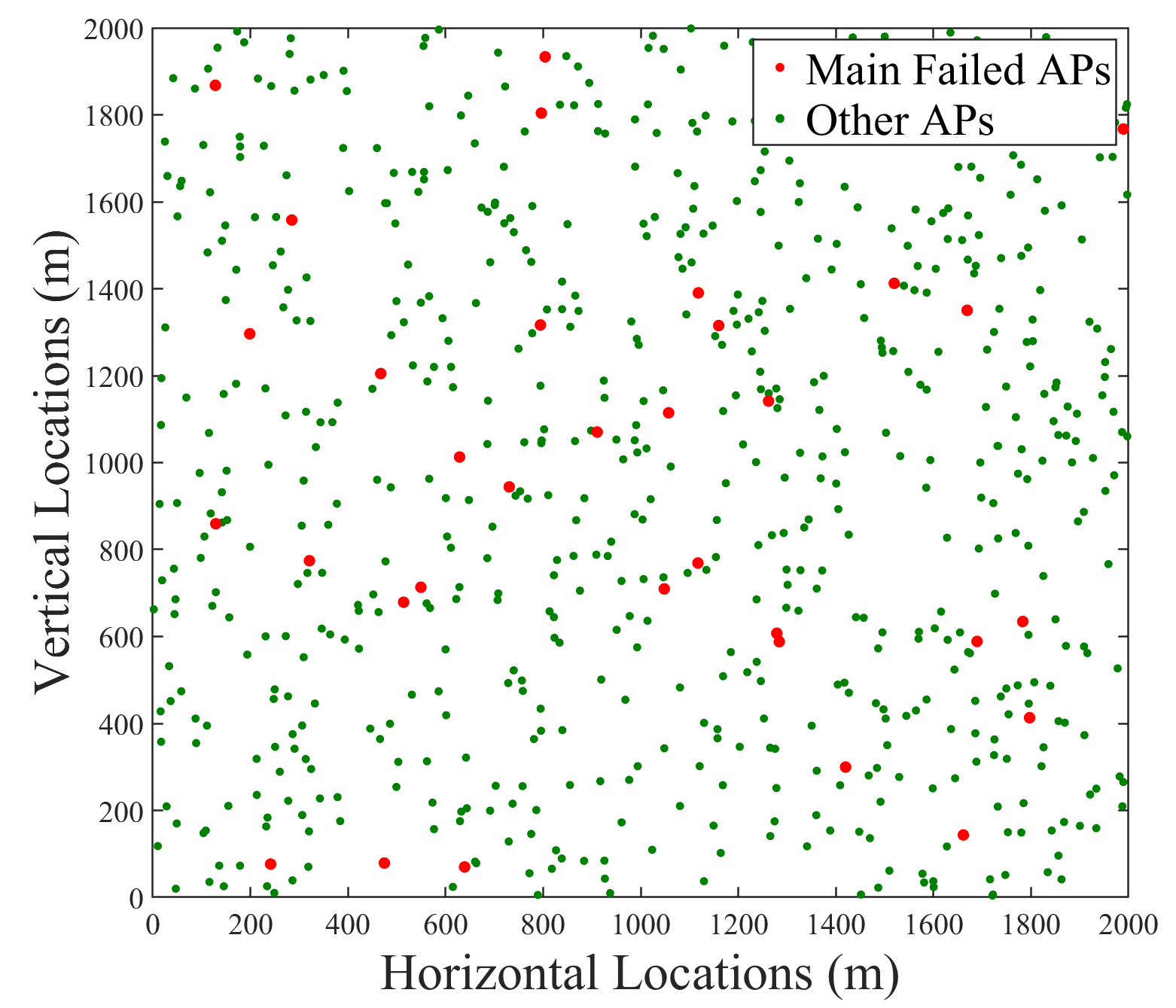}}
    \hfill    
    \subfloat[Sample of cascaded failures in HS.\label{Resilience2HS}]{
        \includegraphics[width=0.485\columnwidth]{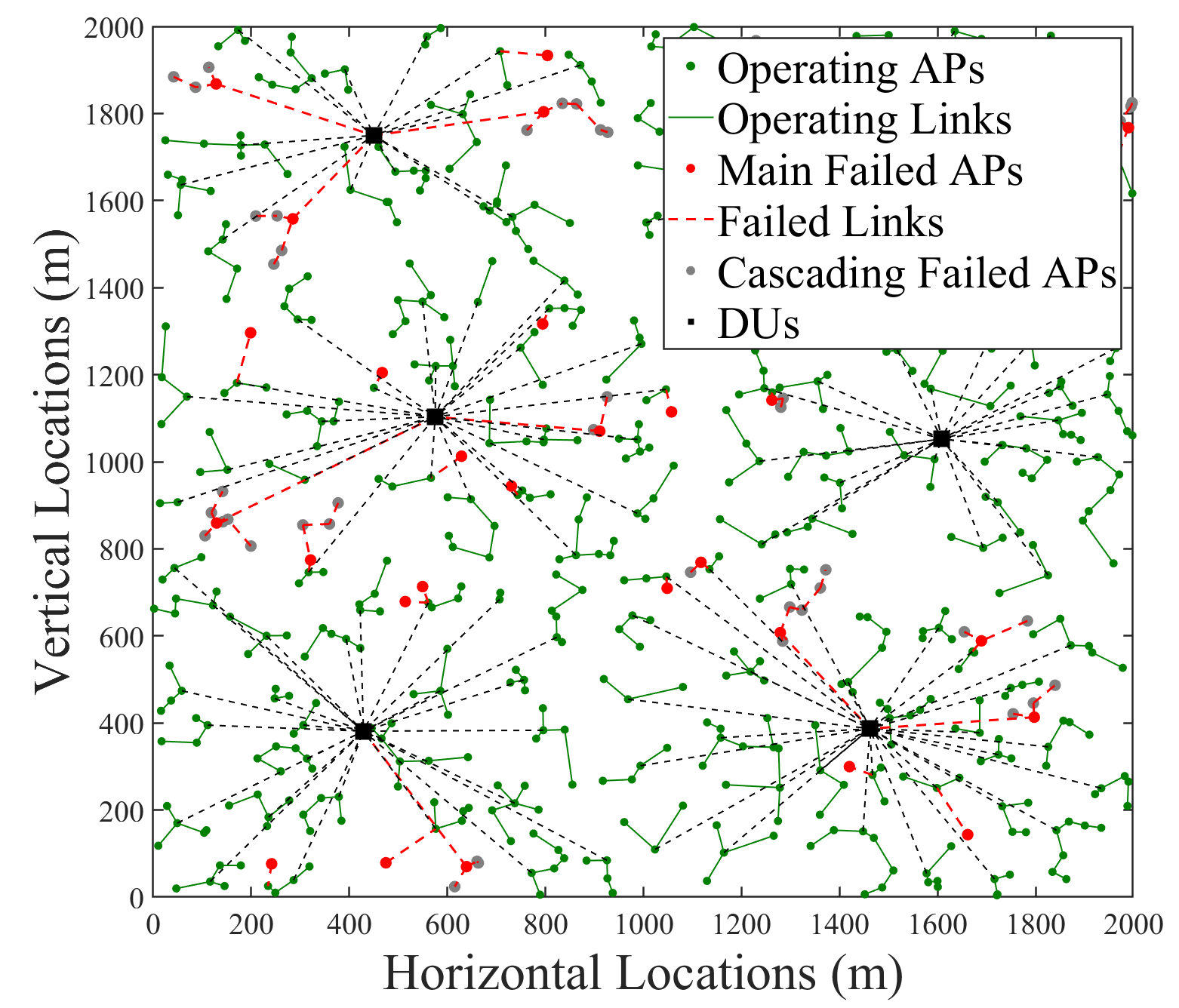}}
    \caption{HS-enabled CF-mMIMO network resilience sample realization for $G = 150$ and $p = 0.05$.}
    \label{CascadedFailuresHS}
\end{figure}

Despite the economic advantages of RS, the serial architecture makes its resilience against link failures questionable. Specifically, the sequential connectivity in RS implies that a single fronthaul link failure could trigger cascading failures affecting the rest of subsequent APs. Conversely, HS uses a branched topology that reduces interdependence among APs within the same group and mitigates the cascading failure effects. Let $p$ represent the fraction of APs that experience an outage due to a failure in the fronthaul link of any deployed AP $\ell \in \mathcal{L}$, forming the main failed APs set $\mathcal{K}_m \subseteq \mathcal{L}$. This failure applies equally to both RS and HS-enabled CF-mMIMO systems. In RS, a single AP failure cascades to subsequent APs in the stripe, owing to the serial fronthaul connection. On the other hand, HS reduces this cascading effect through its branched connections. In both schemes, we denote APs failing due to cascading effects as the set $\mathcal{K}_c \subseteq \mathcal{L}$. The worst-case scenario occurs when the fronthaul link of the leading AP (Tier 2) fails, resulting in the complete loss of all dependent APs within the group, as depicted in Figures \ref{CascadedFailuresRS} and \ref{CascadedFailuresHS}. Hence, the set of all failed APs, $\mathcal{K}$, is the total sum of main and cascaded failures, i.e., $\mathcal{K} = \mathcal{K}_m \cup \mathcal{K}_c \subseteq \mathcal{L}$.


\begin{figure}[t]
  \begin{center}
  \includegraphics[width=0.9\linewidth]{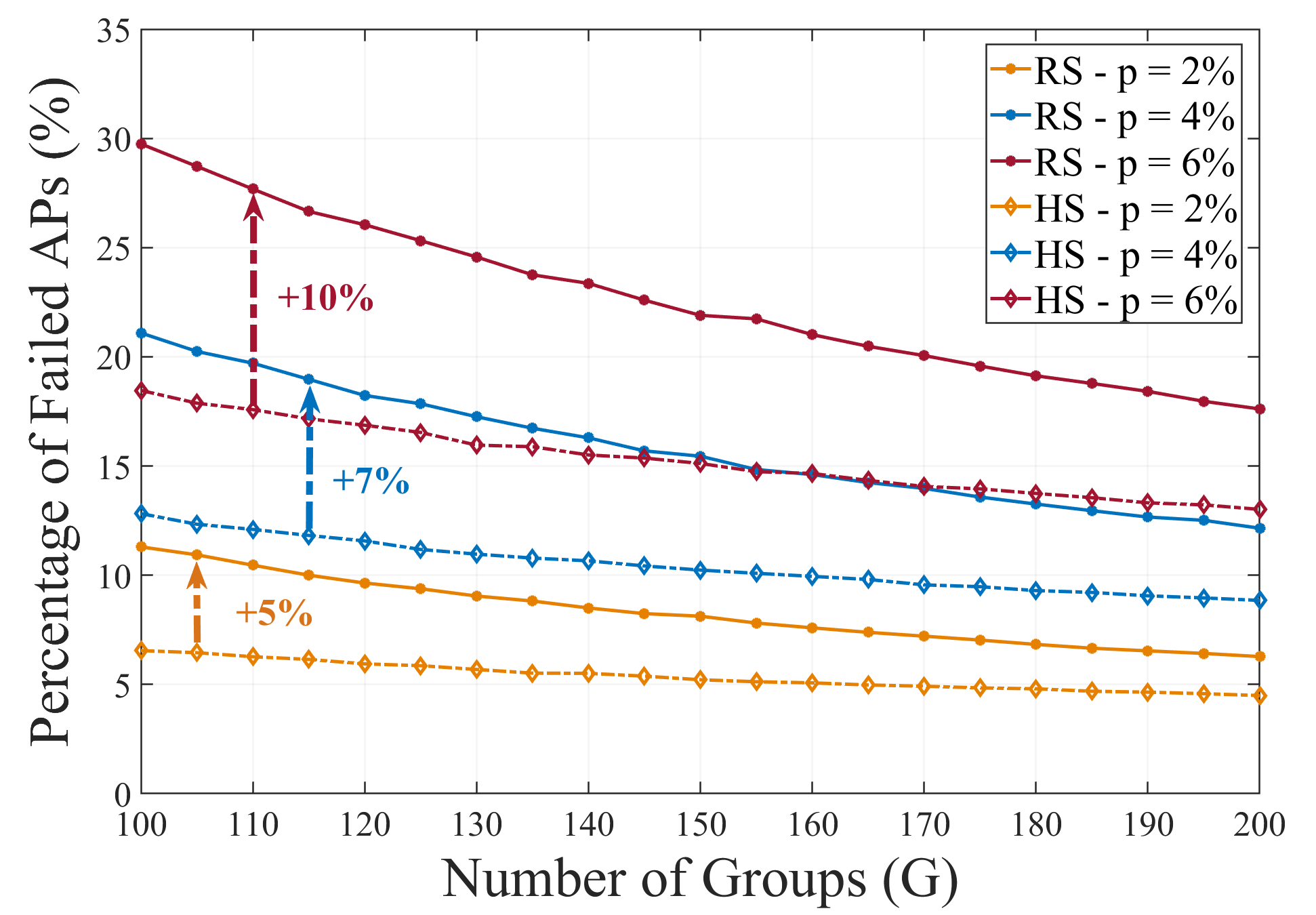}
  \vspace{-1pt}
\caption{Comparison of average total AP failures for varying numbers of groups $G$ and main failure fractions $p$.}\label{ResiliencePlot}
 \end{center}
\end{figure}

The severity of cascading failures depends on the fraction of main AP failures $p$, the number of AP groups $G$, the group size $L_{\mathcal{G}_i}$, and the specific locations of main failing APs. To provide a broader perspective, Figure~\ref{ResiliencePlot} shows the average percentage of total AP failures for various $G$ and $p$ values. We observe that the RS scheme exhibit high vulnerability to cascading failures, with $p = 6\%$ leading to about 30\% total AP outages for $G = 100$. The HS topology, however, shows improved resilience, reducing total outages to around 19\% under similar conditions. Given its comparable cost-efficiency and superior resilience, HS emerges as a compelling fronthaul connectivity solution for future UDNs and CF-mMIMO deployment. Additionally, Figure \ref{ResiliencePlot} reveals that the severity of cascading failures inversely correlates with the number of deployed groups $G$. Increasing the number of groups $G$, while keeping $L$ fixed, naturally decreases the total number of APs per group $L_{\mathcal{G}_i}$, thereby lowering interdependence and limiting the impact of cascading failures. This trend approaches the behavior of P2P small cells, which isolate failures through dedicated links and offer higher overall robustness.

\subsection{Variance Analysis Across Randomized Deployments}
\begin{figure}[t]
  \begin{center}
  \includegraphics[width= 0.9\linewidth]{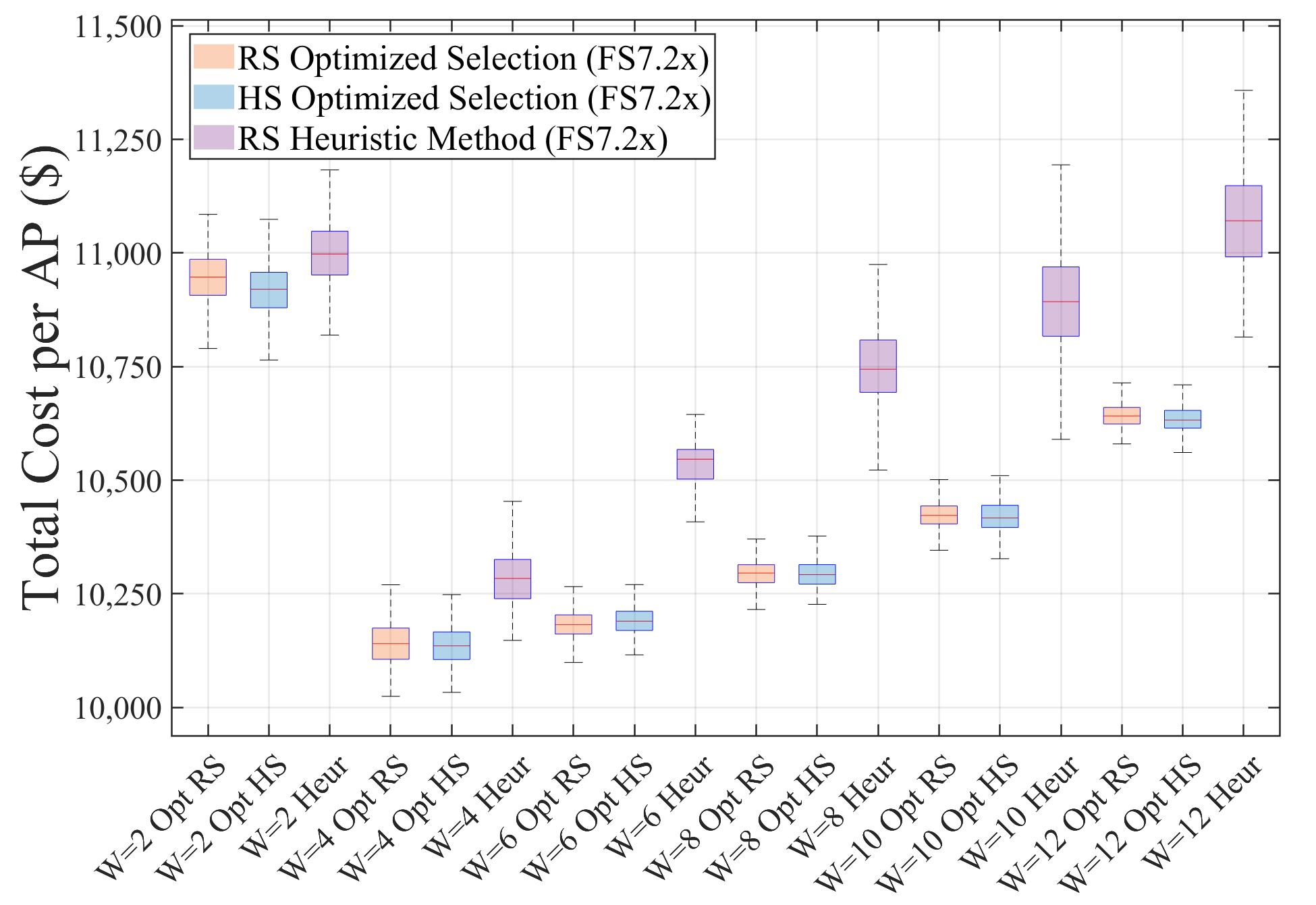}
    \vspace{-1pt}
\caption{Distribution of TCO per AP over $500$ network realizations for different DU counts and $G = 150$, comparing optimized and heuristic solutions.}\label{VariancePlot}
 \end{center}
\end{figure}

To complement the average-case performance results presented earlier in this section, we now investigate the per-realization variability of the proposed optimization framework and compare it against the heuristic method. This analysis is motivated by the need to understand how consistently the solution performs across randomized network topologies. Figure~\ref{VariancePlot} presents a boxplot showing the distribution of total cost per AP across $500$ randomized realizations for both the optimized and heuristic approaches under FS 7.2x,  different DU pooling configurations ($W = 2, 4, ..., 12$), and  $G = 150$. Each $W$ value includes three configurations: the optimized RS method, the optimized HS method, and the heuristic method and shows the spread, median, and interquartile range of deployment cost. 

As observed, the optimized method consistently yields lower median cost and narrower variance across all $W$ values. In contrast, the heuristic method exhibits greater variability and a wider spread, especially at higher $W$. This trend highly supports our results in Figure \ref{fig20:AvgTCO}, highlighting that as the number of DU pools increases, the complexity of the planning problem grows, and the benefits of joint optimization over simple heuristics become more pronounced. The variance gap between these methods increases with $W$, reinforcing the value of global cost-aware planning in more decentralized deployment scenarios. This confirms that the optimized strategy not only lowers average cost but also reduces the risk of extreme deployment scenarios, providing stronger guarantees for real-world planning under uncertainty. The figure also shows that RS and HS architectures have comparable costs across different network realizations; however, the HS advantage lies in the greater resilience to AP failures, as demonstrated in Figure 16.

%\vspace{-0.5cm}

\section{Conclusion}\label{CH6}
This paper presented a two-tiered optimization framework for hybrid fronthaul network planning that minimizes TCO while ensuring scalability and high performance in HS- and RS-enabled CF-mMIMO, as well as broader UDN deployments within O-RAN. The framework jointly optimizes fiber, mmWave, and FSO fronthaul options under capacity and reliability constraints, achieving efficient infrastructure utilization and robust performance. Results demonstrate that hybrid deployments outperform single-technology (all-fiber or all-mmWave) and heuristic schemes by balancing cost, capacity, and resilience. HS-enabled CF-mMIMO further enhances reliability compared to RS-based systems, offering a cost-efficient yet failure-resilient architecture. An important extension of this framework is the explicit incorporation of future traffic growth or reserve capacity requirements into the optimization itself, for example through location-dependent demand scaling or minimum headroom constraints. Such extensions would enable a direct trade-off between cost efficiency and future-proofing and are left for future investigation. These findings underscore the need to balance economic and performance objectives when designing future fronthaul networks, providing actionable insights for service providers in deploying scalable and cost-effective UDNs that meet evolving 6G and O-RAN requirements.

\bibliographystyle{ieeetr}%{IEEEtran}
\bibliography{IEEEabrv}
\vspace{-2.7cm}

\begin{IEEEbiography}
[{\includegraphics[width=1in,height=1.25in,clip,keepaspectratio]{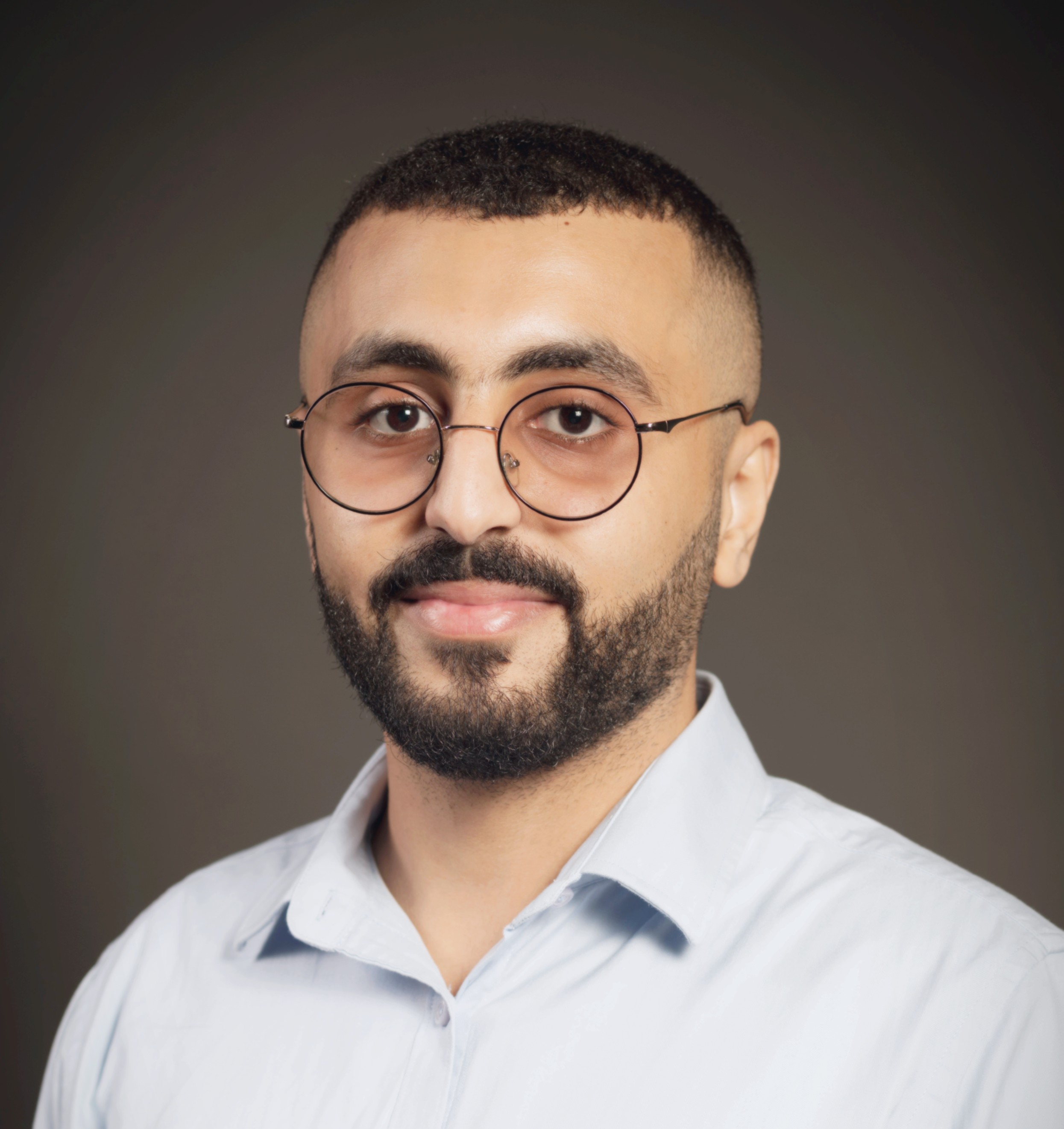}}]
{Anas~S.~Mohammed}  (Student Member, IEEE)
received his B.Sc. (Hons.) in electrical engineering, with a concentration in energy efficiency, from King Fahd University of Petroleum and Minerals (KFUPM), Saudi Arabia, in 2023, and his MASc. in electrical and computer engineering from Queen’s University, Canada, in 2025. During his graduate studies, he was a graduate research fellow with the Telecommunications Research Lab (TRL) and the Smart Wireless Massive Systems Lab (SWIMS), Queen's University, Canada. His background spans both advanced research and industry practice. His research interests lie in the broad areas of electrical engineering, including wireless communication systems, mobile networks, electrical and electronic systems design, power systems, smart grids, networks strategic planning and AI-driven optimization. Over the years, he conducted several hands-on research projects in the broad fields of electrical engineering, resulting in high-quality publications, and participation in global conferences, events and demonstration sessions.
\end{IEEEbiography}
\vspace{-3.5cm}
\begin{IEEEbiography}
[{\includegraphics[width=1in,height=1.25in,clip,keepaspectratio]{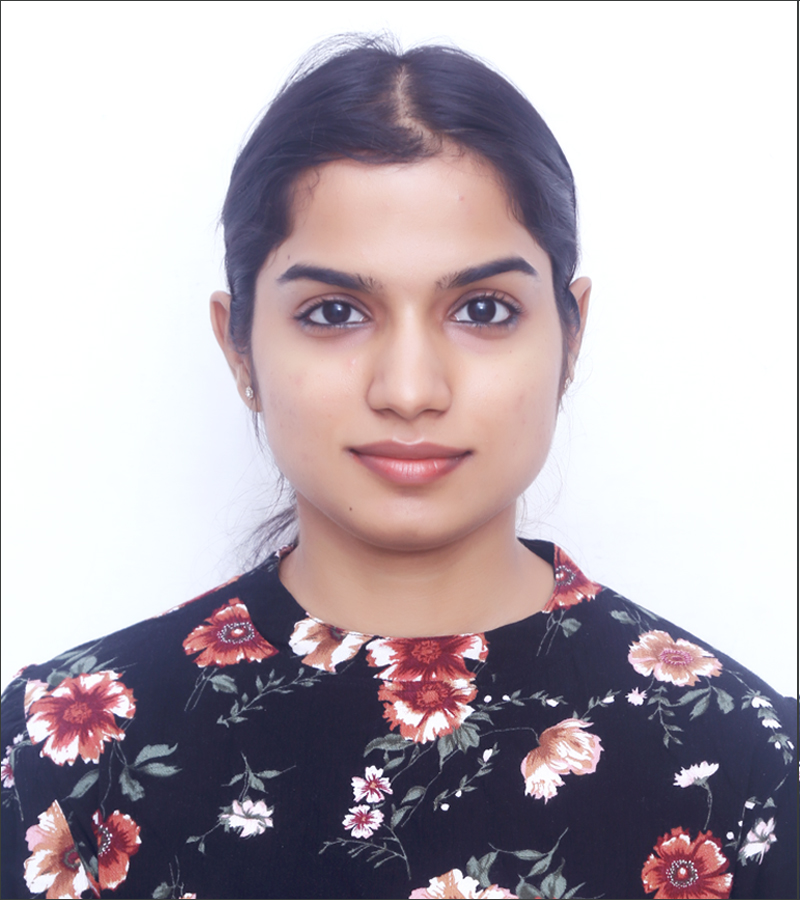}}]
{Krishnendu S. Tharakan}  (Member, IEEE)
received her B.Tech. degree in electronics \& communications Engineering from the National Institute of Technology Calicut, India, in $2016$, and the PhD degree from the Electrical Engineering Department, Indian Institute of Technology Indore, India in $2023$. From $2016-17$, she worked as an Engineer with Tata Elxsi, Trivandrum, India. She was a recipient of Visvesvaraya PhD Fellowship from MeitY, Government of India. She is currently a Post-Doctoral Fellow with the School of Electrical Engineering and Computer Science, KTH Royal Institute of Technology, Stockholm, Sweden. Prior to that, she was a Post-Doctoral fellow in the School of Computing at Queen's University, Kingston, Canada. Her current research interests include wireless communications, distributed optimization, federated learning, and statistical learning theory. She was
recognized as an Excellent Reviewer by IEEE TRANSACTIONS ON NETWORK
SCIENCE AND ENGINEERING in 2024.
\end{IEEEbiography}
\vspace{-1.4cm}
\begin{IEEEbiography}
[{\includegraphics[width=1in,height=1.25in,clip,keepaspectratio]{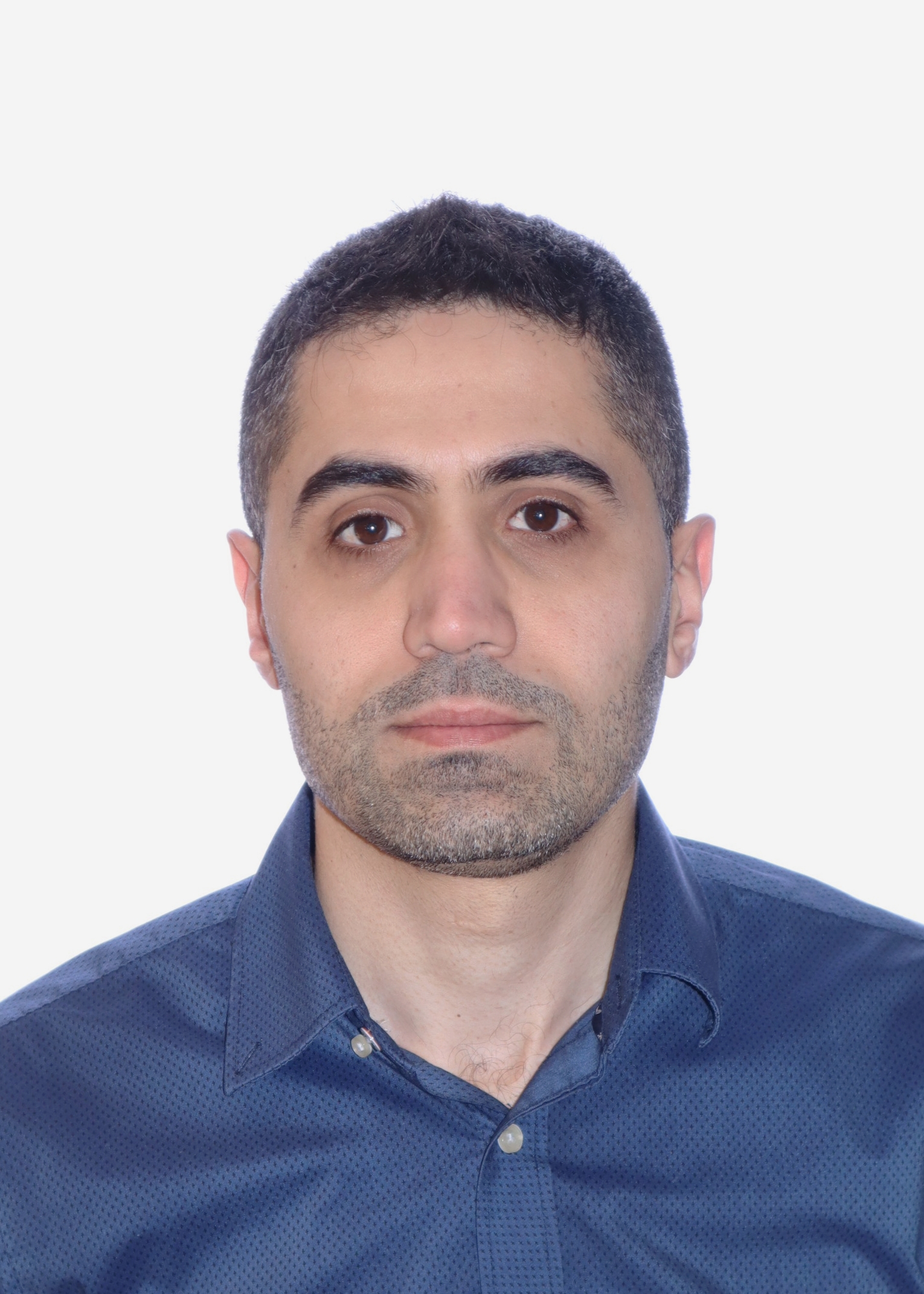}}]
{Hussein~A.~Ammar} (Member, IEEE)
received his Ph.D. degree in Electrical and Computer Engineering from the University of Toronto in 2023 and his M.A.Sc degree in Electrical and Computer Engineering from the American University of Beirut (AUB) in 2017. In 2023, he completed a four-month internship with Ericsson on developing artificial intelligence (AI) solutions for wireless communications. In 2018, he worked as a research assistant at the Mobile and Distributed Computing Laboratory at AUB and as an R\&D engineer in the information and communications technology industry. Since 2024 he has been an Assistant Professor in the Department of Electrical and Computer Engineering at the Royal Military College of Canada. Dr. Ammar received the University of Toronto Fellowship and the Edward S. Rogers Sr. Graduate Scholarship. He was recognized as an Exemplary Reviewer by \textsc{IEEE Communications Letters} in 2023. His research interests include wireless communications, AI for wireless networks, scalable and resilient deep reinforcement learning, coordinated distributed MIMO, user-centric cell-free massive MIMO, statistical signal processing, and mathematical optimization.
\end{IEEEbiography}
\vspace{-2cm}
\begin{IEEEbiography}
[{\includegraphics[width=1in,height=1.25in,clip,keepaspectratio]{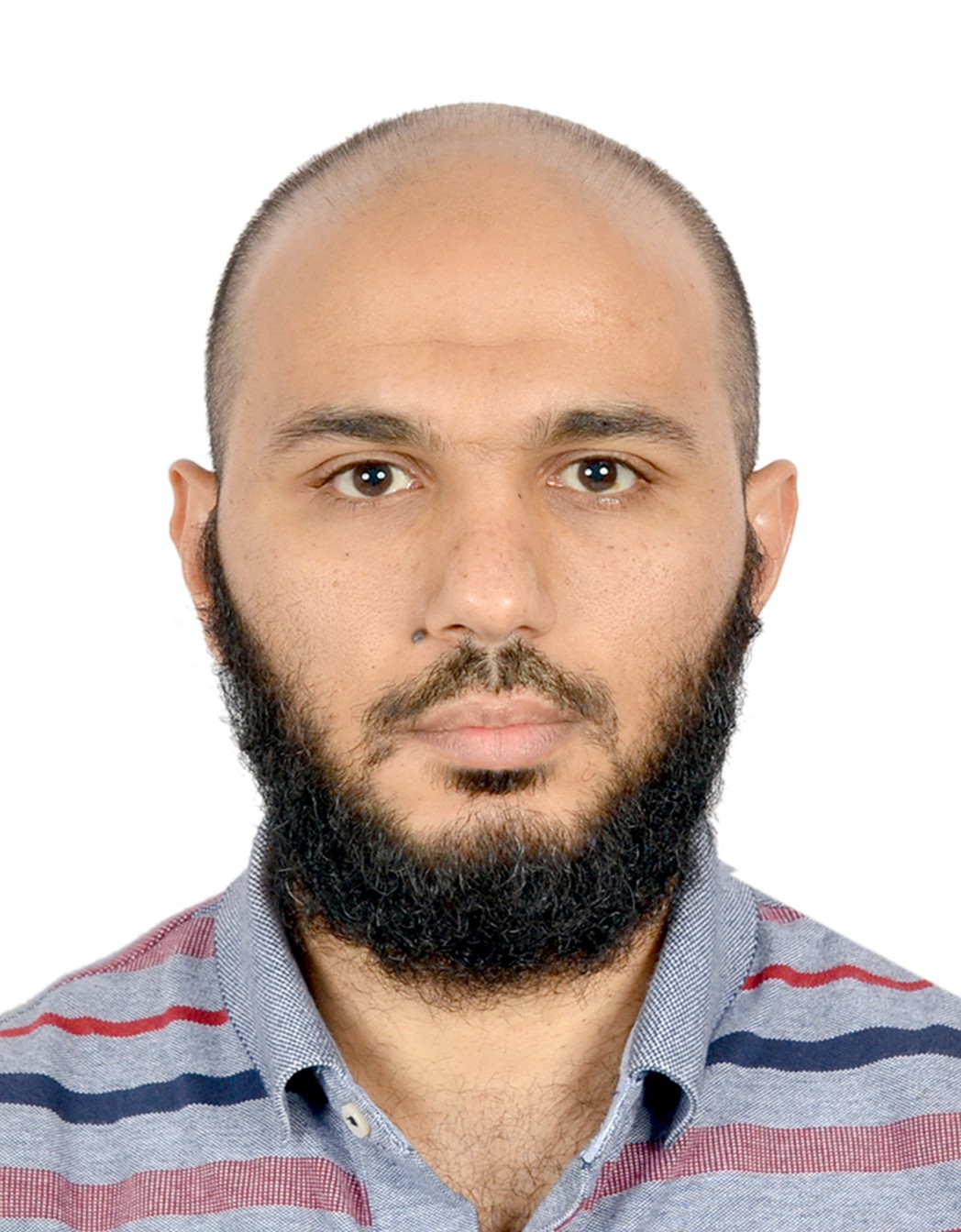}}]
{Hesham ElSawy} (Senior Member, IEEE) an Associate Professor with the School of Computing, Queen's University, Kingston, ON, Canada. Prior to that, he was an assistant professor at King Fahd University of Petroleum and Minerals (KFUPM), Saudi Arabia, a Post-Doctoral Fellow at the King Abdullah University of Science and Technology (KAUST), Saudi Arabia, a Research Assistant at TRTech, Winnipeg, MB, Canada. He received the Ph.D. degree in electrical engineering from the University of Manitoba, Canada, in 2014. He conducts research in the broad area of wireless communications and networking with a special focus on 5G/6G networks, joint communications and sensing, Internet of Things, edge computing, non-terrestrial networks, and wireless security. Dr. Elsawy is a recipient of the IEEE ComSoc Outstanding Young Researcher Award for Europe, Middle East, and Africa Region in 2018. He also received several best paper awards including the IEEE COMSOC Best Tutorial Paper Award in 2020 and IEEE COMSOC Best Survey Paper Award 2017. He is an Editor of the IEEE Transactions on Wireless Communications, the IEEE Transactions on Network Science and Engineering, and the IEEE Communications Letters. 

\end{IEEEbiography}
\vspace{-2cm}
\begin{IEEEbiography}
[{\includegraphics[width=1in,height=1.25in,clip,keepaspectratio]{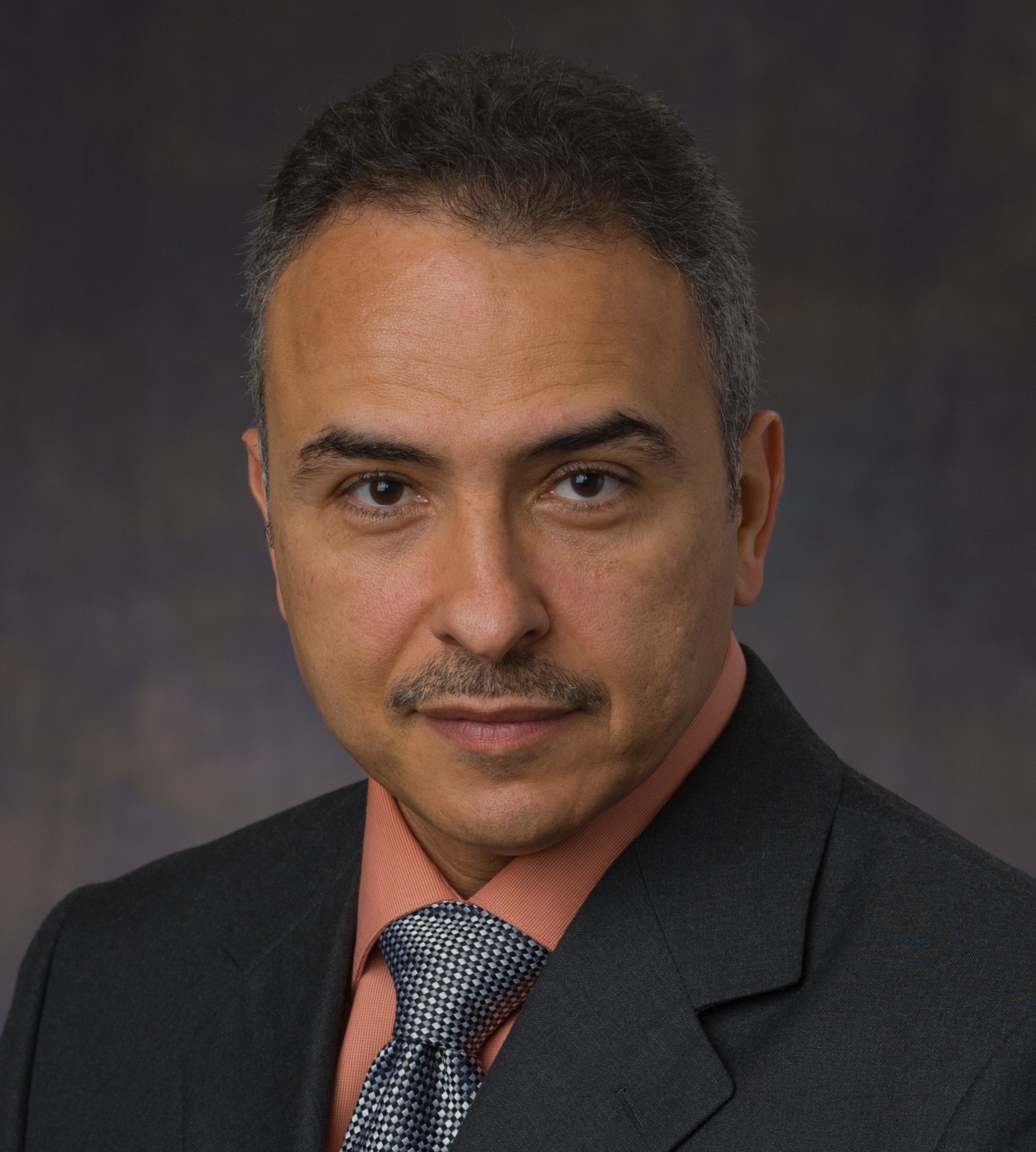}}]
{Hossam~S.~Hassanein}  (Fellow, IEEE) is currently a leading Researcher in the areas of broadband, wireless and mobile networks architecture, protocols, control, and performance evaluation. His record spans more than 700 publications in journals, conferences, and book chapters, in addition to numerous keynotes and plenary talks in flagship venues. He has received several recognition and best paper awards at top international conferences. He is the Founder and the Director of the Telecommunications Research Laboratory (TRL), School of Computing, Queen’s University, with extensive international academic and industrial collaborations. He is a recipient of the 2016 IEEE Communications Society Communications Software Technical Achievement Award for outstanding contributions to routing and deployment planning algorithms in wireless sensor networks and the 2020 IEEE IoT, Ad Hoc and Sensor Networks Technical Achievement and Recognition Award for significant contributions to technological advancement of the Internet of Things, ad hoc networks, and sensing systems. He is the former Chair of the IEEE Communication Society Technical Committee on Ad hoc and Sensor Networks (TC AHSN). He is an IEEE Communications Society Distinguished Speaker [a Distinguished Lecturer (2008–2010)].
\end{IEEEbiography}
\end{document}